\documentclass[twocolumn,english,aps,prx,10pt,superscriptaddress,floatfix,reprint]{revtex4-2}

\usepackage{times}
\usepackage{bm}% bold math
\usepackage{graphicx}
\usepackage{amssymb,amsfonts,amsmath,amsbsy,bm,t1enc,latexsym}
\usepackage{float}

\usepackage[colorlinks=true,citecolor=blue,linkcolor=magenta]{hyperref}

\usepackage[english]{babel}
\usepackage{url}%

\usepackage{booktabs}
\usepackage{multirow}

\usepackage{color,soul}
\usepackage{textcomp}
\usepackage[]{lipsum}
\usepackage[flushleft]{threeparttable}

\usepackage[markup=blue, authormarkupposition=left]{changes}

\usepackage{layouts}
\newcommand{\BTO}[0]{BaTiO$_3$}
\newcommand{\LN}[0]{LiNbO$_3$}
\newcommand{\STO}[0]{SrTiO$_3$}

\newcommand{\SiO}[0]{SiO$_2$}

\usepackage{placeins,enumerate}

\usepackage{xr}
\usepackage{braket}

\usepackage[separate-uncertainty=true]{siunitx}
\usepackage{physics}

\begin{document}

\title{
	Bidirectional microwave-optical conversion with an integrated soft-ferroelectric barium titanate transducer
	%Bidirectional microwave-optical conversion using an integrated barium-titanate transducer
}

\author{Charles M\"{o}hl}
\altaffiliation{The authors contributed equally to this work.}
\affiliation{
	IBM Research Europe, Zurich, S\"{a}umerstrasse 4, CH-8803 R\"{u}schlikon, Switzerland
}
%\affiliation{
%	Department of Physics, ETH Zürich, Zurich, CH-8093 Switzerland}

\author{Annina Riedhauser}
\altaffiliation{The authors contributed equally to this work.}
\affiliation{
	IBM Research Europe, Zurich, S\"{a}umerstrasse 4,  CH-8803 R\"{u}schlikon, Switzerland
}
%\affiliation{
%	Institute of Physics, EPF Lausanne, Lausanne, CH-1015 Switzerland}

\author{Max Glantschnig}
\altaffiliation{
	Present address: Infineon Technologies Austria AG, Siemensstra\ss e 2, 9500 Villach, Austria
}
\affiliation{
	IBM Research Europe, Zurich, S\"{a}umerstrasse 4, CH-8803 R\"{u}schlikon, Switzerland
}

\author{Daniele Caimi}
\affiliation{
	IBM Research Europe, Zurich, S\"{a}umerstrasse 4, CH-8803 R\"{u}schlikon, Switzerland
}

\author{Ute Drechsler}
\affiliation{
	IBM Research Europe, Zurich, S\"{a}umerstrasse 4, CH-8803 R\"{u}schlikon, Switzerland
}

\author{Antonis Olziersky}
\affiliation{
	IBM Research Europe, Zurich, S\"{a}umerstrasse 4, CH-8803 R\"{u}schlikon, Switzerland
}

\author{Deividas Sabonis}
\affiliation{
	IBM Research Europe, Zurich, S\"{a}umerstrasse 4, CH-8803 R\"{u}schlikon, Switzerland
}

\author{David I. Indolese}
\affiliation{
	IBM Research Europe, Zurich, S\"{a}umerstrasse 4, CH-8803 R\"{u}schlikon, Switzerland
}

\author{Thomas M. Karg}
\altaffiliation{Contact author: thomas.karg@ibm.com}
\affiliation{
	IBM Research Europe, Zurich, S\"{a}umerstrasse 4, CH-8803 R\"{u}schlikon, Switzerland
}

\author{Paul Seidler}
\altaffiliation{Contact author: pfs@zurich.ibm.com}
\affiliation{
	IBM Research Europe, Zurich, S\"{a}umerstrasse 4, CH-8803 R\"{u}schlikon, Switzerland
}

\begin{abstract}
	Efficient, low-noise, and high-bandwidth transduction between optical and microwave photons is key to long-range quantum communication between distant superconducting quantum processors.
	Recent demonstrations of microwave-optical transduction using the broadband direct electro-optic (Pockels) effect in optical thin films made of AlN or LiNbO$_3$ have shown promise.
	To improve efficiency and added noise, materials with larger Pockels coefficients, such as the soft ferroelectrics BaTiO$_3$ or SrTiO$_3$, are required.
	However, these materials require adapted designs and fabrication approaches due to their nonlinear and, in some cases, hysteretic electro-optic response.
	Here, we engineer an on-chip, triply resonant transducer comprising low-loss BaTiO$_3$-on-SiO$_2$ waveguides monolithically integrated with a superconducting microwave resonator made of Nb.
	We demonstrate bidirectional microwave-optical transduction and reach total off-chip efficiencies of $1\times10^{-6}$ using pulsed pumping.
	Our novel device concept permits in-situ poling of the ferroelectric material without introducing excess microwave loss, using a fully subtractive fabrication process with superconducting air bridges.
	In addition, we investigate optically induced heating, revealing fast thermalization and quasiparticle resilience of the microwave resonator. 
	Our transducer concept and fabrication process are applicable to other materials with a large bias-induced Pockels effect and pave the way for efficient, low-power quantum interconnects.
\end{abstract}

\maketitle

%\the\linewidth\\
%\printinunitsof{cm}\prntlen{\linewidth}

\section{Introduction}

\begin{figure*}[]
	\centering
	\includegraphics{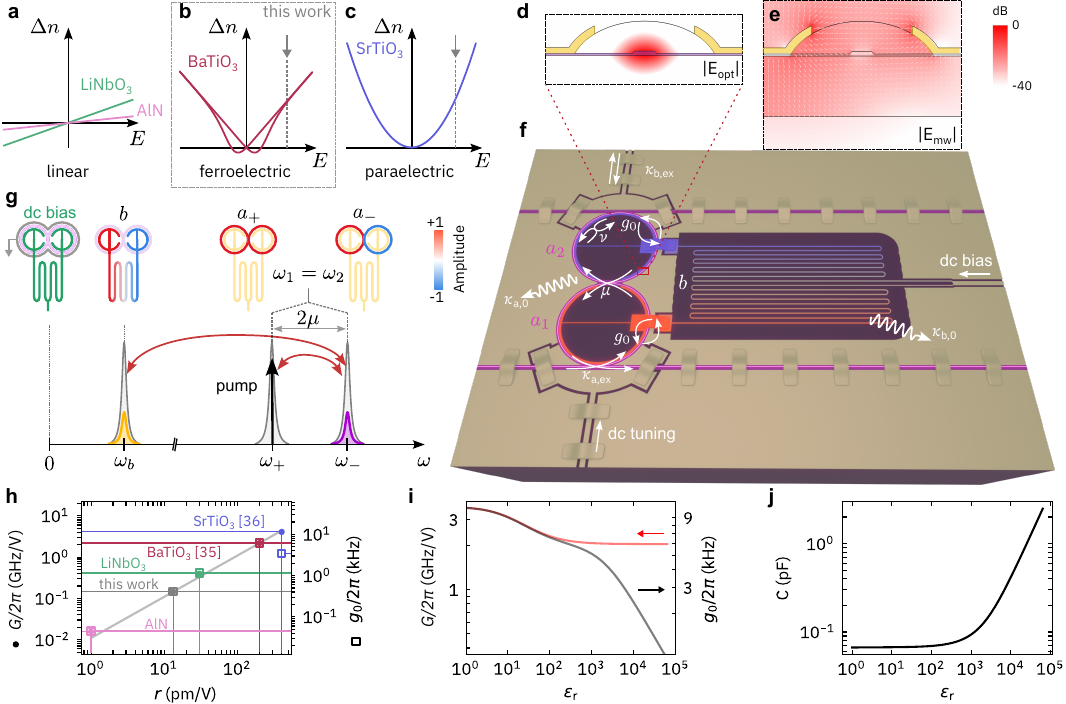}
	\caption{\textbf{Electro-optic transducer design for materials with nonlinear Pockels effect.}
		(a-c) Diagrams of refractive index change $\Delta{n}$ vs. applied dc electric field $E$ for (a) linear electro-optic materials AlN, \LN, and non-linear electro-optic materials: ferroelectric \BTO\ (b), and \STO\ in the quantum paraelectric phase (c).
		Grey arrows and dashed lines in (b) and (c) indicate working points at non-zero bias fields.
		(d, e) Simulated optical and microwave field amplitudes, respectively, in a cross section of \BTO\ ridge waveguide and superconducting electrodes.
		%(d) Cross-section of an optical ridge waveguide confining the optical mode to an electro-optic material and superconducting electrodes across which dc and microwave fields are applied to induce a refractive index modulation $\Delta n$.
		(f) Schematic of the integrated, triply resonant transducer consisting of a photonic molecule formed by two evanescently coupled optical ring resonators ($a_{1}$, $a_2$) and a $\lambda/2$-type microwave resonator ($b$).
		%. Two optical ring resonators ($a_1$, $a_2$) made of the electro-optic material are placed in the capacitive gap of a $\lambda/2$-type superconducting microwave resonator $b$.
		A voltage applied at the dc bias port poles the electro-optic material, and a voltage at the dc tuning port tunes the two ring resonators into resonance.
		(g) Scattering picture of triply resonant microwave-optical transduction.
		The hybridized photonic molecule modes $a_{\pm}$ produce an antisymmetric beating that is phase-matched to the microwave field distribution $b$.
		Pumping on the lower optical resonance coherently scatters photons between $b$ and $a_{-}$ when the microwave resonance frequency $\omega_\mathrm{b}$ matches the optical splitting $2\mu$.
		Shown is also the symmetric dc bias field poling the material.
		(h) Simulation of electro-optic frequency pulling factor $G = \partial\omega_\mathrm{a}/\partial{V}$ (circles) and single-photon coupling strength $g_\mathrm{0}$ (rectangles) as a function of the Pockels coefficient $r$ for different materials.
		(i) Simulated $G$ (red) and $g_\mathrm{0}$ (black) for $r=200$~pm/V as function of $\epsilon_\mathrm{r}$.
		(j) Total microwave resonator capacitance $C$ as function of $\epsilon_\mathrm{r}$.
		For (i) and (j) $\omega_\mathrm{b}/2\pi=6.8$~GHz is kept constant.
	}
	\label{fig1}
\end{figure*}

%Taming quantum mechanics for useful quantum computation is one of the great challenges of our time and 
Quantum computing promises to revolutionize many areas across academia and industry \cite{grumbling_quantum_2019}.
Because of their long coherence times, fast gates, and scalability, on-chip superconducting qubits represent a front runner for universal quantum computation  \cite{bravyi_future_2022}, and superconducting quantum computers are rapidly approaching practical quantum advantage.
%Beyond scaling to larger numbers of qubits in order to achieve practical quantum advantage, 
Eventually, additional functionality may be provided by the ability to send and receive quantum information between distant quantum computing units, allowing them to be interconnected in a quantum network.
However, operating at frequencies of $\sim5$~GHz, superconducting qubits must be cooled to temperatures of $\sim10$~mK in order to suppress thermal decoherence and ensure high-fidelity computation.
Despite efforts to create cryogenically cooled, low-loss, microwave quantum links over distances as long as 30 m \cite{storz_loophole-free_2023}, 
an entire system maintained at millikelvin temperatures is not a viable solution for long-range interconnects between quantum processing units.
%is required to the strict requirements to limit thermal noise and considerable propagation losses in waveguides at microwave frequencies pose a significant challenge for long-range interconnects between quantum processing units \cite{magnard_microwave_2020}.
\par
Optical photons on the other hand offer negligible thermal noise, high bandwidth, and orders-of-magnitude lower propagation losses at room temperature, using either optical fibers \cite{stolk_metropolitan-scale_2024, liu_creation_2024} or free-space transmission \cite{ren_ground--satellite_2017, huang_vacuum_2024}.
Optical quantum systems have thus been employed to pioneer entanglement distribution over hundreds of kilometers \cite{lago-rivera_telecom-heralded_2021, van_leent_entangling_2022, stolk_metropolitan-scale_2024, knaut_entanglement_2024, liu_creation_2024}, paving the way towards a quantum internet \cite{kimble_quantum_2008, wehner_quantum_2018}.
%theoretically enabling long-range interconnects over thousands of kilometers with large bandwidth and without the need for quantum repeaters \cite{huang_vacuum_2024}.
Recently, demonstrations of efficient, low-noise, microwave-optical photon interconversion \cite{chu_perspective_2020, lauk_perspectives_2020} have sparked the idea of networking superconducting quantum systems by means of optical quantum communication \cite{krastanov_optically_2021}.
%has sparked the field of quantum transduction \cite{lauk_perspectives_2020}. 
Central to reaching this goal is a quantum transducer capable of entangling microwave and optical photons with high fidelity and large bandwidth.
\par
%In the pursuit of realizing an integrated quantum transducer, 
Currently, the most advanced approaches to an integrated quantum transducer are piezo-opto-mechanical \cite{mirhosseini_superconducting_2020, honl_microwave--optical_2022, jiang_optically_2023, weaver_integrated_2024, meesala_quantum_2024, zhao_quantum-enabled_2024} and direct electro-optic \cite{fan_superconducting_2018, mckenna_cryogenic_2020, holzgrafe_cavity_2020, warner_coherent_2023} systems.
Whereas the former relies on an intermediary mechanical resonance serving as an  interface to both microwave and optical modes, the latter directly couples microwave and optical photons via the Pockels effect. 
Classically, direct electro-optic coupling can be described as a modulation of the optical refractive index %$\Delta\left(n_\mathrm{opt}^{-2}\right)=r\cdot E$
proportional to the Pockels coefficient $r$ and an applied microwave field.
% $E$
State-of-the art thin-film electro-optic platforms offer only comparatively low vacuum coupling rates $g_0$ on the order of $\sim1$~kHz  as opposed to $\sim1$~MHz for piezo-opto-mechanical systems.
Nevertheless, the direct electro-optic approach is appealing because it avoids the complexity and potential added noise of an intermediary mode and offers the possibility of larger bandwidth.
In bulk devices, electro-optic transduction has been demonstrated with efficiency close to unity and added noise below one photon \cite{sahu_quantum-enabled_2022}.
First proof-of-concept demonstrations have shown microwave-optical entanglement \cite{sahu_entangling_2023} as well as optical control \cite{warner_coherent_2023} and readout of a superconducting qubit \cite{arnold_all-optical_2023}.
In addition to scalability and higher efficiency for a given pump power, integrated devices offer the prospect of GHz bandwidths \cite{shen_photonic_2023}. 
\par
Following initial demonstrations using AlN ($r= 1$~pm/V) \cite{fan_superconducting_2018,fu_cavity_2021}, most transducer devices today employ optical waveguides made of the ferroelectric \LN\ ($r = 30$~pm/V) \cite{holzgrafe_cavity_2020,mckenna_cryogenic_2020,xu_bidirectional_2021,xu_light-induced_2022,warner_coherent_2023} (Fig.~\ref{fig1}a).
The primary route to improving efficiency has so far been the enhancement of the optical quality factor, with internal quality factors recently reaching $2.9\times 10^7$ ($1.93\times 10^8$) in optimized integrated \cite{zhu_twenty-nine_2024} (bulk \cite{hease_bidirectional_2020}) devices.
Although the transduction efficiency scales quadratically with loaded optical quality factor $Q_\mathrm{a}$, this gain comes at the cost of reduced optical bandwidth.
If a transduction bandwidth of $\sim 10$ MHz is to be maintained, $Q_\mathrm{a}$ is limited to about $\sim 2\times10^7$ at an optical vacuum wavelength of 1550~nm.
Beyond this point, more complex designs are required, such as mode-selective bus-coupling schemes \cite{hu_high-efficiency_2022}, that provide at best a linear dependence of efficiency on quality factor.
%is required to benefit from merely linear improvement by increasing the circulating pump power at the cost of pump bandwidth.
%An increase in intrinsic optical quality reduces dissipation at the device and thus permits the use of larger pump powers until dissipation at the fiber-chip interface limits further increase.
There is also a trade-off between optical quality factor and device footprint \cite{zhu_twenty-nine_2024}.
Hence, further advances using established materials in integrated devices will be challenging. 
%and lead to increasingly difficult trade-offs between reliability, manufacturability, device footprint, bandwidth, and added noise.
\par
There is therefore increasing interest in integrating alternative materials with larger Pockels coefficients, due to the quadratic scaling of transduction efficiency with $r$ \cite{tsang_cavity_2010}.
Stronger electro-optic materials would enable higher vacuum coupling rates and thus more efficient and broadband transducers operating at lower optical pump powers.
Promising candidates are the soft ferroelectric perovskites \BTO\ and \STO.
For \BTO, effective Pockels coefficients of $r=500$~pm/V and $200$~pm/V have been reported in thin films at 300~K  and 4~K, respectively \cite{eltes_integrated_2020}, while for \STO, values of up to $r=400$~pm/V were demonstrated in a bulk device at 4~K \cite{anderson_quantum_2023}.
Whereas single-crystalline \LN, a hard ferroelectric with a Curie temperature over 1300~K  \cite{smolenskii_curie_1966}, features a permanent polarization, a stable ferroelectric phase, and a linear electro-optic response in the temperature range of interest for quantum applications  (10 mK - 300 K) (Fig.~\ref{fig1}a), the large electro-optic response of \BTO\  and \STO\ stems from softening of the ferroelectricity in the vicinity of phase transitions, usually marked by a divergence in permittivity.
Indeed, the Curie temperature of \BTO\ is 420~K \cite{harwood_curie_1947}, and a phase transition occurs at around 240~K, accompanied by a peak in the electro-optic response ($r\approx700$~pm/V) \cite{eltes_integrated_2020}.
In \BTO, the Pockels effect is hysteretic
%accompanied by a hysteresis of the $\Delta{n}$-vs.-$E$ relationship 
due to ferroelectric domain poling (Fig.~\ref{fig1}b),
%In particular,
and maximizing the electro-optic response of \BTO\ requires a static electrical bias field to orient the domains \cite{abel_large_2019}.
In contrast, the Curie temperature of \STO\ is close to 0~K, giving rise to a quantum paraelectric phase near this temperature \cite{muller_srti_1979} and, in turn, a large quadratic electro-optic response \cite{fujii_interferometric_2003, anderson_quantum_2023} (Fig.~\ref{fig1}c).
Thus, both \BTO\  in its ferroelectric phase (Fig.~\ref{fig1}b) and \STO\ in its quantum-paraelectric phase (Fig.~\ref{fig1}c) require a dc-bias field to control and harness their non-linear electro-optic response for applications such as quantum transduction.
\par
Despite the promising electro-optic properties of soft ferroelectric materials, their use in integrated devices faces several challenges:
%The integration and use of this material class for electro-optic transduction is thus challenging for several reasons:
i)~There are often no commercially available (single-)crystalline thin films with the necessary optical quality.
ii)~The material properties depend on growth parameters, film thickness, strain, temperature and environmental conditions, such as the presence of oxygen or water vapor \cite{abel_large_2019,fredrickson_strain_2018}.
iii)~An electrical microwave circuit designed for dc-biasing of the active material is required to enable control of the non-linear electro-optic response.
\par
Here, we take a first step to address these challenges by developing the design and fabrication processes for an integrated triply resonant electro-optic transducer for soft ferroelectric thin films.
%Our triply resonant transducer design enhances the efficiency of the three-wave-mixing process, %includes electrical bias ports to pole the electro-optic material and tune the three-wave frequency-matching condition.
The transducer enables in-situ ferroelectric domain poling
%of the electro-optic material 
without incurring microwave losses, as well as independent tuning of the photonic resonances without sacrificing electro-optic mode overlap.
Using a device fabricated with \BTO, we show bidirectional continuous-wave (cw) and pulsed transduction with a peak efficiency of $1\times10^{-6}$ and linear behavior over a wide range of pump powers.
Further, we investigate optically induced heating of the microwave circuit.
Due to their different dynamics, we can distinguish fast in-cavity dielectric heating and slower substrate heating, as well as quasiparticle effects.
We find that in-cavity dielectric heating dominates over the other mechanisms at optical pump powers relevant for quantum transduction.

\section{Transducer design}

Our on-chip transducer consists of a photonic molecule integrated with a quasi-lumped-element superconducting microwave resonator (Fig.~\ref{fig1}f). 
Two optical ring resonators comprising \SiO-clad ridge waveguides (Fig.~\ref{fig1}d) are coupled evanescently at rate $\mu$ to form the photonic molecule \cite{holzgrafe_cavity_2020,mckenna_cryogenic_2020,xu_bidirectional_2021,xu_light-induced_2022,warner_coherent_2023}.
Using a voltage applied at the dc-tuning port (Fig.~\ref{fig1}f) the ring modes $a_{1}$ and $a_{2}$ are tuned in resonance ($\omega_\mathrm{1}=\omega_\mathrm{2}$) and hybridize into symmetric and antisymmetric normal modes $a_{\mathrm{\pm}} = (a_1 \pm a_2)/\sqrt{2}$ with frequencies $\omega_\mathrm{\pm} $, respectively (Fig.~\ref{fig1}g).
The microwave resonator is composed of two capacitors enclosing the optical waveguides (Fig.~\ref{fig1}e) and a connecting meandered inductor.
It is dimensioned to have an open-ended half-wavelength ($\lambda/2$) microwave field distribution (mode $b$) matching the antisymmetric optical beating $a_{+}^\dag a_{-} + a_{-}^\dag a_{+}$(Appendix Sec.~\ref{sec:mw_equiv_circuit}).
To achieve triply resonant frequency conversion, the microwave resonance frequency $\omega_\mathrm{b}$ should match the photonic molecule splitting $\omega_\mathrm{-}-\omega_\mathrm{+} = 2\mu$.
The dc-bias port (Fig.~\ref{fig1}f) is connected to the microwave resonator at the midpoint where the voltage amplitude vanishes. 
Designing the microwave mode to be mirror symmetric about a line through the bias port enables biasing and orientation of the ferroelectric domains without incurring excess microwave loss.
The symmetry of the layout is further maintained by having two optical bus waveguides pass along the upper and lower ring resonators.
While the lower bus waveguide is designed for critical optical coupling, the upper one is displaced a few microns from the upper ring resonator to avoid optical coupling.
Instead, it couples to another device on the chip.
For microwave input and output, a coplanar waveguide capacitively couples to the microwave resonator at the top.
In contrast to all previous integrated designs \cite{holzgrafe_cavity_2020,mckenna_cryogenic_2020,xu_bidirectional_2021,xu_light-induced_2022,warner_coherent_2023}, the dc-tuning electrode is capacitively coupled to the transducer electrode inside the rings instead of tuning  with respect to ground.
This means that the tuning section is also part of the microwave resonator and contributes to transduction, resulting in a record-high usable electrode coverage of 91~\%.
The tuning electrode as well as the microwave port electrode on the opposite side of the transducer are purposely wide to ensure both, strong optical dc-tuning and a large external coupling rate $\kappa_\mathrm{b,ex}$ to the microwave bus.

%%%%%% EO coupling
\par
The electro-optic interaction in our device is described by the Hamiltonian $H_\mathrm{int} = \hbar g_\mathrm{0} (a_{+}^\dag a_{-} + a_{-}^\dag a_{+}) (b+b^\dag)$ with vacuum electro-optic coupling constant $g_\mathrm{0} = G V_\mathrm{zpf}$.
Here, $G = \partial \omega_\mathrm{a}/\partial V$ is the frequency pulling factor and $V_\mathrm{zpf} = \sqrt{\hbar \omega_\mathrm{b} / (8 C)}$ is the microwave resonator's zero-point voltage across the ring capacitors (Appendix Sec.~\ref{sec:eo_coupling}).
Critical parameters determining the strength of $g_\mathrm{0}$ are the Pockels coefficient $r$ and the relative permittivity $\epsilon_\mathrm{r}$ of the electro-optic material, the microwave-optical mode overlap, and the effective transducer capacitance $C$.
We can estimate $g_\mathrm{0}$ via numerical simulations. First, an equivalent circuit is derived for the microwave resonator that is motivated by the $\lambda/2$ field distribution obtained from a fully three-dimensional finite-element simulation (Sec.~\ref{sec:mw_equiv_circuit}).  
This breaks down the problem into simulations of the electro-optic mode overlap and individual capacitances and inductances.
%We can then determine the effective capacitance of the transduction electrodes and electro-optic mode overlap as a function of the relative permittivity $\epsilon_\mathrm{r}$.
% to simulate different electro-optic materials.
Exploiting angular symmetry, the overlap integral is performed on a two-dimensional cross-section of the optical waveguide and transduction electrodes (Fig.~\ref{fig1}d, e).
\par
We thus compute $g_\mathrm{0}$ and $G$ as a function of $r$ for different electro-optic materials (Fig.~\ref{fig1}h).
The simulation predicts $g_\mathrm{0}/2\pi=0.05$~kHz for AlN ($r=1$~pm/V, $\epsilon_\mathrm{r}=9$) \cite{tian_piezoelectric_2024}, $g_\mathrm{0}/2\pi = 1.2$~kHz for \LN\ ($r = 30$~pm/V, $\epsilon_\mathrm{r} = 25$) \cite{tian_piezoelectric_2024}, and  $g_\mathrm{0}/2\pi = 6.1$~kHz for \BTO\ ($r=200$~pm/V, $\epsilon_\mathrm{r} = 200$ \cite{eltes_integrated_2020}).
For quantum paraelectric \STO, a value of $g_\mathrm{0}/2\pi \approx 3.4$~kHz is obtained from bulk values ($r=400$~pm/V, $\epsilon_\mathrm{r} = 2\times 10^4$) \cite{anderson_quantum_2023}, though the significantly larger biasing fields possible in integrated devices may enable larger values of $r$ and $g_0$.
The obtained value of $g_0$ for \LN\ is comparable to other published devices \cite{holzgrafe_cavity_2020, mckenna_cryogenic_2020,xu_bidirectional_2021,warner_coherent_2023}.
As expected, there is an approximate linear relationship between $g_\mathrm{0}$ or $G$ and $r$ (solid gray line in Fig.~\ref{fig1}h computed with $\epsilon_\mathrm{r} = 200$). Deviations from the linear relation are caused by the vastly different material permittivities, in particular that of \STO.
Large permittivity reduces the electro-optic mode overlap by suppressing the microwave field penetration of the optical ridge waveguide and increases the transducer capacitance.
\par
To elucidate the dependence on $\epsilon_\mathrm{r}$, we simulate $g_\mathrm{0}$ and $G$ as a function of $\epsilon_\mathrm{r}$ between $1$ and $10^5$ for the expected $r = 200$~pm/V for \BTO\ (Fig.~\ref{fig1}i), revealing a first drop ($\epsilon_\mathrm{r}\approx$100) for both and second ($\epsilon_\mathrm{r}>10^3$) for $g_\mathrm{0}$ while $G$ plateaus. Here, the microwave resonance frequency is kept constant by adapting the inductance to compensate the change in capacitance.
%which is usually large in strong-Pockels materials, e.g. $\epsilon_{r,\mathrm{LiNbO_3}}\approx 25$ \cite{}, $\epsilon_{r,\mathrm{BaTiO_3}}\approx 200$ \cite{}, and $\epsilon_{r,\mathrm{SrTiO_3}}\approx2\times10^4$) \cite{}.
At low $\epsilon_\mathrm{r} \approx 1$, the field penetrates the waveguide core both directly through the slab in contact with the electrodes and through the optical cladding ($\epsilon_\mathrm{r,SiO_2}\approx4$).
With increasing permittivity, the field penetration through the cladding becomes insignificant and plays virtually no role for permittivities above $10^2$ (Fig.~\ref{fig1}i), where the field strength in the waveguide core saturates to half of that in the slab, roughly corresponding to the ratio of the slab and total waveguide thicknesses.
This means that $G$ and $g_\mathrm{0}$ reduce by the same amount and reach a plateau at large $\epsilon_\mathrm{r} \sim 10^2$ (Fig.~\ref{fig1}i), a fact that should be considered in waveguide optimization.
For even larger $\epsilon_\mathrm{r}$ exceeding $2\times 10^3$ we see another drop in $g_\mathrm{0}$ caused by the increase of transducer capacitance (Fig.~\ref{fig1}j).
In this regime the capacitance of the transduction electrodes dominates over the stray capacitance and lowers $V_\mathrm{zpf}$.
This effect can be mitigated to some extent by making the device smaller.
\par
Our simulations show that this novel transducer design can leverage the large electro-optic effect offered by soft ferroelectrics to enable devices with significantly larger vacuum coupling strengths than achievable with established materials - that is, up to $5\times$ larger than with \LN, assuming reported values.
Despite the large permittivities of these materials, strong microwave-optical overlap is maintained by the use of a ridge waveguide with low ridge-slab contrast and electrodes in direct contact with the slab.

\section{Device fabrication and characterization}
\label{sec:implementation}

\begin{figure*}[]
	\centering
	\includegraphics{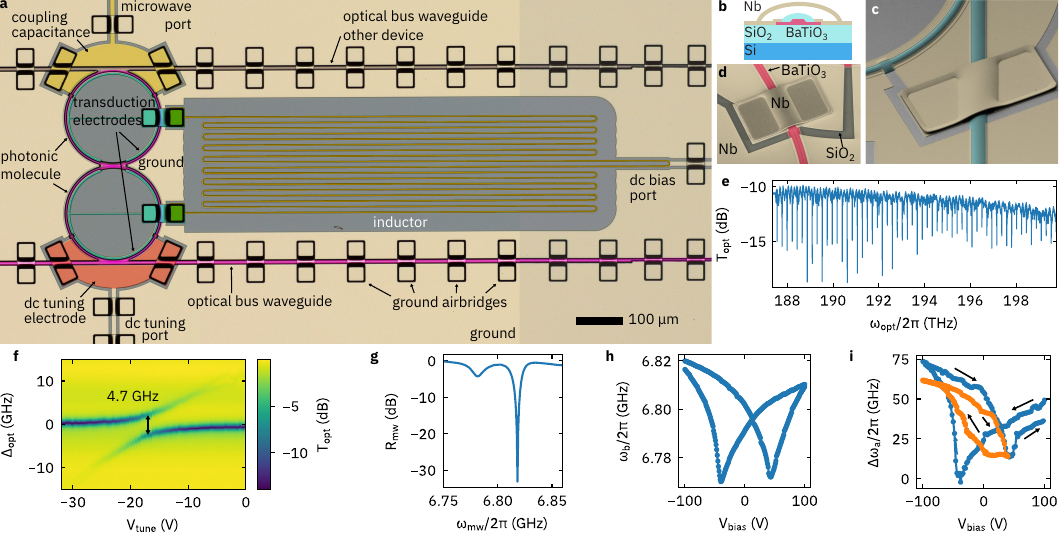}
	\caption{\textbf{DC-biased triply resonant photonic molecule transducer.}
		(a) False-colored optical micrograph of the photonic molecule transducer. 
		Optical signals are evanescently coupled from the optical bus waveguide to the lower ring of the photonic molecule, which is evanescently coupled to the upper ring at its opposite side. 
		Symmetric electrodes for dc-tuning (orange) of the photonic molecule and microwave bus coupling (yellow) are placed on each ring. 
		(b) Schematic cross-section of the measured device, showing \BTO\ ridge waveguides on \SiO\ on Si and \SiO\ cladding. Nb electrodes are in contact with the \BTO\ slab, with Nb air bridges across, where required. 
		(c) False-colored SEM micrograph of a representative Nb air bridge used throughout the design.
		(d) For illustration, an air bridge across an air-clad waveguide on another fabricated chip is shown. 
		(e) Optical spectrum with a free-spectral range of 218 GHz measured in transmission at room temperature after optical packaging.
		(f) Optical avoided crossing as a function of dc-tuning voltage showing photonic molecule hybridization with a ring-ring coupling of $2\mu=2\pi\times$4.7~GHz.
		(g) Microwave reflection measured at 10 mK at $V_\mathrm{bias} = -$100~V revealing the microwave resonance at 6.82 GHz.
		(h) Tuning curve of the microwave resonance frequency as a function of dc-bias voltage showing ferroelectric hysteresis (arrows indicate tuning direction). 
		(i) Tuning curve of the optical resonance measured in reflection at 75 mK. The blue data points show the initial hysteresis curve after cooldown which includes a drift due to material conditioning (open curve at 100~V).
		The orange data points show hysteresis when ramping the bias between $V_\mathrm{bias} = -$100~V and  $V_\mathrm{bias} =+$40~V, corresponding to the coercive field.
	}\label{fig2}
\end{figure*}

% short summary
Our transducer device (Fig.~\ref{fig2}a) is fabricated in a fully subtractive process (see Sec.~\ref{sec:sample_fabrication}).
Optical ridge waveguides are defined by dry-etching of a 225-nm \BTO\ film bonded to an oxidized Si substrate and are later cladded with \SiO.
The microwave circuit, including the air bridges, is made of sputtered Nb. 
%An \SiO\ cladding is deposited on top of the ridge waveguides.
The fabrication process differs from that of other transducer realizations \cite{fan_superconducting_2018,holzgrafe_cavity_2020,mckenna_cryogenic_2020,fu_cavity_2021,xu_bidirectional_2021,xu_light-induced_2022,warner_coherent_2023} in that lithographically defined superconducting air bridges are employed to enable optical waveguide-superconductor and superconductor-superconductor crossings for complex layouts.  The subtractive metallization fabrication in two steps using dry etching preserves sub-micron critical dimensions, \textit{e.g.}, for nanowire inductors in the main metallization layer. The air bridges are compatible with both oxide- (Fig.~\ref{fig2}c) as well as air-clad optical waveguides (Fig.~\ref{fig2}d). 
Note that air bridges are included in the current-carrying portion of the microwave resonator connecting the inductor with the transduction electrodes.
%We demonstrate that high-quality microwave resonators can be realized with air bridges as part of the center conductor by using them to connect the inductor with the transduction electrodes.
This is an advancement over the typical use of air bridges as patches between regions of the ground plane, through which no significant current flows.
A waveguide cross section (Fig.~\ref{fig2}b) showcases the dome-shaped cladding that enables a clean subtractive electrode fabrication instead of a lift-off, allowing thicker superconducting films. 
Without the dome shape, the sputter-deposited Nb layer would not be cleared cleanly from any vertical sidewall during dry-etching, risking the creation of a short-circuit.
\par
%For efficient edge coupling to the optical fiber, 
The \BTO\ optical bus waveguides lead to the chip facets, which are defined by dry etching deep trenches into the Si substrate.
Approaching the facet, the ridge waveguide is tapered to achieve good mode matching with an ultra-high-numerical-aperture (UHNA) optical fiber (Sec.~\ref{sec:packaging}). 
Broadband optical fiber-to-chip coupling with 200~nm bandwidth is achieved, as evidenced by the envelope of the optical transmission spectrum in Fig.~\ref{fig2}e, in contrast to the comparatively narrrowband grating couplers used in most transducer demonstrations to date \cite{mckenna_cryogenic_2020,holzgrafe_cavity_2020,warner_coherent_2023}.
\par
We initially characterize the optical transmission spectrum (Fig.~\ref{fig2}e) at room temperature.
Without bias or tuning voltages applied, a series of resonances of the lower ring resonator coupled to the optical bus waveguide are observed with a free-spectral range of 218 GHz (see Appendix \ref{sec:trandsuction_opt_charact} for further optical characterization).
Monitoring the spectrum of a selected resonance at 1603.5~nm as a function of tuning voltage (Fig.~\ref{fig2}f), we observe an avoided crossing between modes of the two ring resonators.
The coherent coupling rate of the photonic molecule is thus determined to be $2\mu = 2\pi\times 4.7$~GHz.
Because this value of $\mu$ is the largest observed and the closest to the microwave resonance frequency, this pair of coupled optical resonances is selected for transduction measurements.
\par
The chip is packaged for cryogenic measurements by aligning UHNA fibers and gluing them to the chip (Sec.~\ref{sec:packaging}).
Wirebonding the chip to a microwave printed-circuit board and enclosing the assembly inside a copper cavity completes the packaging. 
During sample handling before mounting in the dilution fridge, one of the two optical fiber connections to the chip broke accidentally.
Consequently, the packaged transducer could only be probed optically in reflection.
The reflected signal originates from moderate coherent back-scattering in the ring resonators. 
%Optical transmission spectra for the determination of optical parameters were carried out at room temperature before and after cooldown.
%It should be noted that the optical transmission was only accessible at room temperature in this study due to one of the two optical fiber gluing points breaking during sample handling, meaning that the assumed optical parameters stem from room temperature characterization before and after cooldown of the sample.
\par
The packaged transducer is cooled in a dilution refrigerator to a base temperature of 8 mK without any voltage applied to the device.
The microwave resonance is then probed in reflection (Fig.~\ref{fig2}g), at a bias voltage of $-$100~V, revealing the critically coupled transducer resonance at $\omega_\mathrm{b} = 2\pi\times 6.82$~GHz, close to the simulated value. A spurious mode at 6.78 GHz is also detected, whose frequency does not tune with temperature and might originate from the package (see Appendix~\ref{sec:transduction_mw_charact} for more detailed microwave characterization).
At single-photon microwave probe power levels (\textit{i.e.} $-$120~dBm), we find an internal microwave linewidth of $\kappa_\mathrm{b,0}=2\pi\times 10$~MHz ($Q_\mathrm{b,0} =7.0\times 10^2$), limited by dielectric loss.
The external coupling rate was extracted to be $\kappa_\mathrm{b,ex}=2\pi\times 7.2$~MHz ($Q_\mathrm{b,ex} =9.5\times 10^2$) for coupling to the microwave feedline.
\par
Ramping the bias voltage from $-$100~V to 100~V and back (Fig.~\ref{fig2}h), we observe tuning of the microwave resonance due to the change of permittivity with applied field, with 
pronounced hysteresis due to the ferroelectricity of \BTO.
We observe similar hysteretic tuning for the optical resonance (Fig.~\ref{fig2}i), albeit with more noise due to the measurement being made in reflection, which precludes an unambiguous assignment of the resonance frequency because of interference effects (Sec.~\ref{sec:transduction_mw_temp_dependence}).
The initially recorded optical tuning curve cycling between 100~V and $-$100~V also exhibits drift, as this measurement was taken immediately after cooldown.
The partial tuning curve cycling between $-$100~V and $+$40~V, the latter voltage corresponding to the coercive field at which ferroelectric domains are reoriented, shows a weaker hysteresis.
For transduction measurements, the device was biased at $-$100~V, where the ferroelectric domains are maximally oriented, achieving the strongest Pockels effect \cite{abel_large_2019}.
% Measurement and device parameters
%With the sample wiring shown in Fig.~\ref{fig2}d we can access the microwave reflection ($R_{MW}=|b_{out}/b_{in}|^2$) at cryogenic temperatures at the bias working point of $V_\mathrm{bias}=-100V$ (Fig.~\ref{fig2}e). Please refer to Supporting information \ref{secS:MW_characterization} for more details on the microwave properties their power and bias dependence. The optical transmission at room temperature as a function of tuning voltage is shown in Fig.~\ref{fig2}f, revealing the photonic molecule anticrossing with a resonant splitting $2\mu/2\pi=4.7$ GHz at the wavelength used for the transduction measurements in this study (1603.5 nm). Further characterization of the optical can be found in Supporting information \ref{secS:opt_characterization}.

\section{Bi-directional continuous-wave transduction}
\label{sec:bidirectional}

\begin{figure*}[]
	\centering
	\includegraphics{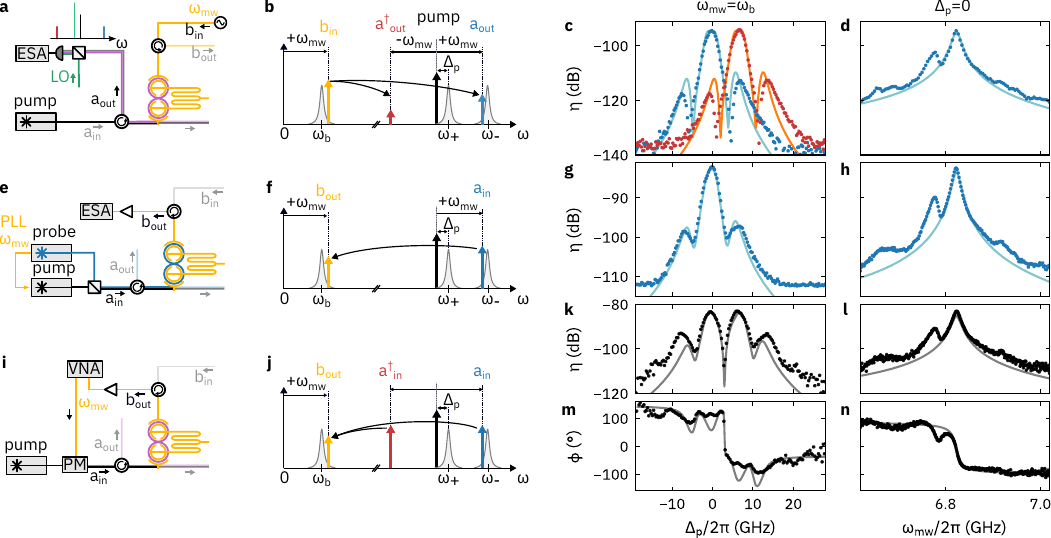}
	\caption{\textbf{Bidirectional coherent triply resonant microwave-optical transduction.}
		(a-d) Microwave-to-optical transduction.
		(a) Setup in which the transducer is driven by a microwave tone $b_\mathrm{in}$ and an optical pump.
		The optical signal $a_\mathrm{out}$ in backward direction is combined with a local oscillator (LO) and detected using heterodyne detection on an electronic spectrum analyzer (ESA).
		(b) Scattering diagram showing the microwave and optical resonances, the  microwave tone ($b_\mathrm{in}$, yellow), and the transduced optical Stokes ($a^\dagger_\mathrm{out}$, red) and anti-Stokes ($a_\mathrm{out}$, blue) sidebands.
		The optical pump is detuned by $\Delta_\mathrm{p}$ from the lower photonic resonance $\omega_\mathrm{+}$.
		(c) Transduction efficiency of the Stokes (red) and anti-Stokes (blue) processes on microwave-resonance ($\omega_\mathrm{mw}=\omega_\mathrm{b}$) as a function of $\Delta_\mathrm{p}$.
		(d) Anti-Stokes transduction efficiency for optical pumping  at $\Delta_\mathrm{p}=0$ as a function of  $\omega_\mathrm{mw}$.
		(e-h) Single-sideband optical-to-microwave transduction.
		(e) Setup in which a probe laser is locked to blue of the pump laser at a frequency offset $\omega_\mathrm{mw}$ and both are driving the device at $a_\mathrm{in}$.
		The transduced signal  $b_\mathrm{out}$ is amplified and detected on an ESA.
		f) Scattering diagram showing probe laser photons $a_\mathrm{in}$ being scattered into the microwave mode at $\omega_\mathrm{mw}$.
		(g) and (h) show the transduction efficiency as a function of $\Delta_\mathrm{p}$ and $\omega_\mathrm{mw}$, respectively.
		(i-n) Coherent optical-to-microwave transduction:
		(i) Setup in which a vector network analyzer (VNA) drives a phase modulator (PM) at frequency $\omega_\mathrm{mw}$, creating coherent sidebands on the pump laser driving the transducer at $a_\mathrm{in}$.
		The transduced signal $b_\mathrm{out}$ is amplified and detected at the VNA. 
		(j) The blue and red phase modulation sidebands are transduced simultaneously to the microwave mode and interfere because of their fixed phase relation.
		(k) and (l) show the efficiency, and (m) and (n) the phase of the transduced signal as a function of $\Delta_\mathrm{p}$ and $\omega_\mathrm{mw}$, respectively.
		For all measurements, the optical pump power is $-$11~dBm in the fiber.
	}
	\label{fig:bidirectional_transduction}
\end{figure*}

Having characterized the optical and microwave resonances individually, we set out to measure microwave-optical transduction.
For these experiments, the device is poled at a bias voltage of $-$100~V. 
To achieve frequency-matching, the photonic-molecule splitting should be tuned using the dc-tuning port to match $\omega_\mathrm{b}$.
Without the availability of optical transmission spectroscopy to determine the photonic-molecule splitting at mK temperatures, we instead directly optimized the transduction efficiency using the dc-tuning voltage (Sec.~\ref{sec:transduction}).
\par
%% MW to optical
For microwave-to-optical transduction (Fig.~\ref{fig:bidirectional_transduction}a), we apply a microwave modulation field $b_\mathrm{in}$ at the microwave coupling port with frequency $\omega_{\mathrm{mw}}$ around $\omega_\mathrm{b}$.
The photonic molecule is pumped with a continuous-wave laser (frequency $\omega_\mathrm{p}$, power $-11$~dBm in fiber) near the resonance pair at $1603.5$~nm.
The electro-optic interaction drives both Stokes scattering $b \to a_{\pm}^\dag$ to a photon with lower frequency $\omega_\mathrm{p} - \omega_{\mathrm{mw}}$ as well as anti-Stokes scattering $b \to a_{\pm}$ to a photon with higher frequency $\omega_\mathrm{p} + \omega_\mathrm{mw}$ (Fig.~\ref{fig:bidirectional_transduction}b).
Using optical heterodyne detection allows us to distinguish both processes.
Due to the misalignment of the second optical fiber, we detect the transduced photons emitted in the backward direction ($a_\mathrm{out}$), \textit{i.e.}, in reflection, relying on residual optical backscattering in our device. 
\par
We first position the microwave tone on resonance $\omega_{\mathrm{mw}} = \omega_\mathrm{b}$ and measure the transduced optical signal as a function of optical pump detuning $\Delta_\mathrm{p} = \omega_\mathrm{p} - \omega_\mathrm{+}$ (Fig.~\ref{fig:bidirectional_transduction}c).
Normalizing the detected signals by the independently calibrated optical and microwave powers, as well as transmission losses, we extract the off-chip transduction efficiency $\eta$ (Appendix Sec.~\ref{sec:efficiency_calibration}).
We note that $\eta$ includes fiber-to-chip loss, estimated at $-5~$dB from room-temperature characterization.
The recorded efficiency curves for the Stokes and anti-Stokes processes are displaced in frequency with respect to each other by the photonic molecule splitting and reach a maximum around $\omega_\mathrm{p} = \omega_\mathrm{-}$ and $\omega_\mathrm{p} = \omega_\mathrm{+}$, respectively.
The traces display the characteristic response of a triply resonant device:
At the central peak, both optical pump and signal are resonant with respective modes of the photonic molecule.
Two side peaks with lower amplitude occur when either the optical pump or the signal are resonant with an optical mode.
These doubly resonant peaks are displaced from the triply resonant peak by the photonic-molecule splitting.
\par
Setting the pump laser to the lower photonic molecule resonance ($\Delta_\mathrm{p} = 0$), we next measure  transduction efficiency of the anti-Stokes sideband as a function of microwave frequency $\omega_{\mathrm{mw}}$ (Fig.~\ref{fig:bidirectional_transduction}d).
As expected, the data reveal the microwave resonance, including the previously observed spurious mode (Fig.~\ref{fig:MW_Vbias_maps}).
The transduction bandwidth is determined from the linewidth to be $\kappa_{\mathrm{b}} = 2\pi\times 14$~MHz.
\par
%% Optical-to-microwave
To measure the optical-to-microwave response of the system, we combine the optical pump with light from a blue-shifted probe laser that is phase-locked to the pump laser with a programmable frequency offset denoted $\omega_{\mathrm{mw}}$ (Fig.~\ref{fig:bidirectional_transduction}f). 
The light from both lasers is injected into the device in the forward direction ($a_\mathrm{in}$, Fig.~\ref{fig:bidirectional_transduction}e). 
The transduced microwave signal is then amplified using a high-electron-mobility transistor (HEMT) amplifier and detected using an electrical spectrum analyzer. 
As before, we measure the transduction efficiency as a function of optical pump detuning, keeping the pump-probe offset fixed at $\omega_\mathrm{mw} = \omega_\mathrm{b}$ (Fig.~\ref{fig:bidirectional_transduction}g). 
The data show a triply resonant response similar to that of the microwave-to-optical experiment.
However, the overall efficiency is more than 10~dB higher.  
We attribute this difference to the low backscattering efficiency in our device reducing the observed microwave-to-optical signal; the backscattering rate has no effect on the optical-to-microwave data. 
Fixing the optical pump signal at $\Delta_\mathrm{p} = 0$, a sweep of the pump-probe frequency offset $\omega_{\mathrm{mw}}$ again traces out the microwave resonance (Fig.~\ref{fig:bidirectional_transduction}h).

\par
%% Model
To understand the electro-optic response of the device, we treat the transducer as a linear system and compute its S-matrix (Appendix Sec.~\ref{sec:full_model}).
We model the data using theory curves determined by the S-matrix coefficients (Fig.~\ref{fig:bidirectional_transduction}), adjusting the photonic-molecule coupling strength $\mu$, the ring-resonator detuning $\Delta_\mathrm{rr}=\omega_1 - \omega_2$, the optical loss rates, as well as the coherent optical forward-backward coupling rates $\nu_{1,2}$.
The microwave loss rates are determined from independent spectroscopy (Fig.~\ref{fig2}g) and the vacuum electro-optic coupling strength $g_\mathrm{0}$ from electro-optic dc tuning (Sec.~\ref{sec:cryo_dc_tune}). 
All system parameters are summarized in Tab.~\ref{tab:transduction_model_parameters} of the appendix. 
\par
We find very good agreement between the theory curves and the data; the triply resonant response as well as the dips in between the doubly and triply resonant peaks are reproduced by the model.
We find that the reduced microwave-to-optical transduction efficiency of the signal emitted in the backward direction is well explained by a low forward-backward coupling strength of $\nu_{1,2} \approx 0.1 \kappa_\mathrm{a}$.
%which independently manifests itself in split optical resonances (Fig.~\ref{figS:optical_transmission_before_packaging}). 
However, since neither the peaks at triple or double resonance are split (Fig.~\ref{fig:bidirectional_transduction}c,g), intra-ring forward-backward coupling cannot clearly be distinguished from residual backscattering that may occur, \textit{e.g.}, at the end facet of the bus waveguide. 
%The slight splitting in the predicted main peak (Fig.~\ref{fig:bidirectional_transduction}c) is not observed experimentally and also depends on the phase of the forward-backward coupling $\nu$ w.r.t. the ring-ring coupling $\mu$.
\par
The measured bidirectional efficiencies
%measured as function of $\Delta_\mathrm{p}$ and $\omega_\mathrm{mw}$ (Fig.~\ref{fig:bidirectional_transduction} ) as well as pump power (Fig.~\ref{fig:pulsed})
are consistent with the vacuum electro-optic coupling rate of $g_\mathrm{0}= 2\pi\times406$~Hz extracted independently from dc-tuning of the optical resonance frequency (see Appendix \ref{sec:cryo_dc_tune}).
%Our simulations show that the observed DC-tunability of 145 MHz/V at cryogenic temperatures corresponds to a vacuum coupling strength of $g_\mathrm{0}= 2\pi\times406$~Hz. 
This corresponds to an effective material Pockels coefficient of $r=13$~pm/V, which is more than an order of magnitude lower than the value $r=200$~pm/V previously reported for \BTO\ thin films at cryogenic temperatures \cite{eltes_integrated_2020} (Fig.\ref{fig1}h).
Possible explanations include variations in material quality and processing but also a difference in strain in the \BTO layer. 
Indeed, it has been reported that the electro-optic response in perovskite titanate films, and strong electro-optic materials in general, depends on strain \cite{hamze_design_2020}.
This is confirmed by first-principles calculations for \BTO\ thin films \cite{fredrickson_strain_2018} and it has also been shown to affect the phase diagram of \STO\ \cite{xu_strain-induced_2020}.
Since those materials typically also feature large thermal expansion coefficients \cite{hamze_design_2020}, strain engineering is likely going to be crucial for optimal performance at cryogenic temperatures.

\par
Finally, we measure phase-coherent optical-to-microwave transduction using a vector network analyzer (VNA).
An external phase modulator driven by the VNA creates blue and red sidebands at $\pm\omega_\mathrm{mw}$ on the pump light injected into the device (Fig.~\ref{fig:bidirectional_transduction}i).
As the two optical sidebands have a fixed phase relation, \textit{i.e.}, $a_\mathrm{in} \propto i (e^{-i\omega_\mathrm{mw} t} + e^{i\omega_\mathrm{mw} t})$, their simultaneous transduction leads to interference, which becomes apparent when sweeping $\Delta_\mathrm{p}$ at fixed $\omega_\mathrm{mw} = \omega_\mathrm{b}$ (Fig.~\ref{fig:bidirectional_transduction}k). 
%the interference between the phase-modulation sidebands $i(a_\mathrm{in} - a_\mathrm{in}^\dag)$ becomes apparent. 
Now, two triply resonant peaks arise, with destructive interference occurring at a pump detuning symmetric with respect to the two photonic molecule modes due to 
the $\pi$-phase shift between the two phase-modulation sidebands, \textit{i.e.}, $a_\mathrm{in} = -a_\mathrm{in}^\dagger$.
In the phase response (Fig.~\ref{fig:bidirectional_transduction}m), a $\pi$ phase shift at the point of symmetric detuning is also observed and well captured by the model.
The smaller features, however, are exaggerated by the simulation, suggesting a slight deviation of the parameters from those of the actual device.
When the pump is tuned to $\omega_\mathrm{+}$ ($\Delta_\mathrm{p}=0$ ), the microwave frequency dependence of the transduced signal (Fig.~\ref{fig:bidirectional_transduction}l) is dominated by the blue sideband, as the red sideband is far off resonance, and therefore looks virtually identical to the previous measurement (Fig.~\ref{fig:bidirectional_transduction}h).
Additionally, the phase of the measured signal (Fig.~\ref{fig:bidirectional_transduction}n) agrees well with the predicted oscillator response, confirming that the process is coherent.

\section{Optically induced heating}

\begin{figure*}[]
	\centering
	\includegraphics{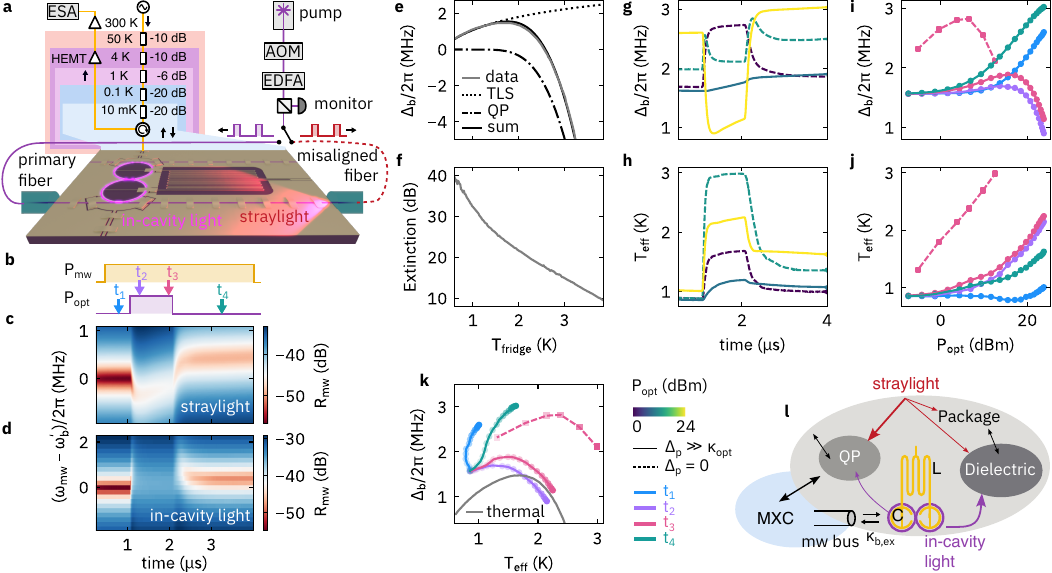}
	\caption{\textbf{Microwave response under optical pump.}
		(a) Experimental setup:
		A pump laser is pulsed using an acousto-optic modulator (AOM) and amplified with an erbium-doped fiber amplifier (EDFA).
		The pulse power is monitored using a fast photodetector and sent to the device either through the primary or misaligned optical fiber.
		Time-resolved microwave spectroscopy is performed using a pulsed high-power microwave source and an electronic spectrum analyzer (ESA).
		Light is guided to the chip either through the primary fiber with good coupling efficiency to the waveguide, or through the second, misaligned fiber, to study effects of in-cavity light or straylight, respectively.
		(b) Pulse sequence with the microwave pulse starting before and ending after the optical pulse.
		Arrows mark times at which microwave resonator properties were evaluated for (i,j,k).
		(c, d) Exemplary time-resolved microwave spectra with (c) off-resonant straylight pulses and (d) resonant, in-cavity light pulses, using in-fiber optical powers of 18 and 19~dBm, respectively.
		During the optical pulse, a shift and reduced extinction of the microwave dip are clearly observed, which partially recovers after the pulse.
		(e) Microwave resonator frequency shift $\Delta_\mathrm{b}$ as a function of cryostat temperature $T_\mathrm{fridge}$ without optical light.
		A theoretical model for $\Delta_\mathrm{b}$ that is a sum of quasiparticle (QP) and two-level system (TLS) contributions is fit to the data.
		(f) Microwave resonator extinction ratio as a function of cryostat temperature used to calibrate effective temperature $T_\mathrm{eff}$ in optical heating experiments.
		(g) Frequency shift $\Delta_\mathrm{b}$ and (h) effective temperature $T_\mathrm{eff}$ determined at each time during the pulse sequence.
		Different traces correspond to selected optical peak powers (colorbar) and off-resonant straylight (solid line) or resonant (dashed line) in-cavity light.
		(i) Frequency shift $\Delta_\mathrm{b}$ and (j) effective temperature $T_\mathrm{eff}$ as a function of optical peak power at different times in the pulse sequence (arrows in (b)).
		(k) Frequency shift vs. effective temperature for optical and thermal experiments.
		(l) Diagram of heating mechanisms:
		The microwave mode is coupled to three baths: quasiparticles (QP), dielectric, and the mixing chamber (MXC) through the microwave bus.
		Optical straylight primarily heats the QP bath, while in-cavity light predominantly causes fast heating of the dielectric at the capacitor gap.
		Both QP and dielectric baths are thermalized to the chip package which itself is cooled by the MXC.
	}\label{fig:heating}
\end{figure*}

We now investigate optical-pump-induced heating of the superconducting circuit, which is known to cause microwave decoherence, add noise, and limit the transduction efficiency \cite{xu_light-induced_2024}.
Tolerance to optical pumping is thus an important aspect of a quantum transducer.
%In this section optical heating effects in the transducer are investigated to shed light on the nature and impact of the heating and dissipation processes under optical pumping. 
%This is an important aspect of a transduction device for quantum applications, as heating usually limits the noise performance.
\par
%% RATIONALE
There are two main pathways through which optical photons can contribute to heating of the microwave circuit (Fig.~\ref{fig:heating}a):
(i) straylight created by scattering at primarily the fiber-chip interface irradiates the circuit, or 
(ii) resonant light coupled to the photonic molecule is absorbed in the dielectric or scattered to the superconductor.
%(ii) light is coupled into the optical bus waveguide and, when resonant, to the photonic molecule where it is absorbed or scattered to the superconductor.
Whereas straylight causes a near homogeneous irradiation of both the capacitors and the inductor, in-cavity light is confined to the optical rings and predominantly affects the capacitors.
Due to these distinct intensity distributions, straylight and in-cavity light result in different heating dynamics. 

\par
%% EXPERIMENT
We probe the possible heating mechanisms using pulsed optical pumping. 
Laser pulses with a duration of $1 ~\mathrm{\mu} s$ and a repetition rate of $1~$kHz are generated by an acousto-optic modulator (AOM), amplified by an erbium-doped fiber amplifier (EDFA) and sent to the device (Fig.~\ref{fig:heating}a).
%The effects of straylight and in-cavity light are distinguished by controlling the optical detuning $\Delta_\mathrm{p}$ and the fiber-chip coupling efficiency.
To investigate straylight effects, we deliver off-resonant light through the second, misaligned optical fiber such that no light is coupled to the on-chip waveguide (Fig.~\ref{fig:heating}a).
We note that the fiber-chip coupling point of this second fiber is located closer to the device than that of the primary fiber with good coupling efficiency.
To study heating from in-cavity light, we couple light through the primary fiber and step the optical detuning $\Delta_\mathrm{p}$ across the resonances.
\par
The microwave resonator's response before, during, and after the optical pulses is probed using microwave pulses synchronized to the optical pulses (Fig.~\ref{fig:heating}b). 
We record reflected microwave power as a function of time and probe frequency $\omega_\mathrm{mw}$.
Microwave spectra for straylight (Fig.~\ref{fig:heating}c) and in-cavity light (Fig.~\ref{fig:heating}d) both show a reduction in the resonant extinction during the optical pulse, but a red-shift versus a blue-shift of the resonance, respectively.
By fitting the spectra at each time bin, we determine the time evolution of both the frequency shift and the extinction of the resonance dip. 
%SI: Figs.~\ref{fig:heating}c and d show the frequency shift and extinction, respectively, measured as a function of time and optical pump wavelength using the primary fiber. During the optical pulse ($t=1.1$ to $2.1~\mu$s), the photonic molecule doublet is observed in terms of two peaks in the microwave shift and dips in the extinction. At large detunings the frequency shift and extinction level off to a plateau. After the pulse, shift and extinction decay to an intermediate state that is largely independent of optical detuning.

%{\color{red} Can we compare detuned pumping using both fibers with same power? It should give more or less the same result} --> Yes

\par
%% CALIBRATION
To translate the observed time dependence of the frequency shift and microwave resonator damping rate into a dependence on effective temperature, we separately measure the microwave resonance shift (Fig.~\ref{fig:heating}e) and extinction (Fig.~\ref{fig:heating}f) as a function of cryostat temperature.
The thermal data show an initial resonance blue-shift with increasing temperature that turns into a red-shift for temperatures above 2~K (Fig.~\ref{fig:heating}e, see Appendix Sec.~\ref{sec:transduction_mw_temp_dependence}  for details).
While a red-shift is expected  from Bardeen-Cooper-Schrieffer (BCS) theory of quasiparticles in a superconductor \cite{peruzzo_surpassing_2020}, the blue-shift is not.
Instead, a blue-shift has been reported to originate from a temperature-dependent contribution to the dielectric constant from two-level systems (TLS) in the dielectric \cite{barends_contribution_2008}.
Here, it most likely stems from TLSs in \BTO\ and the \SiO\ optical cladding in the transducer capacitor gap.
We fit a theory curve to our data that is the sum of the dielectric and the BCS models (Fig.~\ref{fig:heating}e and Appendix Sec.~\ref{sec:transduction_mw_temp_dependence}), showing very good agreement, and also plot the individual contributions. 
A determination of the microwave quality factor was complicated by the hybridization with the spurious mode (Sec.~\ref{sec:cryo_bias_dependence}), and would be subject to systematic errors.
Instead, we use the relationship between the measured extinction of the microwave resonance and the cryostat temperature $T_\mathrm{fridge}$ (Fig.~\ref{fig:heating}f) as a calibration for the effective non-equilibrium temperature ($T_\mathrm{eff}$) during optical pumping.
Since the measured decrease in extinction with temperature is due to a resonance broadening, it is consistent with quasiparticle-induced losses which also tend to increase with temperature, whereas TLS-induced losses tend to decrease \cite{crowley_disentangling_2023}.
This points to $T_\mathrm{eff}$ being dominated by the quasiparticle bath, while the temperature of the dielectric may be different.
In the following, we use the thermal model to identify dielectric and quasiparticle heating with blue and red frequency shifts, respectively. 

%% RESULTS
\par

% Time dependence
%We record time-resolved microwave reflection spectra and extract the microwave frequency shift $\Delta_\mathrm{b}$ (Fig.~\ref{fig:heating}e) and microwave extinction, which we then convert into an effective device temperature $T_{eff}$ (Fig.~\ref{fig:heating}f) by interpolation of Fig.~\ref{fig:heating}b. 
First, we look at the dynamics of the frequency shift (Fig.~\ref{fig:heating}g) and effective temperature (Fig.~\ref{fig:heating}h).
Traces for resonant light (primary fiber) and off-resonant light (secondary fiber) at various optical powers are shown.
Off-resonant light at low power causes a blue shift and a weak temperature increase with a time constant $\sim 1~\mu$s.
At high power, the temperature increase becomes much faster, following essentially the rise of the optical power ($\sim 0.1~\mu$s).
The frequency now shifts strongly to the red starting from a blue-shifted background level. 
Resonant light causes a stronger temperature increase during the optical pulse with fast rise time, quasi instantaneously following the optical pulse shape.
At low power, the frequency is strongly blue-shifted during the pulse.
At high power, a red-shift with slightly lower rise time superimposes on the blue-shift, as seen by the peaks at the beginning and end of the optical pulse. 
All traces show a strong rise of the background temperature after the optical pulse due to in-pulse heating (Fig.~\ref{fig:heating}h), proportional to optical power, which is slowly decaying to a steady-state (i.e. before pulse) at a rate of $\sim 10$~kHz (Appendix Sec.~\ref{sec:slow_heating}) by thermalization with the substrate.
The background frequency shift also increases with optical power (Fig.~\ref{fig:heating}g), but the difference between levels after and before the pulse is less pronounced.
Part of the background level is caused by continuous output noise of the EDFA (average power $-22$~dBm).

% Power dependence
We next analyze the optical-power dependence of $\Delta_\mathrm{b}$ (Fig.~\ref{fig:heating}i) and $T_\mathrm{eff}$ (Fig.~\ref{fig:heating}j) between $-7.5$ and $+24$~dBm at selected times before, during and after the optical pulse (Fig.~\ref{fig:heating}b).
%The in-fiber optical peak power $P_\mathrm{opt}$ is varied between $-7.5$ and $+24$~dBm.
For off-resonant light, we observe increasing blue-shifts up to power levels of $\sim 15$ dBm at all times during the pulse sequence.
Increasing power further turns this into a red-shift when the pulse is on, and enhances the blue shift when the pulse is off. 
Resonant light produces a strong initial blue shift at low powers with a peak at $\sim3$~dBm after which a red-shift starts to set in.
Compared to straylight, in-cavity light creates a 50\% larger peak blue-shift at about 10~dB less power.
The temperature increase for both in-cavity light and sufficiently strong off-resonant light appears to scale with the logarithm of the optical power (Fig.~\ref{fig:heating}j). 
Resonant light causes a more than 10-fold stronger temperature increase from the background level, which seems to be consistent with the peak blue-shift being reached at $1/10$ the optical power. 

From the dynamics and power-dependence we conclude that there are three distinct heating processes in the device (Fig.~\ref{fig:heating}l):
%(i) slow dielectric heating (blue-shift) of the substrate from scattered straylight,
%(ii) fast quasiparticle heating (red-shift) from straylight, and
%(iii) fast dielectric heating from in-cavity light.
(i) Slow dielectric heating (blue-shift) of the substrate due to straylight manifests at low power and causes a raised temperature level after the optical pulse, when quasiparticles have decayed.
(ii) Straylight photons hitting the inductor produce quasiparticle heating (red-shift) with a fast response time due to the high quasiparticle recombination rate in Nb.
(iii) In-cavity light results in dielectric heating (blue-shift) with fast rise- and fall-times, which may be explained by its strong confinement to the ring resonators inside the capacitor gaps.
Because of this confinement, in-cavity light contributes stronger dielectric heating and has a lower probability of generating quasiparticles in the relatively distant inductor where they influence microwave coherence the most.
The latter is aided by the fast quasiparticle recombination rate of Nb.

\par
These data highlight a transition from dielectric heating (blue-shift) dominating at low optical powers to quasi-particle-dominated heating (red-shift) at resonant optical powers above 5-10 dBm.
Such high optical powers already go along with fast in-pulse heating to temperatures well above 1~K Fig.~\ref{fig:heating}j).
These findings are insightful for quantum applications, as it implies that in the quantum-enabled regime ($k_\mathrm{B} T_\mathrm{eff} \ll \hbar\omega_\mathrm{b}$ or $P_\mathrm{opt} \ll 0$~dBm), in-cavity dielectric-heating dominates over quasiparticle heating of the superconductor.
This result is different to a previous study on a device with a 50 nm NbN film, where off-resonant straylight at the chip interface was concluded to be the dominant heating source \cite{fu_cavity_2021}, suggesting that the device studied here is more resilient against straylight, likely due to the use of a thicker superconducting film.
Note, that due to the fast sub~$\mu$s response of the local dielectric heating, it cannot be mitigated by reducing the duty cycle of the pump.
This emphasizes the importance of reducing optical absorption loss as much as possible.

% Shift vs temperature
\par
The different relative weights of dielectric and quasiparticle effects for in-cavity light and straylight are illustrated by plotting $\Delta_\mathrm{b}$ vs. $T_\mathrm{eff}$ (Fig.~\ref{fig:heating}k).
For straylight, the traces for pulse-off show large frequency blue-shift at low temperature increase.
The straylight data traces for pulse-on tend to move towards the curve described by the thermal data.
For resonant in-cavity light, however, the trace is shifted to both larger blue-shift and higher effective temperature.
This can be interpreted in the sense that straylight causes global heating of the chip that corresponds most closely to an equilibrated thermal state.
In-cavity light, however, deviates strongly from the thermal equilibrium because it generates a localized heat source with a stronger dielectric component.
An additional reason for the large temperature offset between dielectric and quasiparticles may be the fact that the quasiparticle bath is cooled by the circulator and isolator at the mixing chamber plate (MXC) via the microwave bus line, acting as a heat sink \cite{xu_radiatively-cooled_2023}.

% heating model
\par
The suggested heating pathways are summarized in Fig.~\ref{fig:heating}l, showing the quasiparticle (QP) bath, which mostly affects the inductive part of the resonator (L) and the dielectric bath, which primarily affects the capacitor (C).
Both baths are driven with different amplitudes by straylight and in-cavity light, and decay to the surrounding chip substrate and package.
Finally, the microwave bus couples a cold bath at the cryostat base temperature to the resonator.
A detailed analysis of the bath temperatures, coupling rates, and resulting added noise of our transducer goes beyond the scope of this work and will be subject of future studies \cite{ren_two-dimensional_2020, xu_light-induced_2024}.
%The effective thermal occupation of the microwave resonator is given by a sum of the different bath spectral densities weighted by the associated coupling rates %\cite{marquardt_quantum_2007}.

\section{Pulsed operation and power dependence}

\begin{figure}[]
	\centering
	\includegraphics{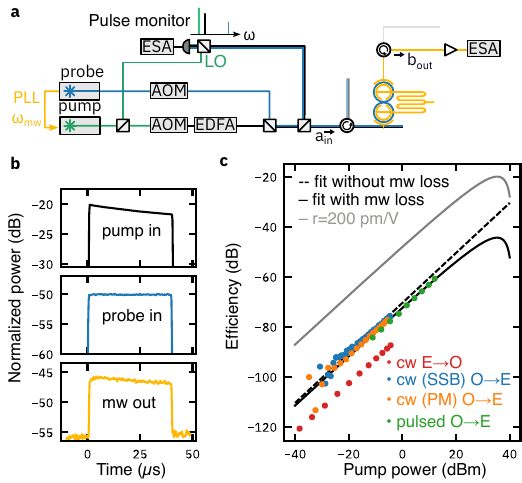}
	\caption{\textbf{Pulsed transduction and power dependence.}
		(a) For pulsed optical-to-microwave transduction, a blue-detuned probe laser is locked to the pump laser at a frequency offset $\omega_\mathrm{mw} = \omega_\mathrm{b}$.
		Each laser is pulsed using an acousto-optic modulator (AOM), and the pump laser is additionally amplified with an erbium-doped fiber amplifier (EDFA).
		Both are combined and sent to the device via $a_\mathrm{in}$.
		The transduced signal $b_\mathrm{out}$ is amplified and detected with a spectrum analyzer (ESA).
		A fraction of the optical input signals is combined with a local oscillator (LO) and detected via a photodetector and an ESA using heterodyne detection for pulse monitoring.
		(b) Pulse monitor time traces for the pump (black), probe (blue) and transduced microwave signal pulse (yellow).
		(c) CW microwave-to-optical (red dots) and optical-to-microwave (blue and orange dots) total efficiency versus optical pump power, using heterodyne detection, two offset-locked lasers and a phase modulator, respectively.
		(green dots) Pulsed optical-to-microwave total efficiency versus optical pump power using two phase-locked lasers.
		The black line is a model prediction of the efficiency vs. optical pump power taking into account the degradation of the microwave quality factor determined in Fig.~\ref{fig:MW_data_and_model}c.
	}\label{fig:pulsed}
\end{figure}

Finally, we investigate the linearity of the transduction efficiency with optical pump power in both pulsed- and continuous wave operation. 
For pulsed optical-to-microwave transduction we combine amplified pump pulses with pulses of a weaker probe laser, that is frequency-locked to the blue sideband of the pump laser (Fig.~\ref{fig:pulsed}a).
A fraction of the pump and probe pulses are measured using a calibrated heterodyne detector to determine their respective power levels (Fig.~\ref{fig:pulsed}b and Appendix Sec.~\ref{sec:efficiency_calibration}). 
In this experiment, with increasing pump power the pulse length is decreased from 1~ms to 40~$\mu$s to both, maintain good signal-to-noise ratio at low pump power and decrease the duty cycle to keep the average power constant.
The repetition rate was fixed at $0.8$~kHz.
The transduced microwave pulses are amplified and recorded on an electronic spectrum analyzer (Fig.~\ref{fig:pulsed}b).
Note that the shapes of the pump and probe pulses are imprinted onto the transduced signal.
\par
Fig.~\ref{fig:pulsed}c shows off-chip conversion efficiency as a function of off-chip optical peak pump power for cw microwave-to-optical and optical-to-microwave transduction, as well as pulsed optical-to-microwave transduction.
Clearly, all optical-to-microwave data agree within 3~dB and follow a linear trend from pump powers of $-30$~dBm to $+12$~dBm.
As discussed in Sec. \ref{sec:bidirectional}, cw microwave-to-optical transduction has a lower total efficiency due to inefficient signal extraction in backward direction.
\par
The pulsed optical-to-microwave transduction data shows clean, linear behavior up to the largest pump power of $+12.5$ dBm reaching an off-chip efficiency of $1\times 10^{-6}$.
Higher pump powers could not be delivered to the device, as the fiber-chip efficiency degraded suddenly and irreversibly, likely due to thermal misalignment of the fiber where it was glued to the chip. 
%Another explanation could be degradation of the optical epoxy at high powers, which was not investigated further.
Damage to the transducer device itself can be ruled out since optical and microwave quality factors did not degrade.
%The time-resolved detected beat signals of the EDFA-amplified optical pump signal (Fig.~\ref{fig:pulsed}b) and signal pulse (c) are shown, as well as the detected transduced microwave signal (d). Note that the pulse shape of the pulse is imprinted onto the transduced signal.
\par
To put the reported efficiencies into perspective, we plot a linear trend (Fig.~\ref{fig:pulsed}c) corresponding to a total-efficiency-per-pump $\eta/P_\mathrm{pump}=-73$~dB/mW, which is about five orders of magnitude lower than the recently reported value for an optimized \LN\ transducer ($-23$~dB/mW) \cite{warner_coherent_2023}.
This difference can be largely attributed to the maturity of the \LN\ platform and its highly optimized optical quality factor, which enters the efficiency quadratically.
Given the relatively low Pockels coefficient found for our device, an improved \BTO\ transducer achieving the previously observed $r=200~$pm/V \cite{eltes_integrated_2020} would result in more-than 20~dB increase in transduction efficiency (Fig.~\ref{fig:pulsed}c).
We use the measured decrease of microwave quality factor with optical power (Fig.~\ref{fig:heating}f and Appendix Sec.~\ref{sec:transduction_mw_charact}) to estimate when optically-induced microwave decoherence limits transduction efficiency, and estimate that this happens only above 30~dBm, an optical power that is far beyond the range where low-noise transduction is possible.

\section{Conclusion}

We have presented an integrated, triply resonant electro-optic transducer using the soft ferroelectric \BTO, exploiting its bias-dependent Pockels effect.
In bidirectional microwave-optical conversion measurements, the efficiencies for both continuous-wave and pulsed pumping showed high linearity up to pump powers of $-$2 and $+$12 dBm, respectively.
Heating of the transducer under pulsed optical pumping was studied using time-resolved microwave spectroscopy, allowing us to distinguish quasiparticle and dielectric heating by their red and blue induced microwave frequency shifts, respectively. 
Overall, optical heating was found to be dominated by in-cavity light, with straylight contributing less than 10\% of the temperature increase.
Interestingly, for in-cavity light, in contrast to straylight, we observed stronger dielectric heating and weaker quasiparticle heating, which we attribute to the different spatial intensity distribution and the use of a thick superconducting film.
\par
Although the demonstrated maximum off-chip pulsed transduction efficiency of $-60$~dB ($-72$~dB per mW pump power) is low in this first demonstration, a number of possible improvements in fabrication processes and device design can significantly improve the efficiency of this novel platform.
%(improvement factors in brackets).
By process optimization or strain engineering \cite{hamze_design_2020, paoletta_pockels_2021}, the effective low-temperature Pockels coefficient of only 13~pm/V could be increased to the level reported previously in the literature of 200~pm/V \cite{eltes_integrated_2020}, enhancing efficiency by $\sim20$~dB.
The intrinsic optical quality factor is limited by sidewall scattering and could potentially be improved from the present value of $1\times 10^5$ to the absorption-limited value of $3.8\times 10^6$ \cite{riedhauser_absorption_2025}, providing a further $\sim30$~dB increase in efficiency.
In addition, adjusting the device geometry should eliminate the mismatch between photonic-molecule resonant splitting and the microwave resonance frequency, providing another $4$~dB improvement). 
A more efficient usage of the angle-dependent Pockels effect can be achieved in a racetrack resonator geometry instead of circular rings, potentially increasing the efficiency by $6$~dB \cite{abel_large_2019}. 
Finally, off-chip efficiency can be improved by reducing fiber-to-chip coupling losses from $-$5 to $-$1~dB through optimized waveguide tapers and chip facets.
Achieving all improvements simultaneously would boost microwave-optical cooperativity by up to $60$~dB and external efficiency by $4$~dB, yielding in principle an off-chip efficiency approaching unity.
\par
Besides improving efficiency, it is crucial for high-fidelity quantum transduction that input-referred added noise be maintained well below one photon \cite{zeuthen_figures_2020, krastanov_optically_2021,zhao_quantum-enabled_2024,sahu_quantum-enabled_2022}.
This study reveals sub-$\mu$s in-cavity dielectric heating as a significant noise source which we distinguish from quasiparticle heating using the observed microwave frequency shift, since they cannot readily be distinguished from their response times alone.
Our analysis of optically induced broadening of the microwave resonance has shown that the effective temperature of the microwave resonator rapidly increases to over 1~K at 0~dBm peak optical power, corresponding to 3 photons of added noise.
To validate this estimate, we plan to carry out a more detailed investigation of added noise in our transducer \cite{ren_two-dimensional_2020, sahu_quantum-enabled_2022, xu_light-induced_2024}.
\par
By providing an integrated means to pole the electro-optic material with a dc-bias field, our approach to cavity electro-optics enables the use of soft ferroelectrics with nonlinear Pockels behavior.
Our numerical simulations show that the large Pockels coefficients of perovskite titanates, despite their high permittivities, can be harnessed to yield vacuum electro-optic coupling rates significantly exceeding those which can be achieved with \LN.
We emphasize that not only our device design but also the fabrication process can be transferred with minor modifications to other material platforms, \textit{e.g.}, \LN\ or \STO, as the waveguide fabrication relies on dry etching alone.
To fully exploit the potential of soft ferroelectrics for electro-optic quantum transduction, future studies should investigate methods to both reduce optical loss and maximize the Pockels response at low temperature \cite{paoletta_pockels_2021}.
This first demonstration opens up a route to microwave-optical quantum transducers with stronger electro-optic nonlinearity, a prerequisite for high-fidelity optical quantum links between superconducting qubits.

%Together with the presented numerical simulations of our novel device design using showing that significantly higher vacuum coupling rates than with \LN\ can be achieved, this first demonstration opens up a route to microwave-optical quantum transducers with superior performance.

%Lastly, it demonstrates that  dc-biasing can successfully used in cavity electro-optics to control material properties, increasing the possible avenues to create more efficient materials and devices.
%\par
%How to improve performance: optical loss, improvement of material (strain, growth, processing)
%\par
%Since \BTO, \STO\ and also other strong electro-optic materials feature large thermal expansion coefficients \cite{hamze_design_2020}, strain engineering is likely going to be a key consideration for cryogenic applications, as the samples are usually fabricated at room temperature or above.
%\par
%Design is generic and can be applied with minor modifications to \LN, or \STO.
%\par
%Thermal response: Observation and distinction of optically induced quasiparticle and two-level system heating
%Could show that two-level system / dielectric heating is relevant for operation at optical powers where heating is restricted to <1 K.

\subsection*{Author Contributions}
C.M. designed the transducer with support from A.R.. 
C.M. and U.D. developed the fabrication process with support from A.R. and A.O..
D.C. and C.M. fabricated the samples with support from U.D., A.O. and A.R.. 
M.G. and C.M. set up and carried out coupling rate calculations with guidance from T.K.. 
C.M. performed the measurements with guidance from T.K. and D.I.. 
C.M. analyzed the data supervised by T.K. and D.I.. 
T.K. developed the theoretical model of the transducer. 
T.K. and D.I. designed and built the experimental setup with support from C.M., A.R. and M.G.. 
C.M. and T.K. wrote the paper with input from all authors. 
P.S. conceived and supervised the project.

\subsection*{Funding}
This work was supported by the European Union Horizon 2020 Programme for Research and Innovation under grant agreement No. 847471 (Marie Curie Co-fund QUSTEC) and SNF QuantEOM (grant No. CRSII5\_186364)

\subsection*{Acknowledgments}
All samples were fabricated at the Binnig and Rohrer Nanotechnology Center (BRNC) at IBM Research Europe, Zurich.
We acknowledge Anel Zulji and Stefan Gamper for manufacturing the sample package and other cryostat components,
Michael Stiefel for scannning electron micrographs,
Ralph Heller and Andreas Fuhrer for printed circuit design, 
Hansruedi Steinauer for wire bonding,
Clarissa Convertino and Stefan Abel for helpful discussions,
and Stephan Paredes for assistance with cryostat wiring.

\subsection*{Disclosures}
The authors declare no conflicts of interest.

\subsection*{Data Availability}
Data supporting the plots within this paper and other findings of this study are available through Zenodo at (to be provided). Further information is available from the corresponding author upon reasonable request.

% Appendix contents

% Numerical simulation of microwave and equivalent circuit
% Electro optic coupling
% Fabrication
% Packaging
% Experimental Setup
% Optical Characterisation
% Microwave characterisation
% Transducer model

\FloatBarrier
\appendix

\section{Electro-optic coupling strength}
\label{sec:eo_coupling}

Here, we derive expressions for the electro-optic coupling in our device and provide details on the numerical simulations.

\subsection{Field quantization}
\label{sec:quantization}

First, we recall that the classical electromagnetic energy density $\rho_\mathrm{em}$ is defined as the absolute value of the Poynting vector \cite{leble_practical_2020},
\begin{equation}
	\rho_\mathrm{em}=\frac{1}{2}\left(\mathbf{E}\cdot\mathbf{D}+\mathbf{H}\cdot\mathbf{B}\right),
\end{equation}
with electric field $\mathbf{E}$, electric displacement field $\mathbf{D}=\epsilon_\mathrm{0}\epsilon_\mathrm{r}\mathbf{E}$, vacuum permittivity $\epsilon_\mathrm{0}$, and relative permittivity $\epsilon_\mathrm{r}$; and magnetic field $\mathbf{H}$, magnetic flux density $\mathbf{B}=\mu_\mathrm{0}\mu_\mathrm{r}\mathbf{H}$, vacuum permeability $\mu_0$, and relative permeability $\mu_r$.
%In a linear, non-magnetic dielectric we have the following relations,
%\begin{eqnarray}
%	\mu_\mathrm{r}&=&1\\
%	\epsilon_\mathrm{r}&=&1+\chi^{(1)}.
%\end{eqnarray}
The vacuum speed of light is $c_\mathrm{0}=1/\sqrt{\epsilon_\mathrm{0}\mu_\mathrm{0}}$. 
%considering only the in-plane field components, we can simplify the energy density to:
Assuming, for simplicity, an isotropic, non-magnetic ($\mu_\mathrm{r} = 1$) dielectric medium with scalar $\epsilon_r$, we can simplify the energy density to
\begin{equation}
	\rho_\mathrm{em}=\frac{1}{2}\epsilon_\mathrm{0}\left(\epsilon_\mathrm{r}|\mathbf{E}|^2+c_\mathrm{0}^2|\mathbf{B}|^2\right).
\end{equation}
which is the sum $\rho_\mathrm{em}=\rho_\mathrm{e}+\rho_\mathrm{m}$ of the electrical and magnetic field contributions, i.e.
\begin{eqnarray}
	\rho_\mathrm{e}&=& \frac{1}{2}\epsilon_\mathrm{0}\epsilon_\mathrm{r}|\mathbf{E}|^2\\
	\rho_\mathrm{m}&=& \frac{1}{2}c_\mathrm{0}^2|\mathbf{B}|^2.
\end{eqnarray}

The total electromagnetic energy can then be obtained by integration, i.e.
\begin{eqnarray}
	W_\mathrm{em}&=& \int_{}^{}\rho_\mathrm{em}\mathrm{d}x\mathrm{d}y\mathrm{d}z\\
	W_\mathrm{e}&=& \int_{}^{}\rho_\mathrm{e}\mathrm{d}x\mathrm{d}y\mathrm{d}z\label{eq:W_e}\\
	W_\mathrm{m}&=& \int_{}^{}\rho_\mathrm{m}\mathrm{d}x\mathrm{d}y\mathrm{d}z.
\end{eqnarray}
For lumped element capacitors and inductors, we have
\begin{eqnarray}
	W_\mathrm{e}&=&\frac{1}{2}CV^2\\
	W_\mathrm{m}&=&\frac{1}{2}LI^2,
\end{eqnarray}
where $C$ is the capacitance, $V$ is the voltage across the capacitor, $L$ is the inductance, and $I$ is the current flowing through the inductor. In the following, we consider time-averaged energy only, for which holds $\langle W_\mathrm{em}\rangle = 2\langle W_\mathrm{e}\rangle$, such that we can focus on the electric energy.
\par
To quantize the cavity electro-optic system, we consider the microwave and optical eigenmodes as harmonic oscillators with ladder operators $a$ and $b$, and resonance frequencies $\omega_\mathrm{a}$ and $\omega_\mathrm{b}$, respectively. Their respective uncoupled Hamiltonians are
\begin{eqnarray}
	H_\mathrm{a}&=&\hbar\omega_\mathrm{a}\left(a^\dagger a+\frac{1}{2}\right), \label{eq:H_a_0}\\
	H_\mathrm{b}&=&\hbar\omega_\mathrm{b}\left(b^\dagger b+\frac{1}{2}\right).\label{eq:H_b_0}
\end{eqnarray}
%Their defining relations are:
%\begin{eqnarray}
%	\left[i,j^\dagger\right]&=&\delta_\mathrm{ij}\\
%	\left[i,i\right]&=&0\\
%	\left[i^\dagger,i^\dagger\right]&=&0\\
%	i\ket{n_\mathrm{i}}&=&\sqrt{n_\mathrm{i}}\ket{n_\mathrm{i}-1}\\
%	i^\dagger\ket{n_\mathrm{i}}&=&\sqrt{n_\mathrm{i}+1}\ket{n_\mathrm{i}+1}.
%\end{eqnarray}
%Here, $\ket{n_\mathrm{i}}$ are the eigenstates of the number operator $i^\dagger i$ with eigenvalues $n_\mathrm{i}$, the commutator $\left[k,l\right]=kl-lk$ and the Kronecker-Delta $\delta_\mathrm{ij}=1$ for $\mathrm{i}=\mathrm{j}$ and zero otherwise.
\par
The quantized electric field operators, in a rotating frame at the optical frequency $\omega_a$, can be defined as \cite{walls_quantum_2008}
\begin{eqnarray}
	\mathbf{E}_\mathrm{a}&=&\sqrt{\frac{\hbar\omega_\mathrm{a}}{2\epsilon_\mathrm{0}U_\mathrm{a}}}\left(a\mathbf{u}_\mathrm{a}e^{-i\omega_\mathrm{a}t} + a^\dagger\mathbf{u}_\mathrm{a}^\ast e^{+i\omega_\mathrm{a}t}\right ), \label{eq:E_a}\\	\mathbf{E}_\mathrm{b}&=&\sqrt{\frac{\hbar\omega_\mathrm{b}}{2\epsilon_\mathrm{0}U_\mathrm{b}}}\left(b\mathbf{u}_\mathrm{b} + b^\dagger\mathbf{u}_\mathrm{b}^\ast\right), \label{eq:E_b}
%	\mathbf{E}_\mathrm{b}=&\sqrt{\frac{\hbar\omega_\mathrm{b}}{2\epsilon_\mathrm{0}U_\mathrm{b}}}\left(b\mathbf{u}_\mathrm{b}e^{-i\omega_\mathrm{b}t} + b^\dagger\mathbf{u}_\mathrm{b}^*e^{+i\omega_\mathrm{b}t}\right), \label{eq:E_b}
\end{eqnarray}
with the spatial mode functions $\mathbf{u}_\mathrm{i}$ and the effective volumes $U_\mathrm{i}=\int \epsilon_\mathrm{r}|\mathbf{u}_\mathrm{i}|^2\mathrm{d}x\mathrm{d}y\mathrm{d}z$ for $\mathrm{i}\in\left[\mathrm{a,b}\right]$.
Here, it is useful to define the field normalization factors
\begin{equation}
	N_\mathrm{i}=\sqrt{\frac{\hbar\omega_\mathrm{i}}{2\epsilon_\mathrm{0}U_\mathrm{i}}}.
	\label{eq:field_norm}
\end{equation}

%The microwave cavity being an electrical circuit, its voltage operator is \cite{tsang_cavity_2010}:
%\begin{equation}
%%	V_\mathrm{b}=\sqrt{\frac{\hbar\omega_\mathrm{b}}{2C}}\left(b e^{-i\omega_\mathrm{b}t} + b^\dagger e^{+i\omega_\mathrm{b}t}\right),
%	V_\mathrm{b}=\sqrt{\frac{\hbar\omega_\mathrm{b}}{2C}}\left(b + b^\dagger \right),
%\end{equation}
%with zero-point voltage $V_\mathrm{zpf}=\sqrt{\frac{\hbar\omega_\mathrm{b}}{2C}}$. 
Inserting the expressions \eqref{eq:E_a} and \eqref{eq:E_b} for the field operators into the time-averaged energy $W_\mathrm{em} = 2 W_e$ yields the Hamiltonians \eqref{eq:H_a_0} and \eqref{eq:H_b_0}, respectively.
%\par
%To confirm that the chosen quantization and normalization are valid, we can now insert the field and voltage operators into the expectation values for the classical expressions of the electromagnetic energy.
%So far we have focused on the electrical field component of the energy, even though any electromagnetic eigenmode has both components.
%We can however account for the magnetic energy by noting that it is the same as the electric energy on average:
%\begin{equation}
%	W_\mathrm{em}=2W_\mathrm{e}\overset{!}{=}\hbar\omega_\mathrm{i}\left(i^\dagger i+\frac{1}{2}\right).
%\end{equation}

\subsection{Electro-optic interaction Hamiltonian}
Based on refs.~\cite{fan_superconducting_2018, mckenna_cryogenic_2020, holzgrafe_cavity_2020}, we first consider the optical electric energy density and express it in terms of the displacement field $D^{i}=\epsilon_\mathrm{r}^{ij}E^{j}$,
\begin{eqnarray}
	\rho_\mathrm{el,opt}&=&\frac{1}{2}\epsilon_\mathrm{0}E_\mathrm{opt}^\mathrm{i}\epsilon_\mathrm{r,opt}^\mathrm{ij}E_\mathrm{opt}^\mathrm{j}\\
	&=&\frac{1}{2}\epsilon_\mathrm{0}D_\mathrm{opt}^\mathrm{i}\left(\epsilon_\mathrm{r,opt}^{-1}\right)^\mathrm{ij}D_\mathrm{opt}^\mathrm{j}.
\end{eqnarray}
Here and in what follows, it is implicit that indices appearing twice are summed over.

We can now introduce the electro-optic interaction as a perturbation of the optical energy density from a microwave field $\mathbf{E}_\mathrm{mw}$ that couples to the optical impermeability tensor $\left(\epsilon_\mathrm{r,opt}^{-1}\right)^\mathrm{ij}\rightarrow\left(\epsilon_\mathrm{r,opt}^{-1}\right)^\mathrm{ij}+r^\mathrm{ijk}E_\mathrm{mw}^k$.
While the first term is just the linear energy density, the second one describes the interaction energy.
We can thus define the electro-optic interaction Hamiltonian
\begin{eqnarray}
	H_\mathrm{eo}&=&\frac{1}{2}\epsilon_\mathrm{0}\int r^\mathrm{ijk}D_\mathrm{opt}^\mathrm{i}D_\mathrm{opt}^\mathrm{j}E_\mathrm{mw}^k\mathrm{d}x\mathrm{d}y\mathrm{d}z.
\end{eqnarray}

%We return to an isotropic medium for ease of derivation, but the analogous holds for the anisotropic case.
%For simplicity, we write here the expression for the in-plane field components only, replacing the Pockels tensor by an effective coefficient $r$. In numerical simulations, we use the full 3d fields and Pockels tensor. 
Moving back to electric fields we have
%\begin{eqnarray}
%	H_\mathrm{eo}&=&\frac{1}{2}\epsilon_\mathrm{0}\int r\mathbf{D}_\mathrm{opt}^*\mathbf{D}_\mathrm{opt}\mathbf{E}_\mathrm{mw}\mathrm{d}x\mathrm{d}y\mathrm{d}z\\
%	&=&\frac{1}{2}\epsilon_\mathrm{0}n_\mathrm{opt}^4r\int_\mathrm{eo}\mathbf{E}_\mathrm{opt}^*\mathbf{E}_\mathrm{opt}\mathbf{E}_\mathrm{mw}\mathrm{d}x\mathrm{d}y\mathrm{d}z,
%\end{eqnarray}
\begin{eqnarray}
	H_\mathrm{eo}&=&\frac{1}{2}\epsilon_\mathrm{0}n_\mathrm{opt}^4 \int_\mathrm{eo} r^{ijk} E_\mathrm{opt}^{i} E_\mathrm{opt}^{j} E_\mathrm{mw}^{k} \mathrm{d}x\mathrm{d}y\mathrm{d}z,
\end{eqnarray}
where $\int_\mathrm{eo}$ denotes the integral over the volume of the electro-optic material, where $r\neq 0$. Inserting the field operators $\mathbf{E}_\mathrm{a}$ and $\mathbf{E}_\mathrm{b}$ from \eqref{eq:E_a} and \eqref{eq:E_b} for the optical and microwave modes, respectively, we obtain
\begin{eqnarray}
	H_\mathrm{eo}&=&\frac{1}{2}\epsilon_\mathrm{0}n_\mathrm{opt}^4\int_\mathrm{eo} r^{ijk} E_\mathrm{a}^{i} E_\mathrm{a}^{j} E_\mathrm{b}^{k} \mathrm{d}x\mathrm{d}y\mathrm{d}z\\
	&=&\epsilon_\mathrm{0} n_\mathrm{opt}^4 N_\mathrm{a}^2 N_\mathrm{b} \cdot\\
	& & \int_\mathrm{eo}  r^{ijk} \left(u_\mathrm{a}^{i}\right)^{\ast} u_\mathrm{a}^{j} \; a^\dagger a \;\left(u_\mathrm{b}^{k} b + \left(u_\mathrm{b}^{k}\right)^{\ast} b^\dagger\right) \mathrm{d}x\mathrm{d}y\mathrm{d}z\nonumber
\end{eqnarray}
In the last equation, we only kept terms conserving the optical photon number. We finally arrive at an expression for the vacuum coupling strength $g_0$. Since the microwave spatial mode function $u_b$ is real-valued, we have $H_\mathrm{eo}= \hbar g_\mathrm{0} a^\dagger a \left(b+b^\dagger\right)$ for a single optical ring resonator, with \cite{mckenna_cryogenic_2020}
\begin{equation}
	\hbar g_\mathrm{0}
	= \epsilon_\mathrm{0} n_\mathrm{opt}^4 N_\mathrm{a}^2 N_\mathrm{b} \int_\mathrm{eo}  r^{ijk} \left(u_\mathrm{a}^{i}\right)^{\ast} u_\mathrm{a}^{j} u_\mathrm{b}^{k} \mathrm{d}x\mathrm{d}y\mathrm{d}z.
	\label{eq:g0_0}
\end{equation}
%with the normalization factors defined in Eq. \ref{eq:field_norm}.
Considering only a single in-plane field component, as is approximately the case in our device, we get 
\begin{equation}
	\hbar g_\mathrm{0}
	= \epsilon_\mathrm{0} n_\mathrm{opt}^4 N_\mathrm{a}^2 N_\mathrm{b} \int_\mathrm{eo} r |u_\mathrm{a}|^2 u_\mathrm{b} \mathrm{d}x\mathrm{d}y\mathrm{d}z.
	\label{eq:g0}
\end{equation}
We note that this describes the coupling between the microwave mode and one of the optical ring resonators. The full electro-optic interaction Hamiltonian contains the contributions of both ring resonators, and takes into account that the $\lambda/2$-type microwave mode function $u_b$ (Fig.~\ref{fig:MW_mode}a) is equal and opposite at both rings, i.e.
\begin{equation}
	H_\mathrm{eo} = \hbar g_0 (a_1^\dag a_1 - a_2^\dag a_2) (b+b^\dag).
\end{equation}

\subsection{Separating longitudinal and cross-sectional integration}

\begin{figure}[t]
	\centering
	\includegraphics{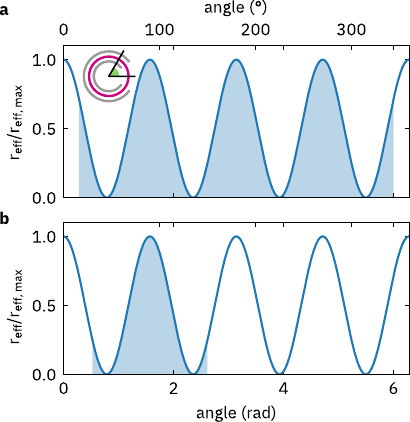}
	\caption[Model for angular dependence of Pockels effect]{\textbf{Model for angular dependence of Pockels effect.}
		Model for the angular dependence of the Pockels effect (blue solid line) and visualization of the overlap factor $\alpha_\mathrm{overlap}$ as the blue shaded area under the curve for (a) the transducer electrode and (b) the tuning electrode, with $\SI{42.0}{\percent}$ and $\SI{13.2}{\percent}$, respectively.
	}
	\label{fig:angle_dependence}
\end{figure}

Since the overlap only happens in the active material of the optical rings, we can separate the integration into a longitudinal ($\mathrm{d}z$) and cross sectional part ($\mathrm{d}x\mathrm{d}y$). For the longitudinal integral along one optical ring resonator with radius $R$, we get (neglecting curvature)
\begin{equation}
	\int r\mathrm{d}z=r\alpha_\mathrm{overlap}L_\mathrm{opt}%=r \; \alpha_\mathrm{overlap}\;2\pi R.
\end{equation}
%Here, the factor of two comes from the fact that the interaction happens in both rings, 
with total optical length of $L_\mathrm{opt}= 2\pi R$ being the ring circumference.
The factor $\alpha_\mathrm{overlap}$ describes the effective overlap fraction of the microwave and optical fields, which takes into account the angle dependence of the Pockels effect.
Motivated by the results reported in \cite{eltes_integrated_2020} and \cite{abel_large_2019}, we model the angular dependence as a sinusoidal function with a maximum every $\SI{90}{\degree}$ (blue sold line in Fig.~\ref{fig:angle_dependence}).
The blue areas under the curves mark the angular interval of the rings which is covered by the transducer electrodes (Fig.~\ref{fig:angle_dependence}a) and tuning electrodes (Fig.~\ref{fig:angle_dependence}b).
The transducer electrodes cover an interval of $2\pi\alpha_\mathrm{coverage}=2\pi\cdot0.91$, but due to the angle dependence the effective overlap is only $\alpha_\mathrm{overlap}=0.42$. We thus obtain
\begin{eqnarray}
	\hbar g_\mathrm{0}
	&=&L_\mathrm{opt} \alpha_\mathrm{overlap}\epsilon_\mathrm{0} r n_\mathrm{opt}^4 N_\mathrm{a}^2 N_\mathrm{b} \nonumber\\
	& &  \cdot \int_\mathrm{eo} |u_\mathrm{a}|^2 |u_\mathrm{b}| \mathrm{d}x\mathrm{d}y.\;
\end{eqnarray}
%with optical length $L_\mathrm{opt}=2\cdot 2\pi R$.
Note, that we replaced the complex amplitude of the microwave field by its absolute value to ensure that no arbitrary phase in the numerical simulation leads to issues with the integration.
Furthermore the $\mathrm{eo}$ subscript for the integral denotes that we only integrate over the electro-optic material cross-sectional surface, as the electro-optic coefficients vanishes everywhere else.
\par
The frequency pulling factor $G=\partial\omega_\mathrm{opt}/\partial V = g_0 / V_\mathrm{zpf}$ can be calculated analogously. We note that $N_\mathrm{b} u_\mathrm{b}$ is the zero-point microwave electric field. Noting that 
\begin{equation}
	\frac{N_\mathrm{b} u_\mathrm{b}}{ V_\mathrm{zpf}} = \frac{E_\mathrm{b}}{V_\mathrm{sim}}
\end{equation}
where for $E_\mathrm{b}$ we use the simulated microwave field for the applied voltage $V_\mathrm{sim}$, we get
\begin{eqnarray}
	G
	&=&\frac{L_\mathrm{opt} \alpha_\mathrm{overlap}\epsilon_\mathrm{0} r n_\mathrm{opt}^4 N_a^2}{\hbar V_\mathrm{sim}}  \int_\mathrm{eo} |u_\mathrm{a}|^2 |E_\mathrm{b}| \mathrm{d}x\mathrm{d}y. \;
\end{eqnarray}

The only quantities left to compute are the field normalization factors.
For the optics this is straightforward by integrating the cross-sectional field distribution,
\begin{equation}
	N_\mathrm{a}=\sqrt{\frac{\hbar\omega_\mathrm{a}}{\left[2\epsilon_\mathrm{0}L_\mathrm{opt}\int n_\mathrm{opt}^2 |\mathbf{u}_\mathrm{a}|^2\mathrm{d}x\mathrm{d}y\right]}}.
	\label{eq:opt_norm}
\end{equation}

\subsection{Microwave equivalent circuit}
\label{sec:mw_equiv_circuit}

\begin{figure}[h]
	\centering
	\includegraphics{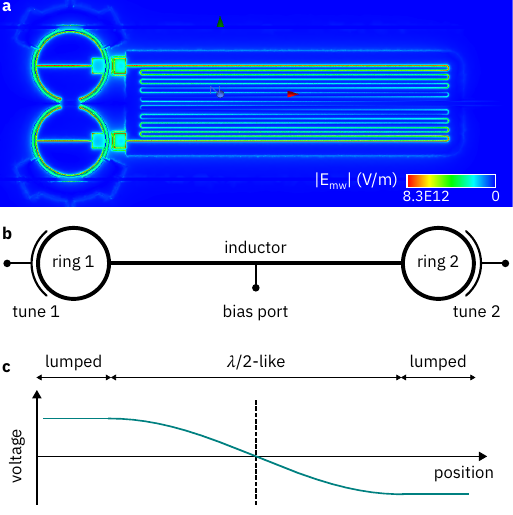}
	\caption[Microwave mode]{\textbf{Microwave mode.}
		(a) 3d-finite element Ansys HFSS eigenmode simulation of the microwave mode field distribution.
		(b) Schematic of the electrical circuit with unfolded meander inductor and with lumped (ring) capacitances at each end. (c) Sketch of the $\lambda/2$-type field distribution in the inductor according to (a) that vanishes at the bias port.
	}
	\label{fig:MW_mode}
\end{figure}

\begin{figure}[b]
	\centering
	\includegraphics{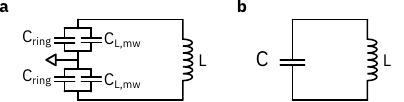}
	\caption[Participation ratio]{\textbf{Participation ratio.}
		(a) Standard equivalent circuit and (b) extended circuit representation of the transducer device.
		The ring capacitances $C_\mathrm{ring}$ go from one inductor end to ground, and the stray capacitance from the inductor is modeled as parallel capacitances $C_\mathrm{L,mw}$, also to ground.
	}
	\label{fig:participation_ratio}
\end{figure}

An analogous normalization factor equation like \eqref{eq:opt_norm} holds for the microwave field as well, but  requires simulating the whole microwave mode (Fig.~\ref{fig:MW_mode}a), which is computationally expensive.
Instead, we only simulate the fraction of the microwave mode in the transducer electrode, and normalize it with the fraction of the energy that is contained in that transducer electrode, i.e. the ring.
We do this by introducing the participation ratio \cite{minev_energy-participation_2021}
\begin{eqnarray}
	\rho_\mathrm{c}=\frac{W_\mathrm{e,ring}}{W_\mathrm{e,total}}.
	\label{eq:rho_c_0}
\end{eqnarray} 
As illustrated by the transducer equivalent circuit (Fig.~\ref{fig:participation_ratio}a), this corresponds to the ratio of the capacitive energies and hence, the capacitances, i.e.
\begin{eqnarray}
	\rho_\mathrm{c}=\frac{C_\mathrm{ring}}{2 (C_\mathrm{ring}+ C_\mathrm{L,mw})}.
	\label{eq:rho_c}
\end{eqnarray} 
where $C_\mathrm{ring}$ is the electrode capacitance at each ring, and $C_\mathrm{L,mw}$ is the parasitic capacitance of each inductor arm between ring electrode and bias port.
Additionally, we need to take into account that the transducer electrode only overlaps with the fraction $\alpha_\mathrm{coverage}$ of the optical ring resonator. 
This gives the corrected normalization
\begin{equation}
	N_\mathrm{b}=\sqrt{\frac{\rho_\mathrm{c}\hbar\omega_\mathrm{b}}{2\epsilon_\mathrm{0}L_\mathrm{opt}\alpha_\mathrm{coverage}\int \epsilon_\mathrm{r} |\mathbf{u}_\mathrm{b}|^2 \mathrm{d}x\mathrm{d}y}}.
	\label{eq:N_b}
\end{equation}

\begin{table}
	\begin{tabular}{c|cccccc}
		\hline
		$\si{\femto\farad}$& ring 2 &ind. & ring 1 & ground & tune 1 & tune 2 \\
		\hline
		ring 2 	  &$57.3$&$-2.1$&$-1.6$&$-36.8$&$-0.4$&$-16.4$\\
		ind.        &$-2.1$&$177.4$&$-2.1$&$-170.5$&$-1.4$&$-1.4$\\
		ring 1    &$-1.6$&$-2.1$&$57.3$&$-36.8$&$-16.5$&$-0.4$\\
		ground &$-36.8$&$-170.5$&$-36.8$&$343.2$&$-49.5$&$-49.5$\\
		tune 1   &$-0.4$&$-1.4$&$-16.5$&$-49.5$&$67.8$&$-0.1$\\
		tune 2   &$-16.4$&$-1.4$&$-0.4$&$-49.5$&$-0.1$&$67.7$\\
	\end{tabular}
	\caption[Simulated dc-capacitances]{\textbf{Simulated dc capacitances.}
		3d-finite element simulation of dc-capacitance matrix of the electro-optic transducer using Ansys Maxwell. Conductor labels correspond to those in Fig.~\ref{fig:MW_mode}b.
		The active material permittivity is set to $\epsilon_\mathrm{r}=200$ in this case.
	}
	\label{tab:dc_capacitances}
\end{table}

To obtain the participation ratio $\rho_\mathrm{c}$, we first determine the capacitance of each ring electrode $C_\mathrm{ring}$.
On the one hand, we can do this in an electrostatic simulation of the whole device using the finite-element solver Ansys Maxwell.
The resulting dc-capacitances of the different circuit elements are listed in Table \ref{tab:dc_capacitances}.
The ring capacitance is about $\SI{57.3}{\femto\farad}$.
On the other hand, we can directly compute the capacitance from the field distribution $\mathbf{u}_\mathrm{b}$ of a Comsol finite-element simulation by equating the capacitive energy with the applied simulation voltage to the energy of the simulated field,
\begin{eqnarray}
	W_\mathrm{e, ring}&=&\frac{1}{2}C_\mathrm{ring}V_\mathrm{sim}^2\\
	&=&\frac{1}{2}\epsilon_\mathrm{0}\alpha_\mathrm{coverage}2\pi R\int \epsilon_\mathrm{r} |\mathbf{u}_\mathrm{b}|^2 \mathrm{d}x\mathrm{d}y.
\end{eqnarray}

%It is important to note here, that we now have a factor of $1/2$ for the electrical energy term, as we are only looking at the capacitive part of the energy.
This can be solved for the ring capacitance,
\begin{equation}
	C_\mathrm{ring}= \frac{\epsilon_\mathrm{0}\alpha_\mathrm{coverage}2\pi R}{V_\mathrm{sim}^2} \int \epsilon_\mathrm{r} |\mathbf{u}_\mathrm{b}|^2\mathrm{d}x\mathrm{d}y.
	\label{eq:C_ring}
\end{equation}
Using a Comsol cross-sectional simulation, we obtain $C_\mathrm{ring} = \SI{61.1}{\femto\farad}$ for the ring electrode, which is fairly close to the value obtained from the three-dimensional dc finite-element simulation (Tab.~\ref{tab:dc_capacitances}).

\begin{table}
	\begin{tabular}{cc}
		\hline
		$C$ ($\si{\femto\farad}$)&$L$ ($\si{\nano\henry}$)\\
		\hline
		$69.0$&$7.4$\\
	\end{tabular}
	\caption[Simulated microwave equivalent circuit parameters]{\textbf{Simulated microwave equivalent circuit parameters.}
		Fit results of the fit of equation \ref{eq:MW_freq_vs_parallel_capacitance} to the data obtained from 3d-finite element eigenmode simulations of the the electro-optic transducer with a paralell capacitance with Ansys Maxwell (Fig.~\ref{fig:MW_eigenmode_with_series_capacitance}a).
	}
	\label{tab:participation_ratio}
\end{table}

\begin{figure}[h]
	\centering
	\includegraphics{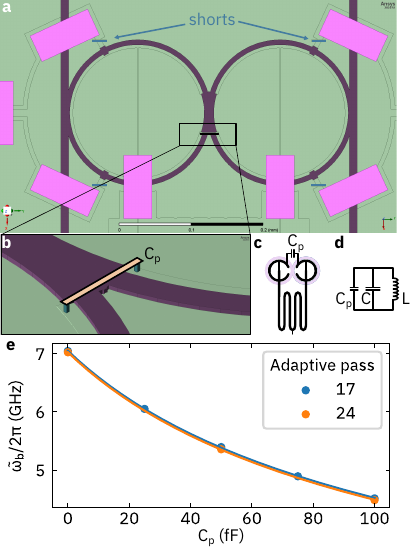}
	\caption[Geometric inductance and capacitance extraction using a parallel capacitance]{\textbf{Geometric inductance and capacitance extraction using a parallel capacitance.}
		(a) Ansys HFSS simulation setup view of the capacitive part of the circuit with patches (blue) to short the coupling electrodes to ground.
		(b) Out-of-plane pins (blue) connected with a 2D-sheet (orange) which is defined as a lumped element boundary condition with capacitance $C_\mathrm{p}$.
		(c) Schematic representation of the modified device circuit with the added parallel capacitance $C_\mathrm{p}$ between the ring electrodes.
		(d) Equivalent circuit representation with the parallel capacitance in addition to the equivalent capacitance $C$ and inductance of the device $L$.
		(e) Resonance frequency of the device as a function of parallel capacitance $C_\mathrm{p}$ using the mesh after 17 (blue dots) and 24 adaptive passes (orange dots) and fit to equation \ref{eq:MW_freq_vs_parallel_capacitance}.
	}
	\label{fig:MW_eigenmode_with_series_capacitance}
\end{figure}

We are now only missing the value of the microwave stray capacitance of the inductor arm $C_\mathrm{L,mw}$ which can in principle be estimated from the dc-capacitance matrix listed in Table~\ref{tab:dc_capacitances}. 
Instead, we determine $C_\mathrm{L,mw}$ from the equivalent circuit capacitance $C$ (Fig.~\ref{fig:participation_ratio}b).
An accurate value for the inductance $L$ and the equivalent circuit capacitance $C$ can be obtained by performing a 3d finite-element eigenmode simulation of the full circuit, but with an added parallel capacitance $C_\mathrm{p}$.
The simulation setup is shown in Fig.~\ref{fig:MW_eigenmode_with_series_capacitance}a, where the coupling capacitances were set to ground by shorting with small patches (blue).
The parallel capacitance is then implemented as a bridge between the two ring electrodes at the shortest possible distance (Fig. \ref{fig:MW_eigenmode_with_series_capacitance}b).
The bridge is implemented using two blocks (blue) that connect the ring electrodes to a two-dimensional sheet, on which a lumped element capacitance boundary condition with value $C_\mathrm{p}$ is defined.
A schematic of the modified device circuit is shown in Fig. \ref{fig:MW_eigenmode_with_series_capacitance}c.
Effectively, this represents an additional parallel capacitance $C_\mathrm{p}$ (Fig.~\ref{fig:MW_eigenmode_with_series_capacitance}d) to the equivalent circuit capacitance $C$ (Fig.~\ref{fig:participation_ratio}b).

The resonance frequency of this modified circuit is then simulated as a function of the parallel capacitance $C_\mathrm{p}$, as shown in Fig. \ref{fig:MW_eigenmode_with_series_capacitance}e.
The advantage of using a lumped element is, that the mesh for the simulation only needs to be calculated once, and can then be used to simulate for all values of $C_\mathrm{p}$.
The observed dependence of the resonance frequency on the the parallel capacitance is modeled as
\begin{equation}
	\tilde{\omega}_\mathrm{b}=\sqrt{\frac{1}{L\left(C+C_\mathrm{p}\right)}}, \label{eq:MW_freq_vs_parallel_capacitance}
\end{equation}
with the inductance $L$ and the equivalent capacitance $C$ of the device as fit parameters.
The results are listed in Table \ref{tab:participation_ratio}.
We can then use the fact that the circuit is symmetric, and together with the rules for series and parallel capacitances we have
\begin{equation}
	C = \frac{C_\mathrm{ring}+C_\mathrm{L,mw}}{2},
\end{equation}
which we can solve for the inductor's capacitance
\begin{equation}
	C_\mathrm{L,mw} = 2C-C_\mathrm{ring},
\end{equation}
which equates to $C_\mathrm{L,mw}=\SI{80.7}{\femto\farad}$ using the value of $C_\mathrm{ring}$ from Table \ref{tab:dc_capacitances}. With this, we arrive at a participation ratio of $\rho_\mathrm{c}=0.43$.

Finally, we evaluate \eqref{eq:N_b} replacing $u_\mathrm{b}$ with the simulated field $E_\mathrm{b}$ and using \eqref{eq:C_ring} and \eqref{eq:rho_c}, which gives
\begin{eqnarray}
	N_b &=& \sqrt{\frac{\rho_c \hbar\omega_{\mathrm{b}}}{2 C_\mathrm{ring} V_\mathrm{sim}^2}} =   \sqrt{\frac{\hbar\omega_{\mathrm{b}}}{4 (C_\mathrm{ring} + C_\mathrm{L,mw})}} \frac{1}{|V_\mathrm{sim}|} \nonumber\\
			&=& \frac{V_\mathrm{zpf}}{|V_\mathrm{sim}|}.
\end{eqnarray}
We thus obtain the zero-point voltage 
\begin{equation}
	V_\mathrm{zpf}= \sqrt{\frac{\hbar\omega_\mathrm{b}}{8C}},
\end{equation}
dropping across one ring electrode. The effective microwave voltage operator \cite{tsang_cavity_2010} can be written as
\begin{equation}
	V_\mathrm{b}=V_\mathrm{zpf} \left(b + b^\dagger \right).
\end{equation}

\subsection{Incorporating scaling with material permittivity}
In order to fairly compare the vacuum coupling strength of the same device design and cross-sectional geometry but for different active materials with similar optical properties but sometimes vastly different permittivities ($\epsilon_{r,\mathrm{LiNbO_3}}=30$, $\epsilon_{r,\mathrm{BaTiO_3}}=2\times 10^2$, $\epsilon_{r,\mathrm{SrTiO_3}}\approx 2\times10^4$), we incorporate the scaling of the inductor, which affects both the inductance and stray capacitance of the circuit, and thereby also the total capacitance and participation ratio.
We do this, by requiring a constant resonance frequency $\omega_\mathrm{b}$ of the transducer
\begin{equation}
	\omega_\mathrm{b}=\frac{1}{\sqrt{LC}}.
\end{equation}

We then assume that by varying the length of the meander inductor wire by the factor $\alpha_\mathrm{ind}$, its inductance and capacitance vary proportionally, such that the resonance frequency scales as
\begin{equation}
	\omega_\mathrm{b}=\frac{1}{\sqrt{\alpha_\mathrm{ind}L\left(C_\mathrm{ring}+\alpha_\mathrm{ind}C_\mathrm{L,mw}\right)/2}}.
\end{equation}
We can solve this quadratic equation for $\alpha_\mathrm{ind}$ of the form $0=\mathrm{a}x^2+\mathrm{b}x+\mathrm{c}$ with the coefficients
\begin{align}
	\mathrm{a}&=LC_\mathrm{L,mw}\\
	\mathrm{b}&=LC_\mathrm{ring}\\
	\mathrm{c}&=-\frac{2}{\omega_\mathrm{b}^2}.
\end{align}
The inductor scaling parameter is then determined via
\begin{equation}
	\alpha_\mathrm{ind}=\frac{-\mathrm{b}\pm\sqrt{\mathrm{b}^2-4\mathrm{ac}}}{2\mathrm{a}},
\end{equation}
where we use the positive solution.

\section{Linear system response}
\subsection{General picture}
\label{sec:open_coupled_system}

%\begin{figure}[b]
%	\centering
%	\includegraphics{figures_appendix/in_out_coupled_theory_v3}
%	\caption[Open coupled mode system]{\textbf{Open coupled mode system.}
%		\hl{Change figure to show actual modes.}
%		Resonant modes $\mathrm{i}\in\left\{\mathrm{a,b,c,d,}...\right\}$ are modeled as quantum harmonic oscillators with intrinsic loss rates $\kappa_\mathrm{i,0}$ and external coupling rates $\kappa_\mathrm{i,ex}$ to the input and output modes $i_\mathrm{in/out}$.
%		Additionally, they are coupled to each other in a pairwise manner as indicated by the black arrows.
%	}
%	\label{fig:in_out_coupled_system}
%\end{figure}

In this section we give an introduction to the input-output formalism and coupled-mode theory to describe the linearized response of the cavity electro-optic system studied in this work.
%As illustrated in Fig.~\ref{fig:in_out_coupled_system}, 
We describe the system as a collection of resonant modes with annihilation operators $A_i$, resonance frequencies $\omega_i$, and internal loss rates $\kappa_{i, 0}$.
Some of the modes are coupled to external bus modes $A_{i, \mathrm{in/out}}$ for input/output at the external coupling rates $\kappa_{i, \mathrm{i,ex}}$. 
Total linewidths are denoted $\kappa_i = \kappa_{i,0} + \kappa_{i,\mathrm{ex}}$.
Additionally, there are pairwise coherent couplings characterized by (possibly complex-valued) couplings $g_{ij}$ ($f_{ij}$) for counter-rotating (co-rotating) terms. 
The generic Hamiltonian reads
\begin{eqnarray}
	H &=& \sum_i \hbar\omega_i A_i^\dag A_i \nonumber\\
		& &+ \sum_{i<j} \hbar \left(g_{ij} A_i^\dag A_j + g_{ij}^\ast A_i A_j^\dag \right) \nonumber \\
		& &+ \sum_{i<j} \hbar \left(f_{ij} A_i^\dag A_j^\dag + f_{ij}^\ast A_i A_j \right).
\end{eqnarray}
The Heisenberg equations of motion including damping and input/output read
\begin{eqnarray}
	\dot{A}_i &=& -i \omega_i A_i - \frac{\kappa_i}{2} A_i - \sqrt{\kappa_{i, \mathrm{ex}}} A_{i,\mathrm{in}} \nonumber\\
					&  & -i \sum_{j\neq i} \left( g_{ij} A_j + f_{ij} A_j^\dag \right). \label{eq:heisenberg_eom_generic}
\end{eqnarray}
Using the vector notation
\begin{eqnarray}
	\mathbf{q} = \left[A_1, A_1^\dag, A_2, A_2^\dag,\ldots \right]^T,
\end{eqnarray}
we can write \eqref{eq:heisenberg_eom_generic} in matrix form, i.e.
\begin{equation}
	\dot{\mathbf{q}} = M' \mathbf{q} - K\mathbf{q}_\mathrm{in}, \label{eq:time_domain_eom}
\end{equation}
where $M'$ is the time-domain dynamical matrix and $K$ is the input-output matrix defined by \eqref{eq:heisenberg_eom_generic}. The input-output relations $A_{i,\mathrm{out}} = A_{i,\mathrm{in}}+ \sqrt{\kappa_{i,\mathrm{ex}}} A_{i}$ have the matrix form
\begin{eqnarray}
	\mathbf{q}_\mathrm{out} = \mathbf{q}_\mathrm{in} + K^T \mathbf{q}. \label{eq:input_output_matrix_form}
\end{eqnarray}
In the frequency domain (with Fourier frequency $\omega$), the equations of motion become
\begin{equation}
	\left(-i\omega - M'\right) \mathbf{q}(\omega) = - K\mathbf{q}_\mathrm{in}(\omega).  \label{eq:freq_domain_eom}
\end{equation}
where
\begin{equation}
	 \mathbf{q}(\omega) = \left[A_1(\omega), A_1^\dag(-\omega), A_2(\omega), A_2^\dag(-\omega),\ldots \right]^T.
\end{equation}
The solution to \eqref{eq:freq_domain_eom} can be found by matrix inversion, i.e.
\begin{equation}
	 \mathbf{q}(\omega) = - M(\omega)^{-1} K\mathbf{q}_\mathrm{in}(\omega)
\end{equation}
where $M(\omega) =  -i\omega - M'$.
Inserting into the input-output relation \eqref{eq:input_output_matrix_form} yields 
\begin{eqnarray}
	\mathbf{q}_\mathrm{out}(\omega) &=& \mathbf{q}_\mathrm{in}(\omega) - K^T M(\omega)^{-1} K\mathbf {q}_\mathrm{in}(\omega) \\
																&=& S(\omega)  \mathbf{q}_\mathrm{in}(\omega)
\end{eqnarray}
where we defined the S-matrix
\begin{equation}
	\label{eq:S_matrix}
	S(\omega) = 1 - K^T M(\omega)^{-1} K.
\end{equation}

\subsection{Photonic molecule with forward-backward scattering}
\label{sec:ph_mol_back}

\begin{figure}[b]
	\centering
	\includegraphics{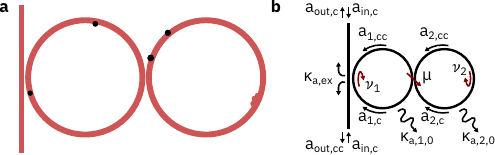}
	\caption[Photonic molecule with backscattering]{\textbf{Photonic molecule with backscattering.}
		(a) Physical representation of photonic molecule circuit, consisting of two coupled optical rings.
		Imperfectins leading to backscattering such as scattering site on or in the waveguide (black dots) or waveguide cross-section irregularities (bottom right) are illustratd.
		(b) Coupling and loss rates between the involved clockwise and counterclockwise propagating resonator modes $a_\mathrm{i,c}$, $a_\mathrm{i,cc}$ ($\mathrm{i}\in\left\{1,2\right\}$) and in- and output modes $a_\mathrm{in}$, $a_\mathrm{out}$.
	}
	\label{fig:optical_mode_coupling}
\end{figure}

First, we apply the formalism derived above to describe the coupled optical resonators in the transducer.
As shown in Fig.~\ref{fig:optical_mode_coupling}b, the two ring resonators are coupled evanescently at rate $\mu$, and resonator $1$ is coupled to the optical bus waveguide at rate $\kappa_{1,\mathrm{ex}}$.
Moreover, the fabricated waveguides exhibit imperfections such as particles on the surface, surface roughness or other scattering sites on the inside (black dots), that cause scattering to both unguided and guided modes (Fig.~\ref{fig:optical_mode_coupling}a).
In particular, these imperfections can couple the clockwise (c) and counterclockwise (cc) propagating optical modes at rates $\nu_i$  \cite{li_backscattering_2016}.
We note that $\nu_i$ can in principle be complex-valued to account for incoherent, non-Hamiltonian scattering, which manifests in asymmetric linewidths of forward-backward split optical resonances.
The system thus consists of the four modes $a_{1,\mathrm{c}}$, $a_{1,\mathrm{cc}}$, $a_{2,\mathrm{c}}$, $a_{2,\mathrm{cc}}$ described by the Hamiltonian
\begin{eqnarray}
	H_\mathrm{a} &=& - \sum_{i=1,2} \sum_{\sigma = \mathrm{c, cc}} \hbar\Delta_{\mathrm{p}, i} a^\dagger_{i, \mathrm{\sigma}} a_{i, \mathrm{\sigma}} \nonumber\\
	& & -\hbar\mu \left(a_{1,\mathrm{c}}^\dag a_{2, \mathrm{cc}} + a_{1,\mathrm{cc}}^\dag a_{2, \mathrm{c}}\right) + \mathrm{h.c.} \nonumber\\
	&& - \sum_{i=1,2} \hbar \Re{\nu_i} a_{i,\mathrm{c}}^\dag  a_{i,\mathrm{cc}} + \mathrm{h.c.}, \label{eq:H_a}
\end{eqnarray}
in the rotating frame of the pump laser, where $\Delta_{\mathrm{p},i} = \omega_\mathrm{p} - \omega_i$ are the detunings of the pump laser from the optical resonances.
Since \eqref{eq:H_a} does not couple $a_{i,\sigma}$ and $a_{i',\sigma'}^\dag$ in the equations of motion, we define the state vectors
\begin{align}
	\mathbf{q}_\mathrm{a} =&\begin{bmatrix}
		a_\mathrm{1,c}&
		a_\mathrm{1,cc}&
		a_\mathrm{2,c}&
		a_\mathrm{2,cc}&
	\end{bmatrix}^T
	\\
	\mathbf{q}_\mathrm{a,in} =&\begin{bmatrix}
		a_\mathrm{in,c}&
		a_\mathrm{in,cc}&
	\end{bmatrix}^T
	\\
	\mathbf{q}_\mathrm{a,out} =&\begin{bmatrix}
		a_\mathrm{out,c}&
		a_\mathrm{out,cc}&
	\end{bmatrix}^T\text{.}
\end{align}

To set up the equations of motion as in \eqref{eq:time_domain_eom}, we define the optical input-output matrix
\begin{equation}
	K_\mathrm{a} =\begin{bmatrix}
		\sqrt{\kappa_\mathrm{a,ex}}&0\\
		0&\sqrt{\kappa_\mathrm{a,ex}}\\
		0&0&\\
		0&0&\\
	\end{bmatrix},
\end{equation}
and the optical time-domain dynamic matrix
\begin{eqnarray}
	M'_\mathrm{a}&=&
	\begin{bmatrix}
		+i\Delta_\mathrm{p,1}&+i\nu_1&0&+i\mu\\
		+i\nu_1&+i\Delta_\mathrm{p,1}&+i\mu&0\\
		0&+i\mu&+i\Delta_\mathrm{p,2}&+i\nu_2\\
		+i\mu&0&+i\nu_2&+i\Delta_\mathrm{p,2}
	\end{bmatrix}
	\nonumber\\
	&&+
	\begin{bmatrix}
		-\frac{\kappa_\mathrm{1}}{2}&0&0&0\\
		0&-\frac{\kappa_\mathrm{1}}{2}&0&0\\
		0&0&-\frac{\kappa_\mathrm{2}}{2}&0\\
		0&0&0&-\frac{\kappa_\mathrm{2}}{2}
	\end{bmatrix}
	\text{,}
\end{eqnarray}
where the imaginary part of $\nu_i$ is included \cite{li_backscattering_2016}.

The frequency-domain dynamical matrix can then be written as
\begin{equation}
	M_\mathrm{a}(\omega) =\begin{bmatrix}
		\frac{1}{\chi_\mathrm{1}(\omega)}&-i\nu_1&0&-i\mu\\
		-i\nu_1&\frac{1}{\chi_\mathrm{1}(\omega)}&-i\mu&0\\
		0&-i\mu&\frac{1}{\chi_\mathrm{2}(\omega)}&-i\nu_2\\
		-i\mu&0&-i\nu_2&\frac{1}{\chi_\mathrm{2}(\omega)}
	\end{bmatrix} \label{eq:M_optical}
	\text{,}
\end{equation}
with the optical response functions
\begin{eqnarray}
	\chi_{i}(\omega) &=& \frac{1}{\kappa_{i}/2 - i \left( \omega+\Delta_{\mathrm{p},i} \right)}
\end{eqnarray}

For optical spectroscopy in transmission or reflection, we set $\omega=0$, and $\omega_\mathrm{p}$ corresponds to the optical probe frequency.
With \eqref{eq:M_optical} we can calculate the S-matrix elements using \eqref{eq:S_matrix}. To fit optical resonances in transmission we use $S_{11}$. In principle, $S_{21}$ corresponds to the reflected pump amplitude. However, real spectra are dominated by other, frequency-dependent optical reflections from various interfaces, e.g. fiber-chip coupler, that make it difficult to reliably fit optical reflection spectra.
\par
In the case when $\nu_i = 0$, solving for the eigenvectors of \eqref{eq:M_optical} yields the symmetric and antisymmetric eigenmodes $a_\mathrm{+}$ and $a_\mathrm{-}$ of the photonic molecule mentioned in the main text, identical for clockwise and counterclockwise propagation. When $\nu_i \neq 0$, these modes are perturbed and further split by about $\nu_1 + \nu_2$. Instead of working with the eigenmodes, we choose to treat the problem in the uncoupled basis $a_{i,\sigma}$ because it simplifies the external couplings.

\subsection{Transduction model}
\label{sec:full_model}

\begin{figure}[b]
	\centering
	\includegraphics{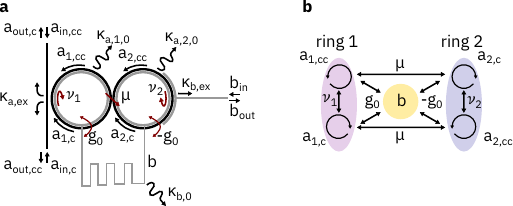}
	\caption[Full transduction model]{\textbf{Full transduction model.}
		(a) Schematic of the device. (b) Effective mode coupling diagram.
	}
	\label{fig:full_model}
\end{figure}

We now use the formalism from Sec. \ref{sec:open_coupled_system} to describe the full transducer by including microwave resonator and the electro-optic interaction (Fig.~\ref{fig:full_model}).
For clarity, we divide the system Hamiltonian into its optical, microwave, and electro-optic parts, i.e.
\begin{equation}
	H = H_\mathrm{a}+H_\mathrm{b}+H_\mathrm{eo},
\end{equation}
with microwave Hamiltonian
\begin{equation}
	H_\mathrm{b}=\hbar\omega_\mathrm{b}b^\dagger b \label{eq:H_b}
\end{equation}
and the electro-optic interaction Hamiltonian
\begin{align}
	H_\mathrm{eo}=
	&-\hbar g_\mathrm{0} \sum_{\sigma = \mathrm{c,cc}} a_\mathrm{1,\sigma}^\dag  a_\mathrm{1,\sigma} \left(b+b^\dag\right)
	\nonumber\\
	&+\hbar g_\mathrm{0} \sum_{\sigma = \mathrm{c,cc}} a_\mathrm{2,\sigma}^\dag  a_\mathrm{2,\sigma} \left(b+b^\dag\right) .
\end{align}
Note that the sign difference in the electro-optic couplings between the two optical rings originates from the antisymmetric microwave mode shape.
Next, we linearize the electro-optic interaction assuming that the optical fields are driven to large amplitude by the pump laser \cite{aspelmeyer_cavity_2014}, i.e. $a_{i,\sigma}\rightarrow\alpha_{i,\sigma}+a_{i,\sigma}$ with $\alpha_{i,\sigma}=\expval{a_{i,\sigma}}$.
This results in the linearized electro-optic interaction
\begin{align}
	H_\mathrm{eo}=
	&-\hbar g_\mathrm{0} \sum_{\sigma = \mathrm{c,cc}} \left( \alpha^{\ast}_\mathrm{1,\sigma} a_\mathrm{1,\sigma} + \alpha_\mathrm{1,\sigma} a_\mathrm{1,\sigma}^\dag  \right) \left(b+b^\dag\right)
	\nonumber\\
	&+\hbar g_\mathrm{0} \sum_{\sigma = \mathrm{c,cc}} \left(\alpha^{\ast}_\mathrm{2,\sigma} a_\mathrm{2,\sigma} + \alpha_\mathrm{2,\sigma} a_\mathrm{2,\sigma}^\dag \right) \left(b+b^\dag\right) .
\end{align}

With the full system Hamiltonian derived, we can now apply the formalism of Sec.~\ref{sec:open_coupled_system} to obtain the scattering matrix.
We thus define the vectors
\begin{align}
	\mathbf{q} =&\begin{bmatrix}
		a_\mathrm{1,c}&
		a_\mathrm{1,c}^\dagger&
		a_\mathrm{1,cc}&
		a_\mathrm{1,cc}^\dagger&
		a_\mathrm{2,c}&
		a_\mathrm{2,c}^\dagger&
		a_\mathrm{2,cc}&
		a_\mathrm{2,cc}^\dagger&
		b&
		b^\dagger
	\end{bmatrix}^T
	\\
	\mathbf{q}_\mathrm{in} =&\begin{bmatrix}
		a_\mathrm{in,c}&
		a_\mathrm{in,c}^\dagger&
		a_\mathrm{in,cc}&
		a_\mathrm{in,cc}^\dagger&
		b_\mathrm{in}&
		b_\mathrm{in}^\dagger
	\end{bmatrix}^T
	\\
	\mathbf{q}_\mathrm{out} =&\begin{bmatrix}
		a_\mathrm{out,c}&
		a_\mathrm{out,c}^\dagger&
		a_\mathrm{out,cc}&
		a_\mathrm{out,cc}^\dagger&
		b_\mathrm{out}&
		b_\mathrm{out}^\dagger
	\end{bmatrix}^T
	\text{.}
\end{align}
\begin{widetext}
The input-output matrix for the full system is given by
\begin{equation}
K =\begin{bmatrix}
	\sqrt{\kappa_\mathrm{a,ex}}&0&0&0&0&0\\
	0&\sqrt{\kappa_\mathrm{a,ex}}&0&0&0&0\\
	0&0&\sqrt{\kappa_\mathrm{a,ex}}&0&0&0\\
	0&0&0&\sqrt{\kappa_\mathrm{a,ex}}&0&0\\
	0&0&0&0&0&0\\
	0&0&0&0&0&0\\
	0&0&0&0&0&0\\
	0&0&0&0&0&0\\
	0&0&0&0&\sqrt{\kappa_\mathrm{b,ex}}&0\\
	0&0&0&0&0&\sqrt{\kappa_\mathrm{b,ex}}\\
\end{bmatrix}
\text{.}
\end{equation}

The time-domain dynamical matrix is
\begin{equation}
	M'=M'_\mathrm{a}+M'_\mathrm{b}+M'_\mathrm{eo},
\end{equation}
with optical part
\begin{eqnarray}
	M'_\mathrm{a}&=&
	\begin{bmatrix}
		+i\Delta_\mathrm{p,1}&0&+i\nu_1&0&0&0&+i\mu&0&0&0\\
		0&-i\Delta_\mathrm{p,1}&0&-i\nu^{\ast}_1&0&0&0&-i\mu&0&0\\
		+i\nu_1&0&+i\Delta_\mathrm{p,1}&0&+i\mu&0&0&0&0&0\\
		0&-i\nu^{\ast}_1&0&-i\Delta_\mathrm{p,1}&0&-i\mu&0&0&0&0\\
		0&0&+i\mu&0&+i\Delta_\mathrm{p,2}&0&+i\nu_1&0&0&0\\
		0&0&0&-i\mu&0&-i\Delta_\mathrm{p,2}&0&-i\nu^{\ast}_2&0&0\\
		+i\mu&0&0&0&+i\nu_2&0&+i\Delta_\mathrm{p,2}&0&0&0\\
		0&-i\mu&0&0&0&-i\nu^{\ast}_2&0&-i\Delta_\mathrm{p,2}&0&0\\
		0&0&0&0&0&0&0&0&0&0\\
		0&0&0&0&0&0&0&0&0&0
	\end{bmatrix}
	\nonumber
	\\
	&& +
	\begin{bmatrix}
		-\frac{\kappa_\mathrm{1}}{2}&0&0&0&0&0&0&0&0&0\\
		0&-\frac{\kappa_\mathrm{1}}{2}&0&0&0&0&0&0&0&0\\
		0&0&-\frac{\kappa_\mathrm{1}}{2}&0&0&0&0&0&0&0\\
		0&0&0&-\frac{\kappa_\mathrm{1}}{2}&0&0&0&0&0&0\\
		0&0&0&0&-\frac{\kappa_\mathrm{2}}{2}&0&0&0&0&0\\
		0&0&0&0&0&-\frac{\kappa_\mathrm{2}}{2}&0&0&0&0\\
		0&0&0&0&0&0&-\frac{\kappa_\mathrm{2}}{2}&0&0&0\\
		0&0&0&0&0&0&0&-\frac{\kappa_\mathrm{2}}{2}&0&0\\
		0&0&0&0&0&0&0&0&0&0\\
		0&0&0&0&0&0&0&0&0&0
	\end{bmatrix},
\end{eqnarray}
%\begin{equation}
%	M'_\mathrm{a}
%	\begin{bmatrix}
%		-\frac{\kappa_\mathrm{1}}{2}+i\Delta_\mathrm{p,1}&0&+i\nu_1&0&0&0&+i\mu&0&0&0\\
%		0&-\frac{\kappa_\mathrm{1}}{2}-i\Delta_\mathrm{p,1}&0&-i\nu^{\ast}_1&0&0&0&-i\mu&0&0\\
%		+i\nu_1&0&-\frac{\kappa_\mathrm{1}}{2}+i\Delta_\mathrm{p,1}&0&+i\mu&0&0&0&0&0\\
%		0&-i\nu^{\ast}_1&0&-\frac{\kappa_\mathrm{1}}{2}-i\Delta_\mathrm{p,1}&0&-i\mu&0&0&0&0\\
%		0&0&+i\mu&0&-\frac{\kappa_\mathrm{2}}{2}+i\Delta_\mathrm{p,2}&0&+i\nu_1&0&0&0\\
%		0&0&0&-i\mu&0&-\frac{\kappa_\mathrm{2}}{2}-i\Delta_\mathrm{p,2}&0&-i\nu^{\ast}_2&0&0\\
%		+i\mu&0&0&0&+i\nu_2&0&-\frac{\kappa_\mathrm{2}}{2}+i\Delta_\mathrm{p,2}&0&0&0\\
%		0&-i\mu&0&0&0&-i\nu^{\ast}_2&0&-\frac{\kappa_\mathrm{2}}{2}-i\Delta_\mathrm{p,2}&0&0\\
%		0&0&0&0&0&0&0&0&0&0\\
%		0&0&0&0&0&0&0&0&0&0
%	\end{bmatrix},
%\end{equation}
microwave part
\begin{equation}
	M'_\mathrm{b}
	=
	\begin{bmatrix}
		0&0&0&0&0&0&0&0&0&0\\
		0&0&0&0&0&0&0&0&0&0\\
		0&0&0&0&0&0&0&0&0&0\\
		0&0&0&0&0&0&0&0&0&0\\
		0&0&0&0&0&0&0&0&0&0\\
		0&0&0&0&0&0&0&0&0&0\\
		0&0&0&0&0&0&0&0&0&0\\
		0&0&0&0&0&0&0&0&0&0\\
		0&0&0&0&0&0&0&0&-\frac{\kappa_\mathrm{b}}{2}-i\omega_\mathrm{b}&0\\
		0&0&0&0&0&0&0&0&0&-\frac{\kappa_\mathrm{b}}{2}+i\omega_\mathrm{b}
	\end{bmatrix},
\end{equation}
and the electro-optic interaction part
\begin{equation}
	M'_\mathrm{eo}
	=
	g_\mathrm{0}
	\cdot
	\begin{bmatrix}
		0&0&0&0&0&0&0&0&+i\alpha_\mathrm{1,c}&+i\alpha_\mathrm{1,c}\\
		0&0&0&0&0&0&0&0&-i\alpha^{\ast}_\mathrm{1,c}&-i\alpha^{\ast}_\mathrm{1,c}\\
		0&0&0&0&0&0&0&0&+i\alpha_\mathrm{1,cc}&+i\alpha_\mathrm{1,cc}\\
		0&0&0&0&0&0&0&0&-i\alpha^{\ast}_\mathrm{1,cc}&-i\alpha^{\ast}_\mathrm{1,cc}\\
		0&0&0&0&0&0&0&0&-i\alpha_\mathrm{2,c}&-i\alpha_\mathrm{2,c}\\
		0&0&0&0&0&0&0&0&+i\alpha^{\ast}_\mathrm{2,c}&+i\alpha^{\ast}_\mathrm{2,c}\\
		0&0&0&0&0&0&0&0&-i\alpha_\mathrm{2,cc}&-i\alpha_\mathrm{2,cc}\\
		0&0&0&0&0&0&0&0&+i\alpha^{\ast}_\mathrm{2,cc}&+i\alpha^{\ast}_\mathrm{2,cc}\\
		+i\alpha^{\ast}_\mathrm{1,c}&+i\alpha_\mathrm{1,c}&+i\alpha^{\ast}_\mathrm{1,cc}&+i\alpha_\mathrm{1,cc}&-i\alpha^{\ast}_\mathrm{2,c}&-i\alpha_\mathrm{2,c}&-i\alpha^{\ast}_\mathrm{2,cc}&-i\alpha_\mathrm{2,cc}&0&0\\
		-i\alpha^{\ast}_\mathrm{1,c}&-i\alpha_\mathrm{1,c}&-i\alpha^{\ast}_\mathrm{1,cc}&-i\alpha_\mathrm{1,cc}&+i\alpha^{\ast}_\mathrm{2,c}&+i\alpha_\mathrm{2,c}&+i\alpha^{\ast}_\mathrm{2,cc}&+i\alpha_\mathrm{2,cc}&0&0
	\end{bmatrix}
	\text{.}
\end{equation}
The frequency-domain dynamical matrix is given by
\begin{equation}
	M(\omega) = M_\mathrm{a}(\omega) + M_\mathrm{b}(\omega) + M_\mathrm{eo}. \label{eq:M_full}
\end{equation}
with optical part
\begin{equation}
	M_\mathrm{a}
	=
	\
	\begin{bmatrix}
		\frac{1}{\chi_\mathrm{1}(\omega)}&0&-i\nu_1&0&0&0&-i\mu&0&0&0\\
		0&\frac{1}{\chi^{\ast}_\mathrm{1}\left(-\omega\right)}&0&+i\nu^{\ast}_1&0&0&0&+i\mu&0&0\\
		-i\nu_1&0&\frac{1}{\chi_\mathrm{1}(\omega)}&0&-i\mu&0&0&0&0&0\\
		0&+i\nu^{\ast}_1&0&\frac{1}{\chi^{\ast}_\mathrm{1}\left(-\omega\right)}&0&+i\mu&0&0&0&0\\
		0&0&-i\mu&0&\frac{1}{\chi_\mathrm{2}(\omega)}&0&-i\nu_2&0&0&0\\
		0&0&0&+i\mu&0&\frac{1}{\chi^{\ast}_\mathrm{2}\left(-\omega\right)}&0&+i\nu^{\ast}_2&0&0\\
		-i\mu&0&0&0&-i\nu_2&0&\frac{1}{\chi_\mathrm{2}(\omega)}&0&0&0\\
		0&+i\mu&0&0&0&+i\nu^{\ast}_2&0&\frac{1}{\chi^{\ast}_\mathrm{2}\left(-\omega\right)}&0&0\\
		0&0&0&0&0&0&0&0&0&0\\
		0&0&0&0&0&0&0&0&0&0
	\end{bmatrix}
	\text{,}
\end{equation}

\end{widetext}
microwave part
\begin{equation}
	M_\mathrm{b}
	=
	\begin{bmatrix}
		0&0&0&0&0&0&0&0&0&0\\
		0&0&0&0&0&0&0&0&0&0\\
		0&0&0&0&0&0&0&0&0&0\\
		0&0&0&0&0&0&0&0&0&0\\
		0&0&0&0&0&0&0&0&0&0\\
		0&0&0&0&0&0&0&0&0&0\\
		0&0&0&0&0&0&0&0&0&0\\
		0&0&0&0&0&0&0&0&0&0\\
		0&0&0&0&0&0&0&0&\frac{1}{\chi_\mathrm{b}(\omega)}&0\\
		0&0&0&0&0&0&0&0&0&\frac{1}{\chi^{\ast}_\mathrm{b}\left(-\omega\right)}
	\end{bmatrix}
	\text{,}
\end{equation}
and $M_\mathrm{eo}=-M'_\mathrm{eo}$.
Here, we also defined the microwave response function
\begin{eqnarray}
	\chi_\mathrm{b}(\omega) = \frac{1}{\kappa_\mathrm{b}/2 - i\left(\omega - \omega_\mathrm{b}\right)}.
\end{eqnarray}

We can now solve the equations of motion for this linearized system.
In the first step, we calculate the optical resonator pump amplitudes using \eqref{eq:M_optical} by
\begin{equation}
	\mathbf{q}_\mathrm{p}= - M_a(0)^{-1} K_a \mathbf{q}_\mathrm{in,p}.
\end{equation}
where 
\begin{equation}
		\mathbf{q}_\mathrm{p} = \left[\alpha_{1,\mathrm{c}}, \alpha_{1,\mathrm{cc}}, \alpha_{2,\mathrm{c}}, \alpha_{2,\mathrm{cc}}\right]^T.
\end{equation}
With the pump field injected in forward (clockwise) direction only, the pump input vector is
\begin{equation}
	\mathbf{q}_\mathrm{in,p}= \sqrt{\frac{P_\mathrm{p}}{\hbar\omega_\mathrm{p}}} \begin{bmatrix}
		1&
		0&
		0&
		0&
	\end{bmatrix}^T,
\end{equation}
with on-chip pump power $P_\mathrm{p}$.
\par
In the second step, we insert the pump resonator amplitudes $\alpha_{i,\sigma}$ into the full matrix, and solve the full problem using \eqref{eq:M_full} and \eqref{eq:S_matrix}.
The optical sideband transmission and the microwave reflection can then be calculated as:
\begin{eqnarray}
	T_\mathrm{opt} &=& \left|\frac{a_\mathrm{out,c}}{a_\mathrm{in,c}}\right|^2 = |S_{11}(\omega)|^2\\
	R_\mathrm{mw} &=& \left|\frac{b_\mathrm{out}}{b_\mathrm{in}}\right|^2 = |S_{99}(\omega)|^2,
\end{eqnarray}
respectively. The transduction efficiencies for the anti-Stokes and Stokes sidebands, respectively, are calculated using
\begin{eqnarray}
	\eta_\mathrm{AS} &=&\left|\frac{a_\mathrm{out,c}}{b_\mathrm{in}}\right|^2 = |S_{19}(\omega)|^2\\
	\eta_\mathrm{S} &=& \left|\frac{a^\dagger_\mathrm{out,c}}{b_\mathrm{in}}\right|^2 = |S_{29}(\omega)|^2 .
\end{eqnarray}

\subsection{Microwave spurious mode coupling}
\label{sec:coupled_MW_mode_model}

\begin{figure}[b]
	\centering
	\includegraphics{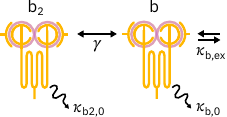}
	\caption[Coupled microwave mode schematic]{\textbf{Coupled microwave mode schematic.}
		The main microwave mode $b_\mathrm{1}$ is coupled to the bus waveguide at a rate $\kappa_\mathrm{b,ex}$ and is additionally coupled to a spurious mode $b_\mathrm{2}$ at a rate $\gamma$.
		The intrinsic loss rates of the two modes are $\kappa_\mathrm{b1,0}$ and $\kappa_\mathrm{b2,0}$, respectively.
	}
	\label{fig:MW_coupled_model_schematic}
\end{figure}

We now derive the linear response for the coupling between the microwave resonator and the spurious mode detailed in Sec.~\ref{sec:transduction_mw_charact}.
The system comprises two microwave resonators $b$ and $b_\mathrm{2}$ with resonance frequencies $\omega_\mathrm{b}$ and $\omega_\mathrm{b,2}$, and intrinsic loss rates $\kappa_\mathrm{b,0}$ and $\kappa_\mathrm{b,2,0}$, respectively.
The main microwave mode $b$ is coupled to the bus waveguide at a rate $\kappa_\mathrm{b,ex}$.
Both resonators are coupled at a rate $\gamma$.
\par
The system Hamiltonian reads
\begin{eqnarray}
	H=
	&+\hbar\omega_\mathrm{b} b^\dag b + \hbar\omega_\mathrm{b,2} b_\mathrm{2}^\dag b_\mathrm{2}\nonumber\\
	&+\hbar\gamma \left(b^\dag b_\mathrm{2} + b_\mathrm{2}^\dag b \right).
\end{eqnarray}
We define the input and output vectors of the MW field
\begin{equation}
	\mathbf{q}_\mathrm{in} =\begin{bmatrix}
		b_\mathrm{in} & b_\mathrm{in}^\dag
	\end{bmatrix}^T
	\text{,}
\end{equation}
and
\begin{equation}
	\mathbf{q}_\mathrm{out} =\begin{bmatrix}
		b_\mathrm{out} & b_\mathrm{out}^\dag 
	\end{bmatrix}^T
	\text{.}
\end{equation}

We infer the scattering matrix elements $S$ using \eqref{eq:S_matrix} with the input-output matrix
\begin{equation}
	K_b =\begin{bmatrix}
		\sqrt{\kappa_\mathrm{b,ex}}&0&\\
		0&\sqrt{\kappa_\mathrm{b,ex}}&\\
		0&0&\\
		0&0&\\
	\end{bmatrix}
	\text{,}
\end{equation}
and the time-domain dynamical matrix
\begin{eqnarray}
	M_b' &=&\begin{bmatrix}
		-i\omega_\mathrm{b}&0&+i\gamma&0\\
		0&+i\omega_\mathrm{b}&0&-i\gamma\\
		+i\gamma&0&-i\omega_\mathrm{b,2}&0\\
		0&-i\gamma&0&+i\omega_\mathrm{b,2}\\
	\end{bmatrix}
	\nonumber
	\\
	+&&
	\begin{bmatrix}
		-\frac{\kappa_\mathrm{b}}{2}&0&0&0\\
		0&-\frac{\kappa_\mathrm{b}}{2}&0&0\\
		0&0&-\frac{\kappa_\mathrm{b,2,0}}{2}&0\\
		0&0&0&-\frac{\kappa_\mathrm{b,2,0}}{2}\\
	\end{bmatrix}
	\text{,}
\end{eqnarray}
from which the frequency-domain dynamical matrix follows, i.e.
\begin{equation}
	M_b(\omega) =\begin{bmatrix}
		\frac{1}{\chi_\mathrm{b}(\omega)}&0&-i\gamma&0\\
		0&\frac{1}{\chi^{\ast}_\mathrm{b}\left(-\omega\right)}&0&+i\gamma\\
		-i\gamma&0&\frac{1}{\chi_\mathrm{b,2}(\omega)}&0\\
		0&+i\gamma&0&\frac{1}{\chi^{\ast}_\mathrm{b,2}\left(-\omega\right)}\\
	\end{bmatrix}
	\text{.}
\end{equation}
From this we determine the S-matrix $S_b$. In Sec.~\ref{sec:transduction_mw_charact} we fit the microwave spectra using
\begin{equation}
	R_\mathrm{mw}(\omega) = \left|\frac{b_\mathrm{out}}{b_\mathrm{in}}\right|^2 = |S_{b, 11}(\omega)|^2.
\end{equation}

\section{Sample preparation}
\label{sec:sample_preparation}

\subsection{Sample fabrication}
\label{sec:sample_fabrication}

\begin{figure}[b]
	\centering
	\includegraphics{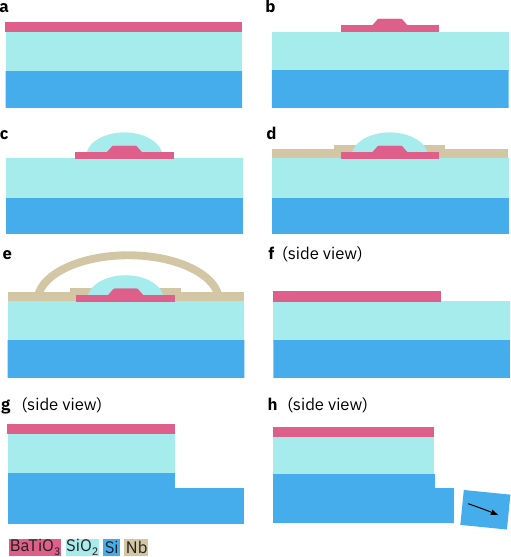}
	\caption{\textbf{Fabrication flow.}
		(a) initial material stack.
		(b) optical ridge waveguide with etched slab.
		(c) Oxide clad ridge waveguide.
		(d) Added main electrode layer.
		(e) Added superconducting air bridges.
		(f) Chip edge prior to facet fabrication.
		(g) Dry-etched facet.
		(h) Diced chip edge.
	}\label{figS:fabrication_process}
\end{figure}

We have developed a process to fabricate low-loss, high-confinement photonic waveguides and resonators in \BTO\ and integrate them with Nb superconducting circuits.
We start with a 225-nm thick \BTO\ thin film bonded on 3 $\mu$m of SiO$_2$ on silicon (Fig.~\ref{figS:fabrication_process}a).
BaTiO$_3$ ridge waveguides are patterned using electron beam lithography and dry etched to produce a 100-nm thick slab and 125-nm thick ridge (b).
The waveguides are protected with a 1.5 -$\mu$m cladding of SiO$_2$ (c).
Electrodes are subsequently fabricated by sputtering Nb and patterning using electron beam lithography (d).
For waveguide-superconductor and superconductor-superconductor crossings, Nb airbridges are fabricated (e) by sputtering Nb and subsequent optical lithography (e).
In the last step, optical facets are etched such that waveguides reach to the edge of the chip for low-loss fiber coupling (f shows a side-view of the chip before optical facet fabrication).
Optical facets are fabricated using optical lithography, SiO$_2$ dry etching and deep reactive-ion etching into the Si substrate (g).
The chip is diced to facilitate access to the facet (h).

\begin{figure}[b]
	\centering
	\includegraphics{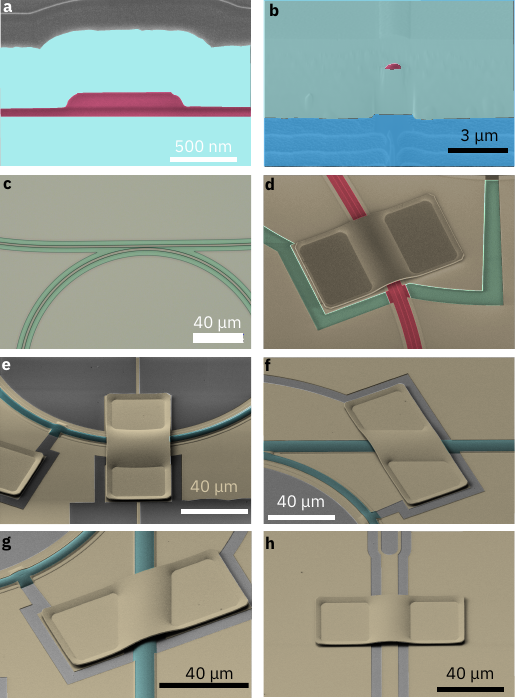}
	\caption{\textbf{Fabrication results.}
		False colored images scanning electron micrographs:
		(a) Cross-section of an oxide clad \BTO\ ridge waveguide.
		(b) Optical chip facet with \BTO\ taper end.
		(c) Optical micrograph of directional coupler section between ring and bus waveguide.
	}\label{figS:fabrication_results}
\end{figure}

Using this process, a variety of elements are fabricated to produce the optical and electronic circuits constituting the transducer.
Fig.~\ref{figS:fabrication_results}a shows a scanning electron micrograph (SEM) of a cross section of an optical ridge waveguide, Fig.~\ref{figS:fabrication_results}b a close-up view of the waveguide facet, Fig.~\ref{figS:fabrication_results}c the optical waveguide coupling region seen from the top, and Fig.~\ref{figS:fabrication_results}d a Nb air bridge of an air-clad device.
Fig.~\ref{figS:fabrication_results} depict the device under test measured in this study, with oxide-clad waveguides.

\subsection{Packaging}
\label{sec:packaging}

\begin{figure}[]
	\centering
	\includegraphics{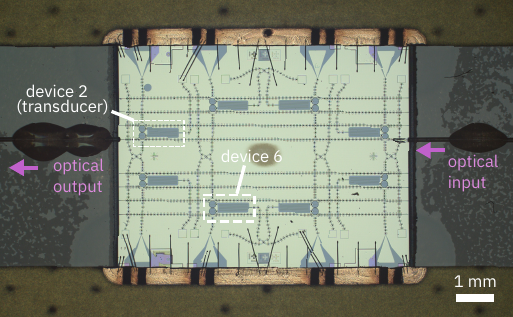}
	\caption{\textbf{Cryogenic microwave-optical co-packaged chip.}
		(a) Sample chip with optical edge-coupling gluing points.
		(b) Electrical on-chip circuit and wirebonds to printed circuit board.
		Lines to the inductor are highlighted in red, and to ring electrodes in yellow.
	}\label{fig:cryo_package}
\end{figure}

In- and output high numerical aperture fibers (UHNA7, mode-field diameter $3.2~\mu$m) are first aligned to the chip by optimizing the optical transmission through the bus waveguide.
The fiber is then glued to the chip facet and supsequently glued to the package further back with additional gluing points as shown in Fig.~\ref{fig:cryo_package}.
\par
Next, the on-chip electrical circuit is connected to the printed circuit board with wire bonds (Fig.~\ref{fig:cryo_package}).
In this case, the transducer device (device 2) is fully packaged with three leads for microwave reflection, photonic molecule tuning and domain biasing.
A second device (device 6) is packaged with connnected microwave reflection and domain biasing port for microwave characterization.

\section{Experimental setups}
\subsection{Cryostat wiring}

\begin{figure}[]
	\centering
	\includegraphics{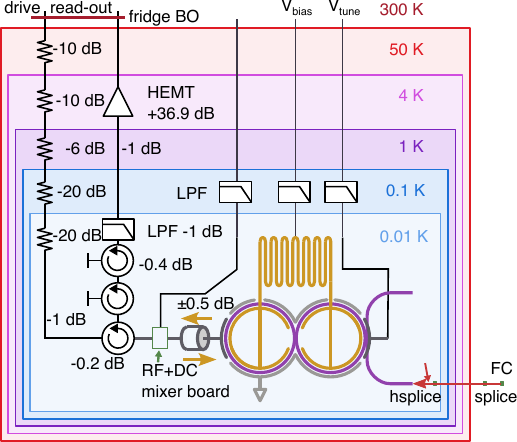}
	\caption[Cryostat wiring with device under test]{\textbf{Cryostat wiring with device under test.}
		Microwave signals are delivered to the device through the drive input at the cryostat breakout (fridge BO) via coaxial cables and a series of attenuators, a circulator and a RF+DC mixer board.
		The reflected microwave signal is passed through two isolators, a low pass filter (LPF) and a high electron mobility amplifier (HEMT) before being delivered to the read-out channel at the fridge breakout.
		Bias and tuning voltages $V_\mathrm{bias}$ and $V_\mathrm{tune}$, respectively, are delivered to the device via twisted wire pairs through low pass filters (LPF).
		Optical signals are delivered through a fiber going from room temperature into the bottom probe of the cryostat, which is spliced to the packaged device fiber before sample mounting.
	}
	\label{fig:DUT_in_fridge}
\end{figure}

The wiring inside the cryostat is shown in Fig.~\ref{fig:DUT_in_fridge}.
Our cryostat (BlueFors LD400) features a fast sample exchange bottom probe mechanism.
While the electrical signals are delivered from the top of the cryostat at the cryostat breakout connectors (fridge BO), the optical fiber is fed through the bottom probe to the device.
\par
For microwave drive signal delivery, SCuNi-CuNi semi rigid coaxial cables lead from the breakout connectors at the top down to the mixing chamber plate of the cryostat.
They are attenuated at the different stages according to the temperature gradient to the previous stage (i.e. from 4 to $\SI{1}{\kelvin}$) only $\SI{-6}{\decibel}$ while from 1 to $\SI{100}{\milli\kelvin}$ by $\SI{-20}{\decibel}$.
At the mixing chamber plate a circulator is connected which in turn passes the signals through a custom mixer board for the ability to combine microwave and direct current (dc) electrical signals up to $\SI{\pm30}{\volt}$.
The reflected microwave signal is sent through two isolators via the circulator, which, importantly, are all thermalized to the mixing chamber plate.
The signal is then filtered with a low pass filter and delivered to a high electron mobility amplifier (HEMT) at the $\SI{4}{\kelvin}$ stage via NbTi superconducting coaxial cables, which are thermalized to each stage with $\SI{0}{\decibel}$ attenuators.
To avoid bulky fiber connectors, the high numerical aperture fibers (UHNA7) of the packaged sample are spliced to the standard single mode fibers (SMF-28) using a hybrid splicing program that creates a smooth transition between the two (hsplice), with typical losses on the order of $\SI{-1}{\decibel}$.
A second standard splice between two SMF-28 fibers connects the cryostat fiber to the fiber connector (FC).

\subsection{Calibration of transduction efficiency}
\label{sec:efficiency_calibration}

The optical input power arriving at the fiber inside the probe is calibrated by comparing cw optical heating of the cryostat to that using the resistive probe heater, from which the factor $+11.3$~dB between the TLPM\_dev\_in power meter reading and power in the fiber is determined (Fig. \ref{fig:CW_setup}).
This also characterizes the transmission of the dilfridge A fiber with its standard (splice) and hybrid (hsplice) splices (Fig. \ref{fig:DUT_in_fridge}) to be $-0.7$~dB using additional loss characterization of the circulator and FPC\_dev\_in.
For pulsed measurements (Fig. \ref{fig:LT_pulsed_setup}), the voltage of the monitoring photodiode (150 MHz PD) is calibrated by measuring the total power output from FPC\_dev\_in. The ratio between optical signal and pump amplitude is measured via the ratio of their respective heterodyne sidebands on the ESA.
Depending on where the optical power is measured, the optical gain ($G_\mathrm{opt}$) is defined.
\par
Next, the detected optical power is calibrated by determining the factor between the power injected into FPC\_dev\_in and the TLPM\_dev\_out power meter reading, which is $+15.7$~dB. The measured heterodyne signal amplitude on the ESA is then calibrated relative to this power meter, giving a factor of $-14.7$~dB between them at constant LO power, where the ratio of the two gives the heterodyne gain ($G_\mathrm{het}$)
\par
The microwave source power (RF source 1+ power amp, Fig. \ref{fig:CW_setup}) is calibrated with respect to the ESA, as well as the room temperature readout and amplification line (RT read amp line). The loss through the cryostat wiring (in Fig. \ref{fig:DUT_in_fridge} from drive to read-out) is determined and compared to the expected losses and gains of all components, and the difference added equally to the drive ($G_\mathrm{drive}$) and read-out ($G_\mathrm{read}$) branch.
\par
The photon number conversion efficiencies are then calculated by multiplying the detected powers with all determined gain factors in the signal chain, i.e.
\begin{eqnarray}
	\eta_\mathrm{opt\leftarrow mw} = \frac{P_\mathrm{ESA}}{G_\mathrm{het}G_\mathrm{opt\leftarrow mw} G_\mathrm{drive} P_\mathrm{mw,in}}
	\\
	\eta_\mathrm{mw\leftarrow opt}=\frac{P_\mathrm{ESA}}{G_\mathrm{read}G_\mathrm{mw\leftarrow opt}G_\mathrm{opt}P_\mathrm{opt,in}}\text{,}
\end{eqnarray}
where $P_\mathrm{ESA}$ is the detected output power on the ESA, and $P_\mathrm{mw,in}$ and $P_\mathrm{opt,in}$ are the microwave and optical input powers, respectively.
The number conversion gains are defined by $G_\mathrm{opt\leftarrow mw}=1/G_\mathrm{mw\leftarrow opt} = \omega_\mathrm{a}/\omega_\mathrm{b}$.

%\subsection{Continuous wave microwave characterization}
%
%\label{sec:MW_CW_setup}
%
%\begin{figure}[]
%	\centering
%	\includegraphics{figures_appendix/MW_CW_setup_v5}
%	\caption[Continuous wave microwave characterization setup]{\textbf{Continuous wave microwave characterization setup.}
%		Electrical network for (a) Low, medium and (b) high power microwave reflection measurements using a vector network analyzer (VNA).
%	}
%	\label{fig:MW_CW_setup}
%\end{figure}
%
%For continuous wave (cw) microwave reflection measurements, a vector network analyzer is used. Depending on the required incident microwave power at the device, the measurement network is set up in a low/medium or high power version. While for the former (Fig.~\ref{fig:MW_CW_setup}a) the return signal from the cryostat is amplified at room temperature, for the latter (Fig.~\ref{fig:MW_CW_setup}b) a power amplifier amplifies the VNA signal before being delivered to the cryostat drive line. Note, that the drive power amplification was only necessary due to the strong attenuation with a total of $\SI{-66}{\decibel}$ in the drive line (Fig.~\ref{fig:DUT_in_fridge}). The quoted transmission values for each cable where either measured or taken from the cable datasheet at the microwave resonance frequency around $\SI{6.8}{\giga\hertz}$.

\subsection{Pulsed microwave characterization}

To characterize the microwave reflection under pulsed optical pumping, we use a setup very similar to the pulsed transduction setup.
The main difference is, that only the pump laser is sent to the device, and instead of measuring the transduced signal, we measure the high power microwave reflection from the device.
This means, that only the pulse power sum measurement with the monitoring diode is necessary.
Due to the high microwave probe power levels, we remove the room temperature amplifier from the detection chain (RT read amp line).
The measurement and pulse sequence is virtually identical to the pulsed transduction measurement described in Sec. \ref{sec:pulsed_transduction_experiment}.
Time-resolved microwave spectra are recorded by running the full sequence for each frequency step.
For time efficiency, only a narrow frequency span of a few $\unit{\mega\hertz}$ because the main parameters of interest are the extinction and frequency shift.
To calculate the extinction, a larger span measurement is taken once using the same routine and setup to obtain the required reference level.
Note, that the setup is readily converted to the pulsed transduction configuration, which can be used to monitor the fiber-to-chip coupling efficiency between experiments.

\begin{figure*}[t]
	\centering
	\includegraphics{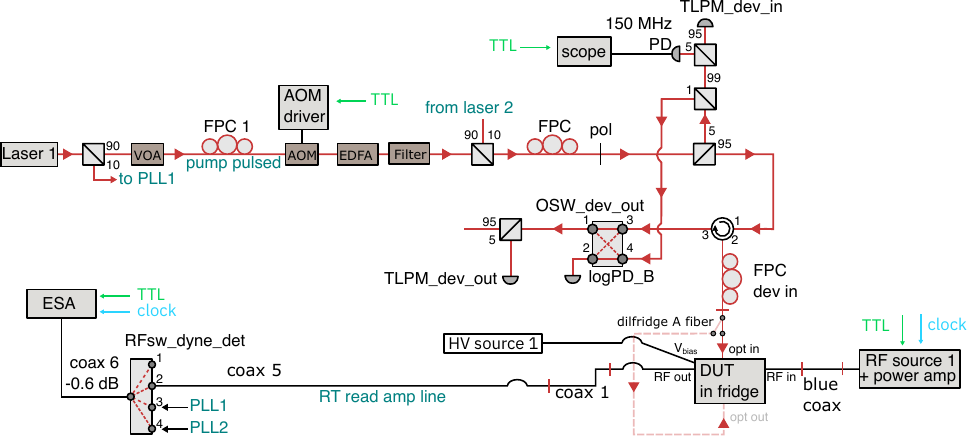}
	\caption[Pulsed microwave characterization setup]{\textbf{Pulsed microwave characterization setup.}
	}
	\label{fig:MW_pulsed_setup}
\end{figure*}

%\subsection{Optical characterizaiton}
\subsection{Continuous wave transduction}

\begin{figure*}[tb]
	\centering
	\includegraphics{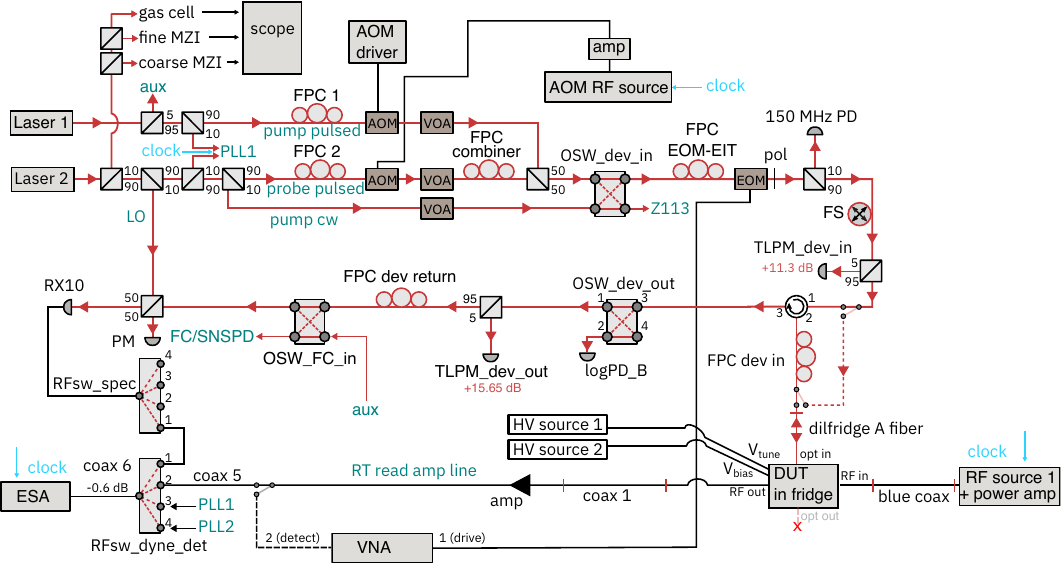}
	\caption[Continuous wave transduction setup]{\textbf{Continuous wave transduction setup.}
	}
	\label{fig:CW_setup}
\end{figure*}

The setup for continuous wave and bidirectional transduction measurements is shown in Fig.~\ref{fig:CW_setup}. Two continuously tunable external-cavity diode lasers (Laser 1 and 2) are used for optical pumping and probing.
The main and pump laser in the cw experiment is Laser 2.
A portion of the light from Laser 2 is sent to the frequency referencing path, where a coarse Mach-Zehnder interferometer (MZI) with a free spectral range of about $\SI{2.5}{\giga\hertz}$ is used to monitor laser detuning sweeps during transduction measurements.
Another fraction of the light is sent as local oscillator (LO) to the heterodyne detection setup.
Next, a fraction of light is combined with that of the auxiliary laser Laser 1 to generate the optical beating to offset-lock Laser 1 with a defined frequency offset to Laser 2 using a phase-locked loop (PLL).
Here, only the pulsed laser paths were used.
Both lasers are polarization controlled using a fiber polarization controller (FPC) to maximize the transmitted power through the acousto-optic modulators (AOM) while minimizing laser frequency dependent power variations, and then sent to variable optical attenuators (VOA) to adjust their power levels.
Both lasers are combined on a 50/50 splitter before which Laser 2's polarization is again optimized using an FPC to align both laser polarizations.
After passing through an optical switch (OSW\_dev\_in), the polarization of the light is optimized to maximize the transmission through the electro-optic modulator (EOM).
A polarization filter (pol) behind the EOM ensures equal polarization for both lasers.
The $\SI{150}{\mega\hertz}$ photodiode (PD) and fiber-stretcher (FS) were not actively used in this experiment.
A fraction of the light is sent to a power meter (TLPM\_dev\_in) to monitor the power sent to the cryostat.
The light passes through a circulator and is sent to the device via a final FPC.
Light reflected from the cryostat fiber is sent to switch OSW\_dev\_out to either monitor the device reflection on the logarithmic photodiode (logPD\_B) or to the heterodyne detection path.
For calibration purposes, the laser signal can be sent into FPC dev in to directly be sent to the detection path.
In the heterodyne detection path, the reflected power is monitored with power meter TLPM\_dev\_out and the polarization optimized using FPC dev return to optimize the overlap with the local oscillator.
The light is then combined with the local oscillator and detected on the $\unit{\giga\hertz}$ photodiode RX10.

Electrical input signals are generated using a radiofrequency source (RF source 1) and the heterodyne transduction signal is detected on an electronic spectrum analyzer (ESA).
Optical input signals generated either as a single sideband from offset-locking Laser 2 to Laser 1 (PLL), or by modulating the EOM using the vector network analyzer (VNA).
The electrical signals from the cryostat are sent through the read out amplification line (RT read amp line) and detected either on the ESA or the VNA.
\par
Note, that the phase-lock (PLL1), Laser 2's AOM source, RF source 1 and the ESA are phase locked to a $\SI{10}{\mega\hertz}$ reference clock.
This allows the lowest possible noise during electrical detection with the ESA.
Since we only had one referenced AOM source, the fixed-frequency AOM source for Laser 1 was kept free-running at a systematic offset of a few $\unit{\kilo\hertz}$ from the nominal $\SI{110}{\mega\hertz}$ of both AOMs.
This slight offset is beneficial for the electrical detection of the optical-to-microwave signal on the ESA, as the PLL creates cross-talk which would otherwise be exactly at the detected microwave frequency.
This noise issue does not exist for the microwave-to-optical direction, as the detected heterodyne signal is shifted by the AOM frequency ($\SI{110}{\mega\hertz}$).

\subsection{Pulsed transduction}
\label{sec:pulsed_transduction_experiment}

\begin{figure*}[tb]
	\centering
	\includegraphics{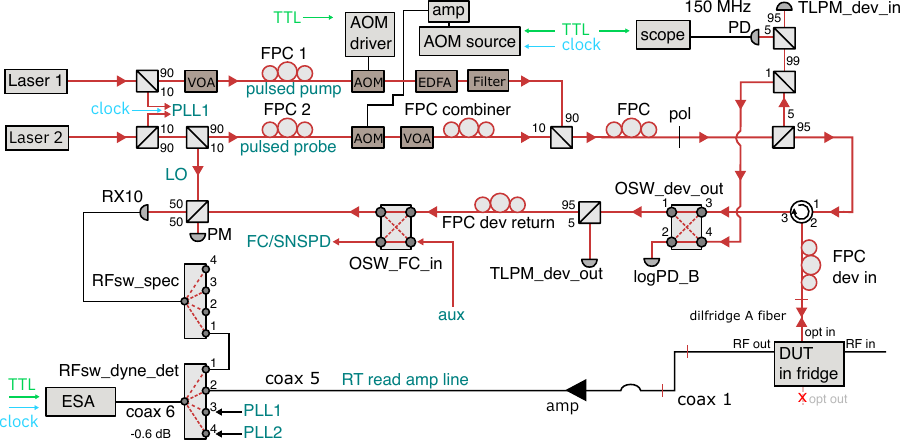}
	\caption[Pulsed transduction setup]{\textbf{Pulsed transduction setup.}
	}
	\label{fig:LT_pulsed_setup}
\end{figure*}

For studying the system in pulsed operation, we only perform optical-to-microwave transduction using two lasers with an offset lock, as the signal is less sensitive to variations in the optical reflection from the device compared to the microwave-to-optical signal.
The setup is shown in Fig.~\ref{fig:LT_pulsed_setup} and is a variation of the continuous wave setup discussed previously.
It differs, however, in certain important points which are mainly related to the generation and monitoring of high power pump pulses.
\par
In this configuration, Laser 2 is the signal laser whose AOM is now pulsed using a transistor-transitor-logic (TTL) signal from a pulse generator.
To achieve large peak powers, the pulsed optical signal of Laser 1, which now acts as the pump, is amplified using an erbium-doped fiber amplifier (EDFA) whose amplified spontaneous emission noise (ASE) is filtered using a tunable filter.
Both laser are then combined, but now using a 90/10 splitter to maximize the pump pulse power.
The combined lasers are again sent through a polarization controller and filter.
\par
Due to the large power range that needs to be covered, the power monitoring had to be modified.
To that end, a portion of the light is sent to a pulse monitoring branch, where the standard power meter measures the average power (TLPM\_dev\_in), while a fast photodiode (bandwidth $\SI{150}{\mega\hertz}$) is used to monitor the optical pulse shape on an oscilloscope, representing the sum of both lasers.
Another fraction of the light is sent to the heterodyne detection arm, which we now use to monitor the power ratio of the optical pulses.
Due to the frequency offset between the two, this detection method also enables recording the pulse shape of each individual pulse with superior signal-to-noise compared to the sum monitoring diode.
\par
Using an electrical switch (RFsw\_dyne\_det) either the optical pulse shapes and ratios are monitored using the heterodyne signal, or the transduction signal from the room temperature (RT) read-out amplification line is measured using an electronic spectrum analyzer (ESA).

\section{Optical and dc characterization}
\label{sec:trandsuction_opt_charact}

\subsection{Room temperature transmission before cooldown}
\label{sec:opt_RT_before}

\begin{figure}[b]
	\centering
	\includegraphics{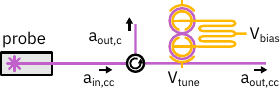}
	\caption[Optical characterization measurement schematic]{\textbf{Optical characterization measurement schematic.}
		Light is injected via the counter clockwise input mode $a_\mathrm{in,cc}$ and the transmission to $a_\mathrm{out,cc}$ or reflection to $a_\mathrm{out,c}$ is measured.
	}
	\label{fig:opt_charact_schematic}
\end{figure}

In this section, the optical response of the device is studied in transmission at room temperature before cooldown, as shown in Fig.~\ref{fig:opt_charact_schematic}.
The transmission coefficient in this case is defined as $T=\left|a_\mathrm{out,cc}/a_\mathrm{in,cc}\right|^2$.
Note, that for all measurements shown in this section no voltage was applied to the device.
\par
\begin{figure}[b]
	\centering
	\includegraphics{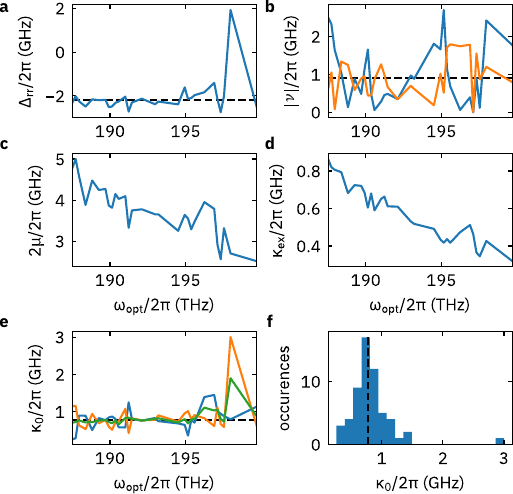}
	\caption[Detuning and loss and coupling rates vs. optical frequency]{\textbf{Detuning and loss and coupling rates vs. optical frequency.}
		(a) Ring-ring detuning and
		(b) back scattering rates as a function of optical frequency.
		(c) Resonant ring-ring and
		(d) external bus coupling rate as a function of optical frequency.
		(e) Intrinsic loss rates as a function of optical frequency (e) and their distribution (f).
		Fit results of transducer (device 2) optical spectrum.
		When more than one trace is shown, blue represents the result for the first ring, orange for the second, and green the average of the two.
		The dashed lines show average ring-ring detuning $\bar{\Delta}_\mathrm{rr}/2\pi=\SI{-2.2}{\giga\hertz}$, intrinsic loss rate $\bar{\kappa}_\mathrm{a,0}/2\pi=\SI{776}{\mega\hertz}$ and back-scattering amplitude $|\nu^{\ast}|/2\pi=\SI{902}{\mega\hertz}$.}
	\label{fig:opt_rates_vs_frequency}
\end{figure}

We apply the photonic molecule with backscattering fit model to each resonance in the spectrum and plot the resulting detuning and rates as a function of optical frequency (Fig.~\ref{fig:opt_rates_vs_frequency}).
Apart from a single outlier at high optical frequencies, the detuning between the rings $\Delta_\mathrm{rr}$ seems to be consistently around $\SI{-2.2}{\giga\hertz}$ (Fig.~\ref{fig:opt_rates_vs_frequency}a).
Note, that no bias or tuning voltage have been applied to the device at that point.
The resonant Rabi splitting $\hbar\Omega=2\mu$ (Fig.~\ref{fig:opt_rates_vs_frequency}b) shows the expected reduction with optical frequency, and reaches almost $\SI{5}{\giga\hertz}$ at the longest wavelengths (lowest frequencies).
The intrinsic loss rates (Fig.~\ref{fig:opt_rates_vs_frequency}c) are scattered around $\SI{776}{\mega\hertz}$, where, as previously mentioned, the average value of the two rings (green trace) seems to be more reliable than the individual ring loss rates.
The forward backward coupling amplitudes (Fig.~\ref{fig:opt_rates_vs_frequency}d) scatter quite a log with no discernible regularity, which is consistent with randomly distributed scattering defects along the ring.
The external bus coupling rate also shows the expected dispersion with more than $\SI{800}{\mega\hertz}$ at the lowest frequencies.
The distribution of intrinsic loss rates is peaked around the mean value indicating a reliable fit with the exception of the outlier at $\SI{3}{\giga\hertz}$ which corresponds to the previously mentioned outlier at large frequencies.

\subsection{Cryogenic temperature optical tuning}
\label{sec:cryo_dc_tune}

\begin{figure}[t]
	\centering
	\includegraphics{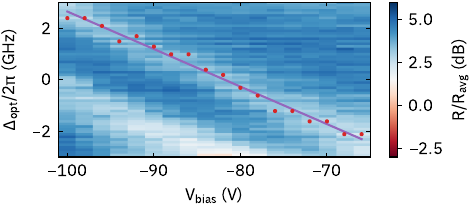}
	\caption[Cryogenic electro-optic tuning efficiency]{\textbf{Cryogenic electro-optic tuning efficiency.}
		Linear fit to the $V_\mathrm{bias}$ tuning dependence of the optical resonance frequency around the standard working point $V_\mathrm{bias}=\SI{-100}{\volt}$.
		The extracted slope is $\partial\omega_\mathrm{a}/\partial V=2\pi\cdot\SI{145}{\mega\hertz\per\volt}$.
	}
	\label{fig:dwdV_mK_fit}
\end{figure}

Next, we extract the resonance frequency $\omega_\mathrm{a}$ of the photonic molecule as a function of bias voltage in the linear range close to the standard working point of $V_\mathrm{bias}=\SI{-100}{\volt}$ (Fig.~\ref{fig:dwdV_mK_fit}).
We then fit a linear function to the trend, and extract an electro optic tuning efficiency of $\partial\omega_\mathrm{a}/\partial V=2\pi\cdot\SI{145}{\mega\hertz\per\volt}$.
This value can later be used to estimate the vacuum coupling strength, as the DC-tunability is typically an upper bound for that in the microwave domain.

\subsection{Room temperature optical transmission after warm-up}
\label{sec:transducer_opt_after_cooldown}

\begin{figure}[b]
	\centering
	\includegraphics{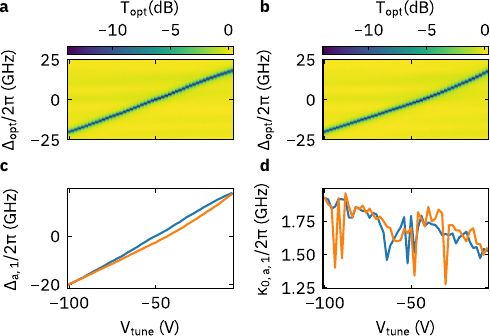}
	\caption[Post-cooldown room temperature tuning hystheresis and loss]{\textbf{Post-cooldown room temperature tuning hystheresis and loss.}
		Optical transmission for a (a) down- and (b) up-sweep of the tuning voltage.
		(c) Optical resonance frequency tuning and (d) intrinsic optical loss rate as a function of tuning voltage for up (orange) and down (blue) sweep.
	}
	\label{fig:opt_DC_hysteresis_loss_RT}
\end{figure}

Since no voltages were applied to the transducer device prior to cooling down the sample, we characterize the sample post-cooldown at room temperature.
Interestingly, the response (Fig.~\ref{fig:opt_DC_hysteresis_loss_RT}) looks vastly different to the expected ferroelectric response of BaTiO$_3$ at room temperature \cite{abel_large_2019}.
While we do not have experimental characterization of this exact device prior to cooldown, other devices fabricated in this platform have shown the expected quadratic-like behavior at room temperature.
Therefore, we suspect that the material was permanently altered compared to before cooldown.
Judging from the fact that the polarity of the tuning is inverted, one may conclude that the ferroelectric domains have a permanent preferential orientation after cooldown.
This may be due to the fact, that the device was usually biased to $V_{bias}=\SI{-100}{\volt}$ during most cryogenic measurements.
While the voltage is brought back to ground before warmup, the domains may still have been oriented and kept their orientation during warmup.

When looking at the optical resonance tuning in the negative branch only (between $0$ and $\SI{-100}{\volt}$), we note that the response is nearly free of hysteresis (Fig.~\ref{fig:opt_DC_hysteresis_loss_RT}c) when comparing the up- and downsweeps.
The dc-tuning efficiency is about $\partial\omega_\mathrm{a}/\partial V=2\pi\cdot\SI{395}{\mega\hertz}$.
Interestingly, the average intrinsic loss rate (averaged over optical frequency) has increased by about a factor of two compared to the initial room temperature characterization prior to application of dc-voltages and before cooldown (Fig.~\ref{fig:opt_rates_vs_frequency}).
Furthermore, there seems to be an increase in optical loss rate as a function of tuning voltage as shown in Fig.~\ref{fig:opt_DC_hysteresis_loss_RT}d.

\section{Microwave characterization}
\label{sec:transduction_mw_charact}

\begin{figure}[b]
	\centering
	\includegraphics{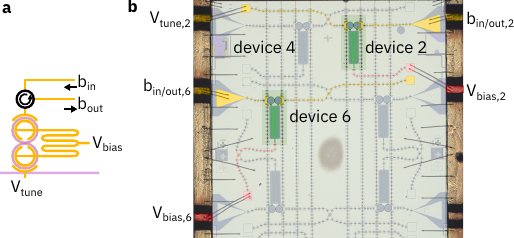}
	\caption[Microwave characterization measurement schematic]{\textbf{Microwave characterization measurement schematic.}
		(a) A microwave tone is injected in $b_\mathrm{in}$, reflected at the device and is sent to $b_\mathrm{out}$ via a circulator.
		(b) Optical micrograph of the two electrically packaged devices (device 2, device 6).
	}
	\label{fig:MW_charact_schematic}
\end{figure}

In this section, the continuous wave microwave response of the device is studied in reflection as shown in Fig.~\ref{fig:MW_charact_schematic}.
The reflection coefficient in this case is defined as $R=\left|b_\mathrm{out}/b_\mathrm{in}\right|^2$.

\subsection{Bias dependence}
\label{sec:cryo_bias_dependence}

\begin{figure}[b]
	\centering
	\includegraphics{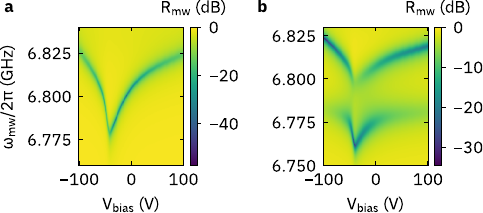}
	\caption[Microwave reflection vs. bias voltage]{\textbf{Microwave reflection vs. bias voltage.}
		(a) Device 6 and (b) device 2 microwave reflection vs. bias voltage $V_\mathrm{bias}$.
	}
	\label{fig:MW_Vbias_maps}
\end{figure}

Analogously to the characterization of the optical response as a function of bias voltage, we now study the microwave reflection with applied bias.
While only one device (device 2) is fully packaged, we wired up a reference device (device 6) for microwave characterization.
The microwave reflection as a function of bias voltage of both devices is shown in Fig.~\ref{fig:MW_Vbias_maps}.
The first thing to note is, that the response shows a significant bias dependence.
Unfortunately, the main transducer device (device 2, Fig.~\ref{fig:MW_Vbias_maps}b) exhibits an anticrossing to an unknown spurious mode at around $\SI{6.79}{\giga\hertz}$.
The observed resonant interaction splitting is $2\gamma=2\pi\cdot\SI{31.9}{\mega\hertz}$.

\begin{figure}[b]
	\centering
	\includegraphics{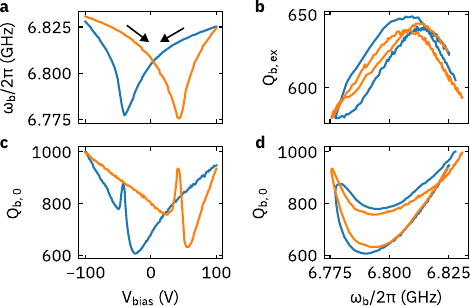}
	\caption[Bias dependence of microwave properties: device 6]{\textbf{Bias dependence of microwave properties: device 6.}
		(a) Microwave resonance frequency $\omega_\mathrm{b}$ and (b) intrinsic quality factor $Q_\mathrm{b,0}$ as a function of bias voltage $V_\mathrm{bias}$.
		(c) External quality factor $Q_\mathrm{b,ex}$ and (d) intrinsic quality factor $Q_\mathrm{b,0}$ as a function of microwave resonance frequency $\omega_\mathrm{b}$.
	}
	\label{fig:MW_Vbias_dev6}
\end{figure}

For characterization of general microwave parameters, we initially focus on device 6 (Fig.~\ref{fig:MW_Vbias_maps}a) for a simplified model description with less error sources and fit parameters than the actual transducer device with spurious mode coupling requires.
To that end, we take a full hysteresis curve and fit a single resonance model (NLQFIT7) to the data using the python package scikit-rf \cite{gregory_q-factor_2022}.
Fig.~\ref{fig:MW_Vbias_dev6}a shows the bias dependence of the microwave resonance frequency, revealing a almost perfectly symmetric hysteresis curve as expected for a ferroelectric material.
Note that, apart from the asymmetry and noise, the microwave hysteresis of device 6 shows a qualitatively similar shape to that of device 2 (\ref{fig2}h).
Importantly, the coercive field values align well.
The bias dependence can most likely be attributed to a bias dependent microwave permittivity of the material, making it a nonlinear capacitor.
Interestingly, the intrinsic quality factor also depends on the applied bias (Fig.~\ref{fig:MW_Vbias_dev6}b).
Generally, bias voltages beyond the coercive field seem to lead to higher quality factors, with the minimal quality factor reached for values close to the coercive field.
Right at the coercive field however, a local peak in the quality factor is observed.
While this behavior may be explained by a bias dependent material absorption of microwave radiation, it may also stem from piezoelectric actuation, since it is known that most electro-optic materials also feature a non-zero piezo-electric effect.
Right at the coercive field, one would expect the ferroelectric domains to have a net vanishing polarization, which would explain the local peaks of the microwave quality factor.
The fact that the largest microwave loss is observed close to the net zero polarization point suggests, that the piezoelectric coupling may be largest there.
For completeness, the external microwave coupling quality factor is plotted versus frequency (Fig.~\ref{fig:MW_Vbias_dev6}c) and shows a slight frequency dependence, which may stem from reflections in the bus line.

\begin{table}
	\begin{tabular}{c|c}
		\hline
		$\omega_\mathrm{b,2}/2\pi$&6.792 GHz\\
		$\kappa_\mathrm{b2,0}/2\pi$&11.4 MHz\\
		$\gamma/2\pi$&15.95 MHz\\
		ref. level&$|S|^2$@6.835 GHz, -100 $V_\mathrm{bias}$\\
		\hline
		$\kappa_\mathrm{b,ex}/2\pi$&7 MHz\\
		\hline
	\end{tabular}
	\caption[Effective dc-tuning model parameters]{\textbf{Effective dc-tuning model parameters.}
		The parameters listed here were fixed when fitting the reflection of device 2.
		$\kappa_\mathrm{b,ex}$ was only fixed for experiments where the resonance frequecy did not shift significantly, that is for microwave power dependence and optical power dependence measurements.
		For the temperature and bias dependence it is a free fit parameter.
	}
	\label{tab:coupled_MW_model_fixed_parameters}
\end{table}

In order to extract the bias dependence of the bare transducer resonance frequency of device 2, we use the coupled microwave mode model derived in Appendix \ref{sec:coupled_MW_mode_model}.
Since a fit of the complex S-parameter response is involved, we implement a fit of the magnitude $|S|$.
To improve the fit stability, we fix certain parameters as listed in Table \ref{tab:coupled_MW_model_fixed_parameters}.

\subsection{Microwave power dependence}

\begin{figure}[b]
	\centering
	\includegraphics{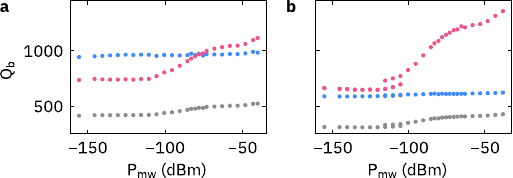}
	\caption[Microwave quality factor vs. microwave power]{\textbf{Microwave quality factor vs. microwave power.}
		(a,b) External (blue), loaded (grey) and intrinsic microwave quality factor for device 2 and 6, respectively, as a function of microwave power $P_\mathrm{mw}$.
		The bias voltage for both measurements is $V_\mathrm{bias}=\SI{-100}{\volt}$.
	}
	\label{fig:MW_power_dependence}
\end{figure}

So far, we have been probing the microwave system in the medium to high power regime.
Typically however, the signal levels in transducers for quantum transduction are just a few or even single photons.
In this low power regime, absorption from two-level systems may become a significant loss channel, as the circulating power is no longer sufficient to saturate the two-level systems in the dielectric for example.
For this reason, we probe the microwave reflection of device as a function of incident microwave power.
By fitting the microwave response of device 2 and 6 as a function of microwave probe power, again using the single resonance model NLQFIT7 from the scikit-rf package \cite{gregory_q-factor_2022}, we can extract the external and intrinsic quality factors $Q_\mathrm{b,ex}$ and $Q_\mathrm{b,0}$, respectively.
As expected, the external coupling quality factor is independent of microwave power.
Interestingly, it is different between both devices, with $960$ and $605$ for device 2 and 6, respectively.
This potentially be attributed to the fact, that the transmission lines from the wire bonding pads to the devices have differnt lengths (see Fig.~\ref{fig:MW_charact_schematic}b).
While device 6 has a very short line without optical crossings, the transmission line of device 2 has two optical crossings and is longer.
The low power quality factors are $740$ and $650$ for device 2 and 6, respectively.

\subsection{Temperature dependence}
\label{sec:transduction_mw_temp_dependence}

In this section, we sweep the temperature of the cryostat ("fridge") by either evaporating the helium cooling mixture (warm up) or recondensing it (cool down) while simultaneously measuring the microwave reflection at high probe power.
Note, that this means that both the bottom probe and mixing chamber temperature are swept simultaneously, which we call $T_\mathrm{fridge}$.
\par
In a first analysis step, we extract the observed center frequency of the reflection dip and its extinction as a function of temperature for both devices.
While for device 6 the observed reflection dip frequency corresponds to the actual bare resonator frequency ($\omega^\prime_\mathrm{b}=\omega_\mathrm{b}$), the observed frequency of device 2 is shifted due to coupling to the spurious mode.
We therefore need to fit the response with the two-mode model to extract the bare frequency.

\begin{figure}[b]
	\centering
	\includegraphics{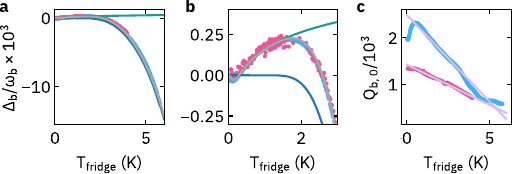}
	\caption[Microwave resonance shift and quality factor vs. temperature]{\textbf{Microwave resonance shift and quality factor vs. temperature.}
		(a) Full and (b) reduced temperature range plot of relative resonance frequency shift $\Delta_\mathrm{b}/\omega_\mathrm{b}$ and intrinsic microwave quality factor $Q_\mathrm{b,0}$ as a function of cryostat temperature $T_\mathrm{fridge}$ for (c) device 6 and (d) device 2, respectively.
	}
	\label{fig:MW_data_and_model}
\end{figure}

To further quantify the observed frequency shift and loss rate, we fit the spectra to extract the resonance frequency shift and quality factor change as a function of temperature.
For device 6 we again use the simple resonance fit NLQFIT7 of the sciki-rf python package \cite{gregory_q-factor_2022}.
For device 2 we use the more complex model that takes into account the coupling to the spurious mode (see Appendix \ref{sec:coupled_MW_mode_model}).
The fixed fit parameters are listed in Table \ref{tab:coupled_MW_model_fixed_parameters}.
The resulting relative frequency shift and intrinsic quality factor as a function of temperature are plotted in \ref{fig:MW_data_and_model} in blue and magenta for device 2 and 6, respectively.
As shown in Fig.~\ref{fig:MW_data_and_model}b, the relative frequency shift is almost identical for both devices.
For both devices, the intrinsic quality factors decrease monotonously as a function of temperature for $T_\mathrm{fridge}>\SI{0.6}{\kelvin}$.
Note, that the quality factor for device 6 only depends on the loaded linewidth and observed extinction, and should therefore be accurate.
The additional spurious mode for device 2 introduces additional parameters such as the loss and coupling rate of the spurious mode which cannot be unambiguously separated, meaning that the extracted intrinsic quality factor shown here for device 2 may have a large systematic error.
\par
While the red-shift and decrease of intrinsic quality factor are a typical signature of quasiparticle (QP) generation in the superconductor, the initial blue shift has been attributed to two-level systems (TLS) often present in dielectrics with defects or amorphous composition \cite{barends_contribution_2008,crowley_disentangling_2023}.
The observed relative frequency shift is then described as a sum of both effects \cite{crowley_disentangling_2023}:

\begin{equation}
	\frac{\Delta\omega_\mathrm{b}}{\omega_\mathrm{b}}=\left(\frac{\Delta\omega_\mathrm{b}}{\omega_\mathrm{b}}\right)_\mathrm{TLS}
	+\left(\frac{\Delta\omega_\mathrm{b}}{\omega_\mathrm{b}}\right)_\mathrm{QP}.
	\label{eq:df_rel_total}
\end{equation}

The individual contributions can be modeled as \cite{gao_physics_2008}:

\begin{eqnarray}
	&&\left(\frac{\Delta\omega_\mathrm{b}}{\omega_\mathrm{b}}\right)_\mathrm{TLS}=\frac{1}{\pi Q_\mathrm{b,0,TLS}}
	\nonumber\\
	&&\cdot\Re{\Psi\left(\frac{1}{2}+i\frac{\hbar\omega_\mathrm{b}}{2\pi k_\mathrm{b}T}\right)-\ln(\frac{\hbar\omega_\mathrm{b}}{2\pi k_\mathrm{b}T})},
	\label{eq:df_rel_tls}
\end{eqnarray}
with the complex digamma function $\Psi$, and

\begin{equation}
	\left(\frac{\Delta\omega_\mathrm{b}}{\omega_\mathrm{b}}\right)_\mathrm{QP}=
	-\frac{\alpha}{2}\left(\frac{\sin(\gamma\phi\left(T\right))}{\sin(\gamma\pi)}\left|\frac{\sigma\left(T\right)/2}{\sigma\left(0\right)}\right|^\gamma-1\right),
	\label{eq:df_rel_qp}
\end{equation}
with the kinetic inductance fraction $\alpha=L_\mathrm{k}/\left(L_\mathrm{g}+L_\mathrm{k}\right)$ and the kinetic and geometric inductance of the circuit $L_\mathrm{k}$ and $L_\mathrm{g}$, respectively.
The kinetic inductance of a thin superconducting film of length $l$ is related to the complex conductivity $\sigma=\sigma_\mathrm{1}-i\sigma_\mathrm{2}$ via $L_\mathrm{k}\approx l/\left(wt\omega_\mathrm{b}\sigma_2\right)$ for $R\rightarrow0$ with the conductor width $w$ and thickness $t$.
For the conductivity, we used the expressions defined in \cite{crowley_disentangling_2023}.

\begin{table}
	\begin{tabular}{ccc}
		\hline
		$Q_\mathrm{b,0,TLS}$&$T_C$&$\alpha$\\
		\hline
		2038&$\SI{8.7}{\kelvin}$&$\SI{7.8}{\percent}$\\
		\hline
	\end{tabular}
	\caption[Thermal microwave frequency shift fit parameters]{\textbf{Thermal microwave frequency shift fit parameters.}
		Fit results for free parameters from equations \ref{eq:df_rel_total}, \ref{eq:df_rel_tls} and \ref{eq:df_rel_qp} fitted to the data shown in Fig.~\ref{fig:MW_data_and_model}a and b.
	}
	\label{tab:MW_thermal_frequency_shift_fit}
\end{table}

We fit the the model to the data of device 2 and plot the TLS, QP and sum of both contributions in green, dark blue and grey, respectively (Fig.~\ref{fig:MW_data_and_model}a,b).
The fit parameters and results are listed in Table \ref{tab:MW_thermal_frequency_shift_fit}.
We note, that the zero-temperature TLS intrinsic quality factor aligns well with the observed quality factor drop in Fig.~\ref{fig:MW_data_and_model}c towards low temperatures, and the critical temperature is also in line with previous experimental studies on the material \cite{rairden_critical_1964}.
\par
Additionally, linear fits of the extracted internal quality factors between $0.6$ and $\SI{6}{\kelvin}$ are shown in Fig.~\ref{fig:MW_data_and_model}c in light purple.

\section{Transduction}
\label{sec:transduction}

\begin{figure}[h!]
	\centering
	\includegraphics{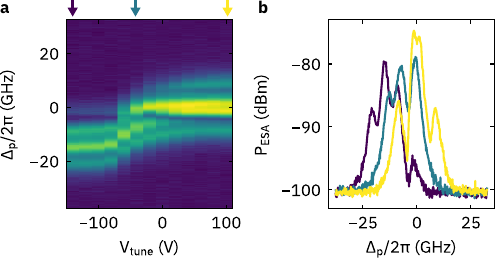}
	\caption[Optical-to-microwave transduction vs. tuning voltage]{\textbf{Optical-to-microwave transduction vs. tuning voltage.}
		(a) Detected microwave-optical signal power as a function of tuning voltage $V_\mathrm{tune}$ and pump detuning $\Delta_\mathrm{p}$.
		(b) Transduction spectra for selected tuning voltages as marked with the colored arrows in a.
	}
	\label{fig:transduction_bias_dependence}
\end{figure}

While it was not possible to directly measure the optical response of the system during cooldown in this study, we can make use of the transduced signal to characterize the optical system.
To that end, we perform an optical-to-microwave measurement as a function of tuning voltage $V_\mathrm{tune}$, which detunes the two rings with respect to each other.
As shown in Fig.~\ref{fig:transduction_bias_dependence}a, the response changes significantly as a function of tuning voltage.
For $V_\mathrm{tune}=\SI{100}{\volt}$, the maximum detected signal is highest (Fig.~\ref{fig:transduction_bias_dependence}c) and the doubly-resonant sattelite peaks are of similar height (Fig.~\ref{fig:transduction_bias_dependence}b, yellow trace).
This suggests, that this is the case where the rings are resonant with each other.
The slight splitting of the triply resonant central peak can either arise from forward-backward scattering, or from a mismatch between microwave frequency and the optical splitting.
The frequency mismatch is the more likely explanation, as otherwise the doubly-resonant peaks should also show a splitting.
In the detuned case (Fig.~\ref{fig:transduction_bias_dependence}b, purple trace), the maximum signal is lower as expected, but now a total of four peaks are clearly visible, indicating strong forward-backward coupling.
Note, that the forward-backward coupling strength is strongly frequency dependent, and may be lower in the resonant case.

\begin{table}[t]
	\begin{tabular}{cc}
		\hline
		Parameter&Value\\
		\hline
		$\Delta_\mathrm{rr}/2\pi$&0.4 GHz\\
		$\omega_\mathrm{b}/2\pi$&6.82 GHz\\
		$2\mu/2\pi$&4.7 GHz\\
		$|\nu_1|/2\pi$&0.35 GHz\\
		$\arg \nu_1$&0\\
		$|\nu_2|/2\pi$&0.15 GHz\\
		$\arg \nu_2$&0\\
		$\kappa_\mathrm{a,0}/2\pi$&2 GHz\\
		$\kappa_\mathrm{a,ex}/2\pi$&0.85 GHz\\
		$\kappa_\mathrm{b,0}/2\pi$&10 MHz\\
		$\kappa_\mathrm{b,ex}/2\pi$&7 MHz\\
		$g_\mathrm{0}/2\pi$&406 Hz\\
		$P_\mathrm{p}$&$\SI{-10}{\deci\bel{m}}$\\
		$\eta_\mathrm{f2c}$&-5 dB\\
		\hline
	\end{tabular}
	\caption{\textbf{System Parameters for transduction model.}}
	\label{tab:transduction_model_parameters}
\end{table}

\section{Slow optical heating}
\label{sec:slow_heating}

\begin{figure}[h!]
	\centering
	\includegraphics{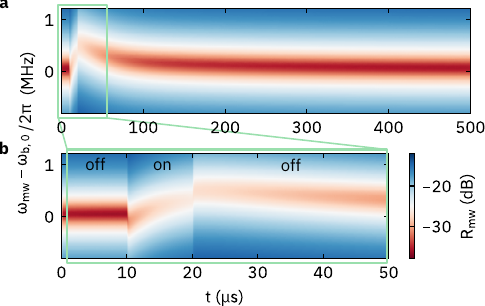}
	\caption[Slow microwave heating under pulsed operation]{\textbf{Slow microwave heating under pulsed operation.}
		(a) Microwave reflection as a function of time and microwave probe frequency $\omega_\mathrm{mw}$ around the pre-pulse dip frequency $\omega_\mathrm{b,0}$.
		(b) Close-up of the first $\SI{50}{\micro\second}$ of a.
	}
	\label{fig:MW_heating_pulsed_slow}
\end{figure}

To characterize the slow thermalization rate in the system, we send a $\SI{10}{\micro\second}$ pulse of off-resonant light with a low repetition rate to the device (Fig.~\ref{fig:MW_heating_pulsed_slow}).
In addition to the fast responding red-shift, which we attribute to quasiparticle generation in the case of off-resonant light, we observe a slow blue-shift and decrease in extinction, indicating heating of the dielectric.
After the pulse, the system recovers to it's pre-pulse state after several tens to hundreds of microsends (Fig.~\ref{fig:MW_heating_pulsed_slow}a), which is significantly more slowly than the on-resonant dielectric and quasiparticle heating.
We attribute this thermalization rate to the thermalization between the sample package and probe with the cryostat mixing chamber stage.

%\clearpage
\FloatBarrier
\bibliography{EOTm_new.bib}

\end{document}